\pdfoutput=1

\documentclass[11pt,twoside,a4paper,cmspaper,final,collab]{cms-tdr}
\def\svnVersion{63855:63952}\def\cmsCernNo{2011-051}\def\cmsCernDate{\today}\def\cmsMessage{Submitted to the Journal of Instrumentation}

\begin{document}\cmsNoteHeader{JME-10-009}

\hyphenation{had-ron-i-za-tion}
\hyphenation{cal-or-i-me-ter}
\hyphenation{de-vices}
\RCS$Revision: 63934 $
\RCS$HeadURL: svn+ssh://alverson@svn.cern.ch/reps/tdr2/papers/JME-10-009/trunk/JME-10-009.tex $
\RCS$Id: JME-10-009.tex 63934 2011-06-24 17:10:59Z alverson $
\newcommand{\RESBOS} {{\textsc{ResBos}}\xspace}
\newcommand{\eslash}{\ensuremath{{\hbox{$E$\kern-0.6em\lower-.05ex\hbox{/}\kern0.10em}}}}
\newcommand{\vecmet}{\mbox{$\vec{\eslash}_{\rm T}$}\xspace} 
\newcommand{\vecet}{\mbox{$\vec{E}_{\rm T}$}\xspace} 
\renewcommand{\MET}{\mbox{$\eslash_{\rm T}$}\xspace}
\newcommand{\met}{\mbox{$\eslash_{\rm T}$}\xspace} 
\newcommand{\mex}{\mbox{$\eslash_{\rm x}$}\xspace} 
\newcommand{\mey}{\mbox{$\eslash_{\rm y}$}\xspace} 
\newcommand{\calomet}{Calo\,\MET}  
\newcommand{\caloIImet}{Calo\,Type\,II\,\MET}  
\newcommand{\tcmet}{TC\,\MET}      
\newcommand{\pfmet}{PF\,\MET}      
\newcommand{\pfsumet}{PF\, \mbox{$\sum E_\text{T}$}\xspace}
\newcommand{\calosumet}{Calo\, \mbox{$\sum E_\text{T}$}\xspace}
\newcommand{\tcsumet}{TC\, \mbox{$\sum E_\text{T}$}\xspace}

\newcommand{\mepar}{\mbox{$\eslash_\parallel$}\xspace}
\newcommand{\meperp}{\mbox{$\eslash_\perp$}\xspace}
\newcommand{\Zmm}{\ensuremath{\cPZ \to \Pgmp\Pgmm}\xspace}
\newcommand{\Zee}{\ensuremath{\cPZ \to \Pep\Pem}\xspace}
\newcommand{\Zll}{\ensuremath{\cPZ \to \ell\ell}\xspace}
\newcommand{\Ztt}{\ensuremath{\cPZ \to \Pgt\Pgt}\xspace}
\newcommand{\Wmn}{\ensuremath{\PW \to \Pgm\cPgn}\xspace}
\newcommand{\Wen}{\ensuremath{\PW \to \Pe\cPgn}\xspace}
\newcommand{\Wtn}{\ensuremath{\PW \to \Pgt\cPgn}\xspace}
\newcommand{\Wln}{\ensuremath{\PW \to \ell\cPgn}\xspace}
\newcommand{\metvec}{\mbox{$\vec{\met}$}\xspace}
\newcommand{\metvecrec}{\mbox{$\vec{\met}^{\rm rec}$}\xspace}
\newcommand{\metvecgen}{\mbox{$\vec{\met}^{\rm gen}$}\xspace}
\newcommand{\metgen}{\mbox{$\met^{\rm gen}$}\xspace}
\newcommand{\metparl}{\mbox{$\mepar^{\rm rec}$}\xspace}
\newcommand{\metperp}{\mbox{$\meperp^{\rm rec}$}\xspace}
\newcommand{\deltamet}{\mbox{$\Delta\met$}\xspace}
\newcommand{\pthat}{\mbox{$\hat{p}_{\rm T}$}\xspace}
\newcommand{\Hslash}{{\hbox{$H$\kern-0.8em\lower-.05ex\hbox{/}\kern0.10em}}}
\newcommand{\MHT}{\mbox{$\Hslash_{\rm T}$}\xspace}
\newcommand{\mht}{\mbox{$\Hslash_{\rm T}$}\xspace}
\newcommand{\sumet}{\mbox{$\sum E_\text{T}$}\xspace}
\newcommand{\scalht}{\mbox{$H_\text{T}$}\xspace}
\newcommand{\etmiss}{\eslash_{\rm T}}
\newcommand{\bigeslash}{{\hbox{$E$\kern-0.38em\lower-.05ex\hbox{/}\kern0.10em}}}
\newcommand{\bigmet}{\mbox{$\bigeslash_{\rm T}$}\xspace}
\newcommand{\bigHslash}{{\hbox{$H$\kern-0.6em\lower-.05ex\hbox{/}\kern0.10em}}}
\newcommand{\bigmht}{\mbox{$\bighslash_{\rm }T$}\xspace}

\newcommand{\MT}{\ensuremath{M_{\mathrm{T}}}\xspace}

\newcommand{\qt}{\ensuremath{{q}_{\rm T}}\xspace}
\newcommand{\ut}{\ensuremath{{u}_{\rm T}}\xspace}
\renewcommand{\pt}{\ensuremath{{p}_{\rm T}}\xspace}

\newcommand{\vqt}{\ensuremath{\vec{q}_{\rm T}}\xspace}
\newcommand{\vut}{\ensuremath{\vec{u}_{\rm T}}\xspace}
\newcommand{\vpt}{\ensuremath{\vec{p}_{\rm T}}\xspace}

\newcommand{\hatqt}{\ensuremath{{\hat q}_{\rm T}}\xspace}
\newcommand{\hatut}{\ensuremath{{\hat u}_{\rm T}}\xspace}
\newcommand{\hatpt}{\ensuremath{{\hat p}_{\rm T}^\ell}\xspace}

\newcommand{\bisec}{\ensuremath{{\hat b}}\xspace}
\newcommand{\upar}{\ensuremath{u_\Vert}\xspace}
\newcommand{\uperp}{\ensuremath{u_\perp}\xspace}

\newcommand{\varet}{\ensuremath{\varepsilon}\xspace}
\newcommand{\varepar}{\ensuremath{\varepsilon_\Vert}\xspace}
\newcommand{\vet}{\ensuremath{\vec{\varet}}\xspace}
\newcommand{\veslt}{\ensuremath{\kern+0.1ex \vec{\not \kern-0.65ex E}_T}\xspace}
\newcommand{\vareslt}{\ensuremath{\kern+0.1ex \vec{\not \kern-0.35ex  \varepsilon}_\
T}\xspace}
\newcommand{\eslt}{\ensuremath{\kern+0.1ex     {\not \kern-0.65ex E}_T}\xspace}
\newcommand{\ersatz}{\ensuremath{\kern+0.1ex \vec{\not \kern-0.65ex E^*} _T}\xspace\
}

\newcommand{\pell}{\ensuremath{\vec{p}_\ell}\xspace}
\newcommand{\pnu}{\ensuremath{\vec{p}_\nu}\xspace}
\newcommand{\pellpl}{\ensuremath{\vec{p}_{\ell^+}}\xspace}
\newcommand{\pellmi}{\ensuremath{\vec{p}_{\ell^-}}\xspace}
\newcommand{\Dphi}{\ensuremath{\Delta\phi}\xspace}
\newcommand{\uparscale}{\ensuremath{\langle u_\Vert \rangle / \qt}\xspace}

\cmsNoteHeader{JME-10-009} 
\title{Missing transverse energy performance of the CMS detector}

\address[neu]{Northeastern University}
\address[fnal]{Fermilab}
\address[cern]{CERN}
\author[cern]{The CMS Collaboration}

\date{\today}

\abstract{
During 2010 the LHC delivered pp
  collisions with a centre-of-mass energy of 7 TeV. In this paper, the
results of comprehensive studies of missing transverse energy as measured
  by the CMS detector are presented.
The results cover the measurements of
the scale and resolution for missing transverse energy,
and the effects of multiple pp interactions
within the same bunch crossings on the scale and resolution.
Anomalous measurements of
missing transverse energy are studied, and algorithms for their identification are
described.
The performances of several
reconstruction algorithms for calculating missing transverse energy are compared.
An algorithm, called missing-transverse-energy significance,
which estimates the compatibility of the reconstructed missing
transverse energy with zero,
is described, and its performance is demonstrated.
}

\hypersetup{%
pdfauthor={CMS Collaboration},%
pdftitle={Missing transverse energy performance of the CMS detector},%
pdfsubject={CMS},%
pdfkeywords={CMS,  missing energy, calorimetry}}

\maketitle 

\section{Introduction}
Neutral weakly interacting particles, such as neutrinos, escape
from typical collider detectors without producing any direct response in the detector
elements. 
The presence
of such particles must be inferred from the
imbalance of total momentum.  The vector momentum imbalance in the plane
perpendicular to the beam direction is particularly useful in pp and
${\rm p\bar p}$ colliders, and is known as missing transverse momentum, here denoted
$\vecmet$.  Its magnitude is called
missing transverse energy, and is denoted \met.  

Missing transverse energy is one of the most important observables for discriminating
leptonic decays of W bosons and top quarks
from background events which do not contain neutrinos, such as multijet
and Drell--Yan events.
It is also an important variable in searches for new 
weakly interacting, long-lived particles.
Many beyond-the-standard-model scenarios, including supersymmetry,
predict events with large \met.
The reconstruction of \vecmet\
is very sensitive to particle momentum mismeasurements, particle
misidentification,
detector malfunctions,
particles impinging on poorly instrumented regions of the detector,
cosmic-ray particles, and beam-halo particles,
which may result in artificial \met.

In this paper, we present studies of \vecmet\ as measured using the 
Compact Muon Solenoid (CMS)
detector at the Large Hadron Collider (LHC), 
based on a data sample corresponding to an integrated luminosity of 36\pbinv.
In Section~\ref{sec:detector}, the CMS detector is briefly described.
In Section~\ref{sec:datas},
particle reconstruction algorithms and identification requirements,
together with the basic sample selection requirements, are given.
In Section~\ref{sec:algorithms}, the different algorithms for 
evaluating \vecmet\ are presented.  
In Section~\ref{sec:tails}, methods for identifying
anomalous \vecmet\ mismeasurements
from known detector artifacts are described.
In Section~\ref{sec:resolution}, the scale and resolution are measured
using events containing photons 
or Z bosons.  The degradation of the resolution due
to the presence of additional
soft pp collisions in the same crossing as the hard scatter (``pile-up'')
is presented.
In Section~\ref{sec:gmet} we present distributions from physics processes containing
genuine \met.
In Section~\ref{sec:performance},
an algorithm, called ``\met\ significance'', which is the
likelihood that the observed \vecmet\ is due to resolution effects,
is described, its performance in jet events is demonstrated, and its efficacy for
separating events containing a  ${\rm W}$ boson decaying to either an electron
and a neutrino or a muon and a neutrino
from multijet backgrounds is shown.
Conclusions are given in Section~\ref{sec:conclusions}.
Finally, in the Appendix,
the optimization of the parameters
used in the correction for the detector response is described.

\section{The CMS detector}
\label{sec:detector}

The central feature of the CMS apparatus is a 
superconducting solenoid, of 6~m internal diameter, providing a field of 3.8~T. 
Within the field volume are the silicon pixel and strip tracker, the crystal 
electromagnetic calorimeter (ECAL), and the brass/scintillator hadron 
calorimeter (HCAL). Muons are measured in gas-ionization detectors 
embedded in the steel return yoke. In addition to the barrel and endcap 
detectors, CMS has extensive forward calorimetry.

CMS uses a right-handed coordinate system, with the origin at the nominal 
interaction point, the $x$-axis pointing to the centre of the LHC, the 
$y$-axis pointing up (perpendicular to the LHC plane), and the $z$-axis 
along the anticlockwise-beam direction. The polar angle, $\theta$, is 
measured from the positive $z$-axis and the azimuthal angle, $\phi$, is 
measured in the $x$-$y$ plane relative to the $x$-axis. 
Transverse quantities, such as ``transverse
momentum'' ($\vec{p}_{\rm T}$), refer to the components in the $x-y$ plane.  The magnitude of $\vec{p}_{\rm T}$ is \pt. 
Transverse energy, $E_{\rm T}$, is defined as
$E \sin{\theta}$. 

The electromagnetic calorimeter (ECAL)
consists of nearly 76\,000 lead tungstate crystals, which provide coverage in pseudorapidity $\vert \eta \vert< 1.479 $ in a barrel region (EB) and $1.479 <\vert \eta \vert < 3.0$ in two endcap regions (EE). A preshower detector consisting of two planes of silicon sensors interleaved with a total of $3X_0$ of lead is located in front of the EE. 
The ECAL 
has an energy resolution of better than 0.5\,\% for unconverted photons with $E_{\rm T}>$
100\,GeV.

The HCAL is comprised of four subdetectors, a barrel detector (HB) covering 
$|\eta|<1.3$, two endcap detectors (HE) covering $1.3<|\eta|<3.0$, two
forward detectors (HF) covering $2.8<|\eta|<5.0$, and a detector outside of the solenoid (HO)
covering $|\eta|<1.3$.
The HCAL, when combined with the ECAL, measures hadrons with a resolution 
$\Delta E/E \approx 100\%\sqrt{E\,[\mbox{GeV}]} \oplus 5\,\%$. 
In the region $\vert \eta \vert< 1.74$, the HCAL cells have 
widths of 0.087 in pseudorapidity and 0.087\,rad in 
azimuth. In the $(\eta,\phi)$ plane, 
and for $\vert \eta \vert< 1.48$, the HCAL cells map onto 
$5 \times 5$ ECAL crystal arrays to form calorimeter towers 
projecting radially outwards from close to the nominal 
interaction point. At larger values of $\vert \eta \vert$, 
the size of the towers increases and the matching ECAL 
arrays contain fewer crystals. 

The muons are measured in the pseudorapidity window $|\eta|< 2.4$, with detection planes made of three technologies: Drift Tubes, Cathode Strip Chambers, and Resistive Plate Chambers. 
A global fit of the measurements from the muon system and the central tracker
results in a \pt resolution between 1 and 5\%, for \pt values up to 1~TeV.

The inner tracker measures charged particles within the $|\eta| < 2.5$ pseudorapidity range. It consists of 1440 silicon pixel and 15\,148 silicon strip detector modules and is located in the 3.8~T field of the superconducting solenoid. It provides an impact parameter resolution of $\sim$\,15~$\mu$m and a 
$p_{\rm T}$ resolution of about 1.5\% for 100~GeV particles.

The first level (L1) of the CMS trigger system, composed of custom hardware processors, uses 
information from the calorimeters and muon detectors to select, in less than 1~$\mu$s, the 
most interesting events. The High Level Trigger (HLT) processor farm further 
decreases the event rate from around 100~kHz to $\sim$\,300\unit{Hz}, before data storage.

The calibrations used in this analysis were those available at
the beginning of the fall of 2010, which are not the final calibrations
for this data sample.  Improvements and updates to the calibrations
can lead to small improvements in resolution.

A much more detailed description of CMS can be found elsewhere~\cite{PTDRI}.

\section{Data sample selection and particle reconstruction}
\label{sec:datas}

The data sets used for the studies presented in this paper were collected
from March through November, 2010, and consist of
pp collisions at a centre-of-mass energy $\sqrt{s}=7$~TeV.
An integrated luminosity with all
subdetectors certified as fully functional of 36 \pbinv was available.
The detailed selection criteria for the individual data samples used for
each study are given throughout the text.
However, all require at least one well-identified primary vertex (PV) whose
$z$ position is less than 24 cm away from the nominal centre of the detector
and whose transverse distance from the $z$-axis is less than 2 cm,
ensuring that particles coming from collisions are well contained
in the CMS detector.

The samples used for the studies in this paper are defined through
selection requirements on reconstructed jets, electrons, photons, muons, and b jets.
We describe the basic identification requirements used for these particles here.

Jet reconstruction and its performance in CMS are discussed in
detail elsewhere~\cite{JME-10-003}.
For the analyses described in this paper, jets are reconstructed
using an anti-${\rm k}_{\rm T}$ algorithm~\cite{antikt} with a jet
radius parameter R of 0.5.
The energy of a jet is corrected, on average,
to that which would have been obtained
if all particles inside the jet cone at the vertex
were measured perfectly (particle-level).
CMS uses three different types of jets.
Calorimeter jets (Calo Jets) are clusters of calorimeter tower energies.
Jet-plus-track jets (JPT Jets) achieve improved response by
supplementing the calorimeter information with tracking
information.
Tracks are associated with Calo Jets if they are within the jet cone at the PV.
The measured momentum of
these tracks is added to the jet.
To avoid double-counting energies, the expected response in the calorimeter
is subtracted from the Calo Jet if the particle is still within the cone
when it impacts the calorimeter.
The response and resolution can be further improved using a global particle-flow
reconstruction.  Details on the CMS particle-flow algorithm and performance can
be found in \cite{PFT-09-001}.
The particle-flow
technique reconstructs a complete, unique list of particles (PF particles)
in each event using an optimized combination of information
from all CMS subdetector systems.
Reconstructed and identified particles include muons,
electrons (with associated bremsstrahlung photons), 
photons (including conversions in the tracker volume), 
and charged and neutral hadrons.
Particle-flow jets (PF Jets) are constructed from PF particles.

Photon candidates are selected from clusters of energy in the ECAL.  They are
required to be isolated.
The ECAL energy  
in an annular region in the $\eta-\phi$ plane with inner radius
0.06 and outer radius 0.4, 
excluding a three-crystal-wide strip 
along $\phi$ from the candidate, is required to be less than 
$4.2+0.006 \, p_{\rm T}^\gamma$~GeV, 
where $p_{\rm T}^\gamma$ is the transverse momentum in GeV
of the photon candidate.
The sum of the $p_{\rm T}$s of the tracks in the same region must be less than 
$2.2+0.0025 \, p_{\rm T}^\gamma$~GeV.
The ratio of the HCAL energy in an annular region with inner radius 0.15 and outer radius
0.4 to the ECAL cluster energy is required to be less than 0.05.
The shape of the 
cluster of calorimeter energies must be consistent with that of an electromagnetic shower.
In addition, to provide strong rejection against misidentification of electrons
as photons, the cluster must not match any track reconstructed in the pixel detector
that is consistent with coming from the primary vertex.  Photon reconstruction and identification are described in detail in \cite{EGM-10-005}.

Electrons are identified using similar criteria.  In addition to similar
shower shape and isolation requirements,
the candidate must match well in both $\phi$ and $\eta$ 
to a charged track, 
but be isolated from
additional tracks. Electron candidates are also required to be in the fiducial portion
of the calorimeter ($|\eta|<1.4442$ or $1.5660<|\eta|<2.5$).
More details are given in  
\cite{EGM-10-004}. 
In addition, photon-conversion rejection is used in some of the analyses
presented in this paper.

Muon candidates consist of a track in the tracker which can be linked to
one reconstructed in the muon system.  
The candidate must be isolated from 
deposits of energy in the ECAL and HCAL that are not consistent with 
having been deposited by the muon. 
The sum of the $p_{\rm T}$s of 
other tracks within an isolation cone centered on the candidate must also be small.

Several algorithms for the identification of b jets have been developed
\cite{SSV,BTV10}. Two of them are used in an analysis
described in this paper.
The \texttt{SimpleSecondaryVertex} (SSV) 
tagging algorithm exploits the significance of the three-dimensional
flight distance between the PV and a 
reconstructed secondary vertex. The \texttt{SoftMuonByPt} (SMbyPt) tagger uses 
the transverse momentum of the muon with respect to the jet axis to construct a discriminant.

The collision data are compared to samples of simulated
events that were generated either using
\PYTHIA 6~\cite{pythia},
with a parameter setting  referred to as tune Z2,
or with \MADGRAPH interfaced with \PYTHIA \cite{MadGraph}.
The generated events are passed through the CMS detector simulation
based on \GEANTfour~\cite{geant4}.
The detector geometry description includes realistic
subsystem conditions such as simulation of nonfunctioning channels.
The samples used in Section~\ref{sec:multint} include pile-up.  
The offline event selection for collision data 
is also applied to simulated events.

\section{Reconstruction of \texorpdfstring{\bigmet}{MET}}
\label{sec:algorithms}

In general, \vecmet is the negative of the vector
sum of the transverse momenta of all
final-state particles reconstructed in the detector.
CMS has developed three distinct algorithms to reconstruct \vecmet\ :
(a) \pfmet, which is calculated using a complete
particle-flow technique~\cite{PFT-09-001};
(b) \calomet, which is based on calorimeter energies and the calorimeter
tower geometry~\cite{METPas07}; and
(c) \tcmet, which corrects \calomet\ by including tracks reconstructed
in the inner tracker after  correcting for the tracks' expected energy
depositions in the calorimeter~\cite{JME-09-010}.

\pfmet is calculated from the reconstructed PF particles.
\pfsumet is the associated scalar sum of the transverse energies of the PF particles.

\calomet is calculated using the energies contained in calorimeter towers and their
direction, relative to the centre of the detector,
to define
pseudo-particles.  The sum excludes energy deposits
below noise thresholds.
Since a
muon deposits only a few GeV on average  in the calorimeter,
independent of its momentum, the muon \pt
is included in the \calomet calculation
while the small calorimetric energy deposit
associated to the muon track is excluded.
\calosumet is the associated scalar sum of the transverse energies of the
calorimeter towers and muons.

\tcmet
is based on \calomet, but also
includes the $p_{\rm T}$s of tracks that have been
reconstructed in the inner tracker, while removing the expected
calorimetric energy deposit  of each track.
The predicted energy deposition for charged pions is used for all
tracks not identified as electrons or muons.
The calorimetric energy deposit is
estimated from simulations of single pions, in intervals of $p_{\rm T}$ and
$\eta$, and an extrapolation of the track in the CMS
magnetic field is used to determine its expected position.
No correction is applied for very high $p_{\rm T}$ tracks
($p_{\rm T}>100~\GeV$), whose energy is already well measured  by the
calorimeters. For low-$p_{\rm T}$ tracks ($p_{\rm T}<2~\GeV$), the measured
momentum is taken into account
assuming
no response from the calorimeter.

The magnitude of the \vecmet
can be underestimated for a
variety of reasons,
including
the nonlinearity of the response of the calorimeter
for neutral and charged hadrons due to its noncompensating nature,
neutrinos from semileptonic decays of particles,
minimum energy thresholds in the calorimeters, \pt thresholds and inefficiencies
in the tracker, and, for \calomet, charged particles that are bent
by the strong magnetic field of the CMS solenoid and whose
calorimetric energies are therefore in a calorimeter cell whose associated
angle
is very different from the angle of the track at the vertex.
The displacement of charged particles with small \pt due to
the magnetic field and the
calorimeter nonlinearity are the largest of these biases,
and thus \calomet\ is affected most.
A two-step correction has been devised in order to remove the bias in
the \vecmet\ scale.  The correction procedure relies on the fact that
\vecmet can be factorized into contributions from jets,
isolated high \pt photons, isolated high \pt electrons, muons, and unclustered energies.
The contribution due to unclustered energies is the difference between the \vecmet
and the negative of the vector sum of the $p_{\rm T}$s of the other objects.
Isolated photons, electrons, and muons are assumed to require no scale corrections.

Jets can be corrected to the particle
level using the jet energy correction~\cite{JME-10-003}.
The ``type-I corrections'' for
$\vecmet$ use these
jet energy scale corrections for
all jets that have less than $0.9$ of their energy in the ECAL
and corrected \pt$>20$~GeV for \calomet,
and for a user-defined selection of jets with \pt$>10$~GeV for \pfmet.
These corrections can be up to a factor of two for \calomet\
but are less than 1.4 for \pfmet \cite{JME-10-014} .
In order to correct the remaining soft jets below this threshold,
and energy deposits not clustered in any jet, a
second correction can be applied to the unclustered energy, which is
referred to as the ``type-II correction''.  This correction is obtained from
$\cPZ\to\Pe\Pe$ events, as discussed in the Appendix.

In this paper, distributions involving \calomet\ include both type-I and type-II corrections,
those involving \pfmet\ include type-I corrections, and those involving
\tcmet\ are uncorrected, as these were the corrections that were available at the time the analyses
presented in this paper were performed and are the versions used most typically
in 2010 physics analyses.
As discussed in the Appendix, type-II corrections have been developed for \pfmet
and can be used in future analyses.
The optimization of both corrections is also discussed in the Appendix.

\section{Large \texorpdfstring{\bigmet}{MET} due to misreconstruction}
\label{sec:tails}
This section describes various instrumental causes of
anomalous \vecmet\ measurements, and the methods used to identify, and sometimes
to correct, \vecmet for these effects.
We also examine the contributions to the tails of
the \met\ distribution from 
non-functioning channels, uninstrumented regions of the detector.
and particles from sources other than pp interactions.

\subsection{Contributions to \texorpdfstring{\bigmet}{MET} from anomalous signals in the calorimeters}

The CMS ECAL and HCAL
occasionally record anomalous signals that correspond to particles
hitting the 
transducers.  Anomalous signals in HCAL can also be produced by rare random discharges
of the readout detectors. Some of these effects had
already been observed during past test beam and cosmic data
taking~\cite{CFT-09-019}.  Detailed studies of these effects have
continued with the 7 TeV data taking, and are
documented in detail in \cite{hcalnoise} for the HCAL, and in
\cite{ecalnoise} for the ECAL.
For some types of anomalous energies, the number of affected channels is
small and the event can still be used in physics analysis after the removal
of the anomaly.  We refer to the removal process as ``cleaning'' the event.
If a large number of channels are effected, ``filters'' instead
tag the event as not suitable for use in physics analysis.

Anomalous energy deposits in EB are associated with particles striking the sensors
and very occasionally interacting to produce secondaries that cause large
anomalous signals through direct ionization of the silicon.
Three main types of noise have been identified in HF:
scintillation light produced in the light guides that carry the light
from the quartz fibres to the photomultipliers, Cherenkov light in the photomultiplier tube (PMT)
windows, and punch-through particles hitting the PMTs.
While the EB, HF scintillation and HF Cherenkov sources typically
affect only a single channel, signals generated in the HF by
particles that exit the back of the calorimeter can affect
clusters of channels per event.
In the HB and HE, electronics noise from the Hybrid Photo
Diode (HPD) and Readout BoX (RBX) occurs, and can
affect from one up to all 72 channels in an RBX. This noise
is not related to interactions with particles from pp interactions but instead
occurs at a low rate and at random times, so
 the overlap with pp interactions is very low at the bunch spacings of the 2010 run.

The basic strategy for the identification and removal of anomalous
signals (cleaning)
is based on information such as unphysical charge sharing between neighbouring
channels in $\eta$-$\phi$ and/or depth, and timing and pulse shape
information. Each of the calorimeters in CMS measures and samples signals
every twenty-five ns and several
samples are saved with the event record.
The shapes of the pulses for signals that
develop from energy deposits in the calorimeters are different
than those from anomalous noise signals.

Once a ``hit'' in an HCAL tower or ECAL crystal is determined to be
unphysical, we exclude it from the
reconstruction of higher-level objects such as jets or \vecmet . We thus arrive at a reconstruction of
jets and \vecmet\ that is consistently ``cleaned'' of anomalous detector effects.
Studies using simulations of a variety of different physics
processes indicate that the amount of energy due to
particles produced in a pp scattering that is removed is negligible.

Some features of anomalous signals can be
used to identify events contaminated by them
most effectively  after higher level objects such as jets have been
reconstructed.  Usually we reject events containing these types of
anomalies using filters
instead of trying to clean them, although some cleaning is available for
\pfmet.
For example, we usually exclude events with HPD or RBX noise affecting many channels
from our data samples
using the requirements described in \cite{CFT-09-019}.
We find that this requirement excludes 
0.003\% of an otherwise good inclusive sample of pp interactions
(minimum-bias events).

\begin{figure}[t!]
  \begin{center}
    \includegraphics[height=0.6\textwidth]{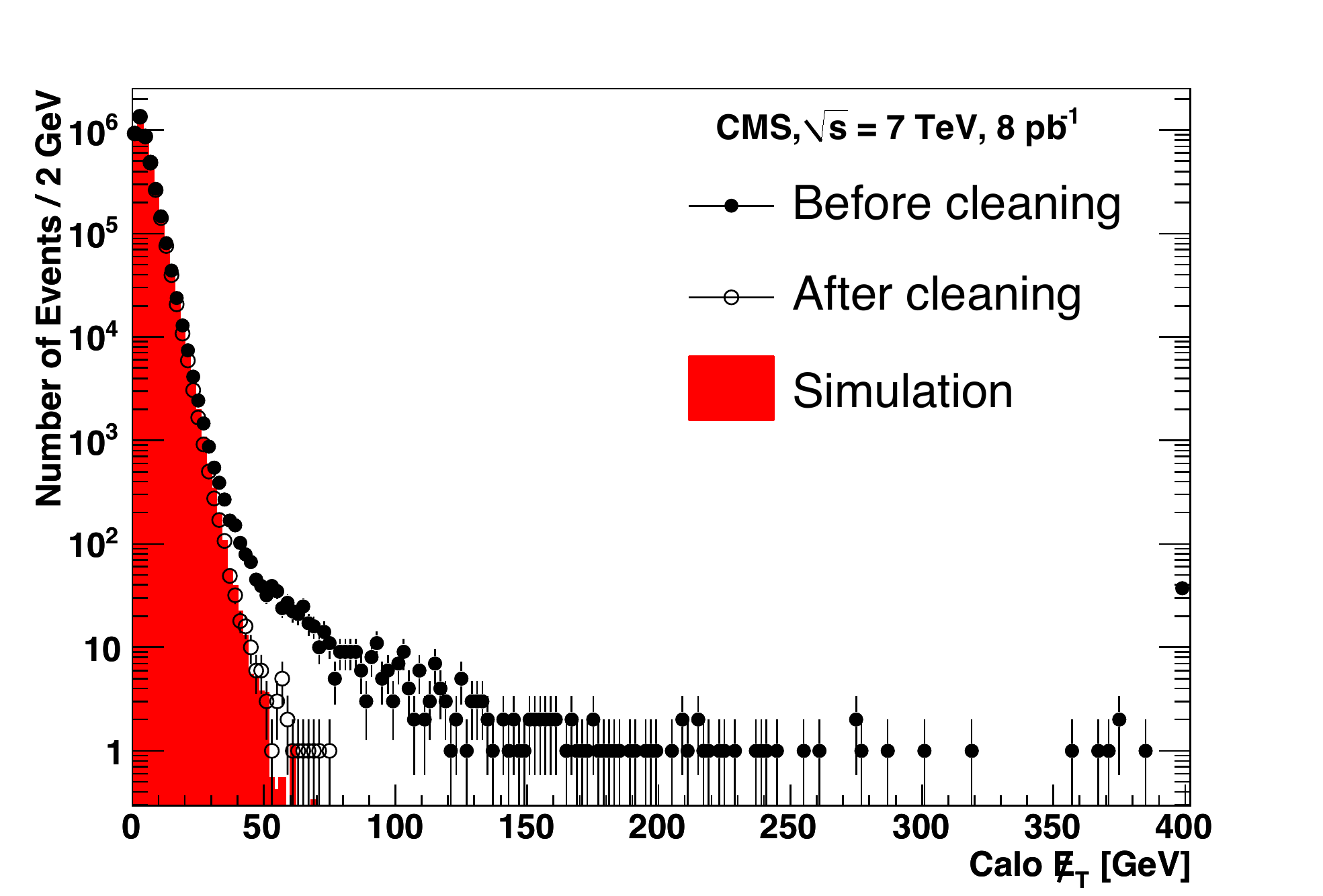}
  \end{center}
 \caption{\calomet\ distributions in a minimum bias data sample without
   (black dots) and with (open circles) cleaning and filters,
   compared to simulation.  Overflows are included in the highest bin.
\label{fig:metAfterCuts}}
\end{figure}

The \calomet\ distribution from a data sample that was collected on a trigger
that requires a coincidence in the beam pick-up monitors and scintillators in
front of the HF calorimeter (minimum bias data)
are shown
before and after removal of the anomalous
signals in
Fig.~\ref{fig:metAfterCuts}, demonstrating the effect of the cleaning
and filters. A
comparison with simulation, which does not include
anomalous energies, shows good agreement.
The effect of the cleaning on the other types of \met is similar.

The minimum bias triggers used to collect the data shown in Fig.~\ref{fig:metAfterCuts}
were prescaled for most of the data-taking period.
Triggers that require large amounts of energy in the detector, such as
\met\ triggers and single jet triggers, are enriched in events with
anomalous energies.
Filters for cosmic rays, other non-collision-related sources
of high \met, and other types of anomalies
have been developed
in the context of specific searches for new particle production~\cite{SUS-10-003}.
An example of a filter for beam-halo muons,
which can produce high energy bremsstrahlung photons in the detector,
is given in the next section.

Hardware modifications to mitigate one of the largest sources of
anomalous energies during the 2010 run, scintillation light produced in
part of the light guide reflective sleeves
in the HF, were implemented during the
winter 2010 shutdown period. During this period,
the material that was producing the scintillation light was replaced with Tyvek.
These modifications reduce the rate of noise
events in HF by an order of magnitude. The HF PMTs will be replaced with
multi-anode PMTs with flat, thinner front glass during 2013/2014 winter shutdown,
reducing the noise from Cherenkov light and punch-through particles. To reduce the noise observed in
HB and HE, HPDs will be replaced by Silicon Photo Multipliers (SiPM)
that do not produce this type of noise.

\subsection{Removal of beam-induced contributions to \texorpdfstring{\bigmet}{MET}}
\label{sec:beamhalo}

Machine-induced backgrounds, especially the production
of muons when beam protons suffer collisions upstream of the detector
(``beam halo''), can cause anomalous, large \met.
The CMS beam-halo event filter
uses trigger and reconstruction-level
information obtained from the Cathode Strip Chambers (CSCs), a subdetector with good
reconstruction performance for both collision and non-collision
muons~\cite{CMS-PAS-MUO-10-002} and can be used to tag events for removal.
The geometry of the CSCs make
it difficult for beam-halo particles, with
mostly parallel-to-beam trajectories, to traverse the barrel
calorimetry without  traversing one or both CSC endcaps.

The filter
can operate in either a ``loose'' or ``tight'' mode.
The former
is designed for high tagging efficiency at the cost of a modest misidentification
probability,
while the
latter tags only well-identified halo candidates and has a smaller misidentification probability.
The tagging efficiencies and misidentification probabilities
have been assessed using simulation.  For
simulated beam-halo particles which impact the calorimeters and produce \calomet $>$ 15 GeV,
the loose (tight) filter is roughly 92\% (65\%) efficient.  The per-event mistag
probability determined from a simulation of inclusive pp interactions (minimum-bias events)
for the loose (tight) filter
is $\sim$10$^{-5}$ (10$^{-7}$). The tagging inefficiency is due in part to
halo muons which do not traverse enough active layers of the CSCs for
a well-measured track to be reconstructed
and in part to muons that do not meet the coincidence requirements of the L1 beam-halo trigger.
Many of the mistagged
events are from extremely soft and forward muons (i.e. $p_{\rm T}<2$ GeV and $|\eta|>1.7$),
from pion decay or from
hadron punch through.

\begin{figure}[b!]
\begin{center}
{\label{fig:Haloa}\includegraphics[height=0.4\textwidth]{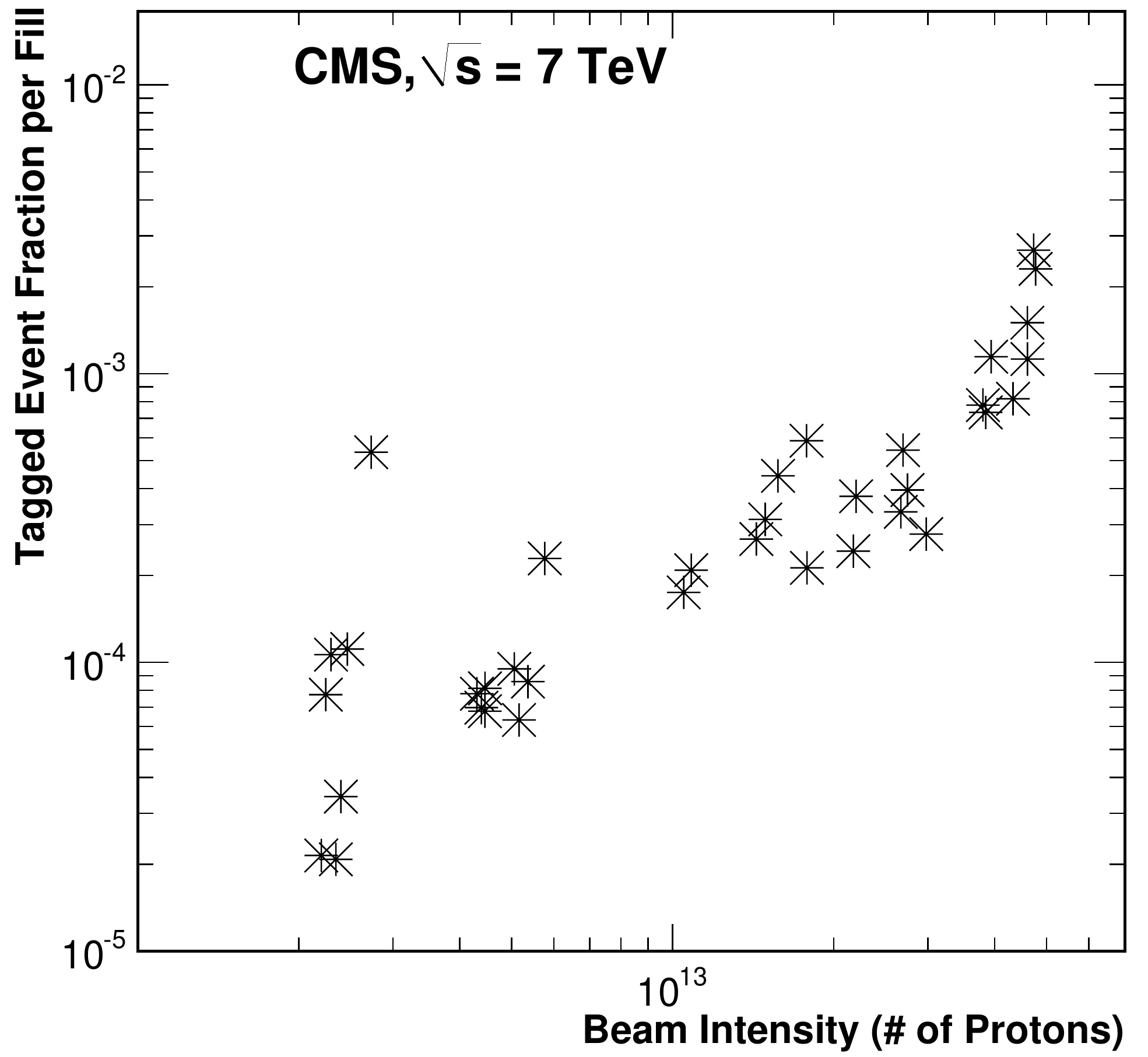}}
{\label{fig:Halob}\includegraphics[height=0.4\textwidth]{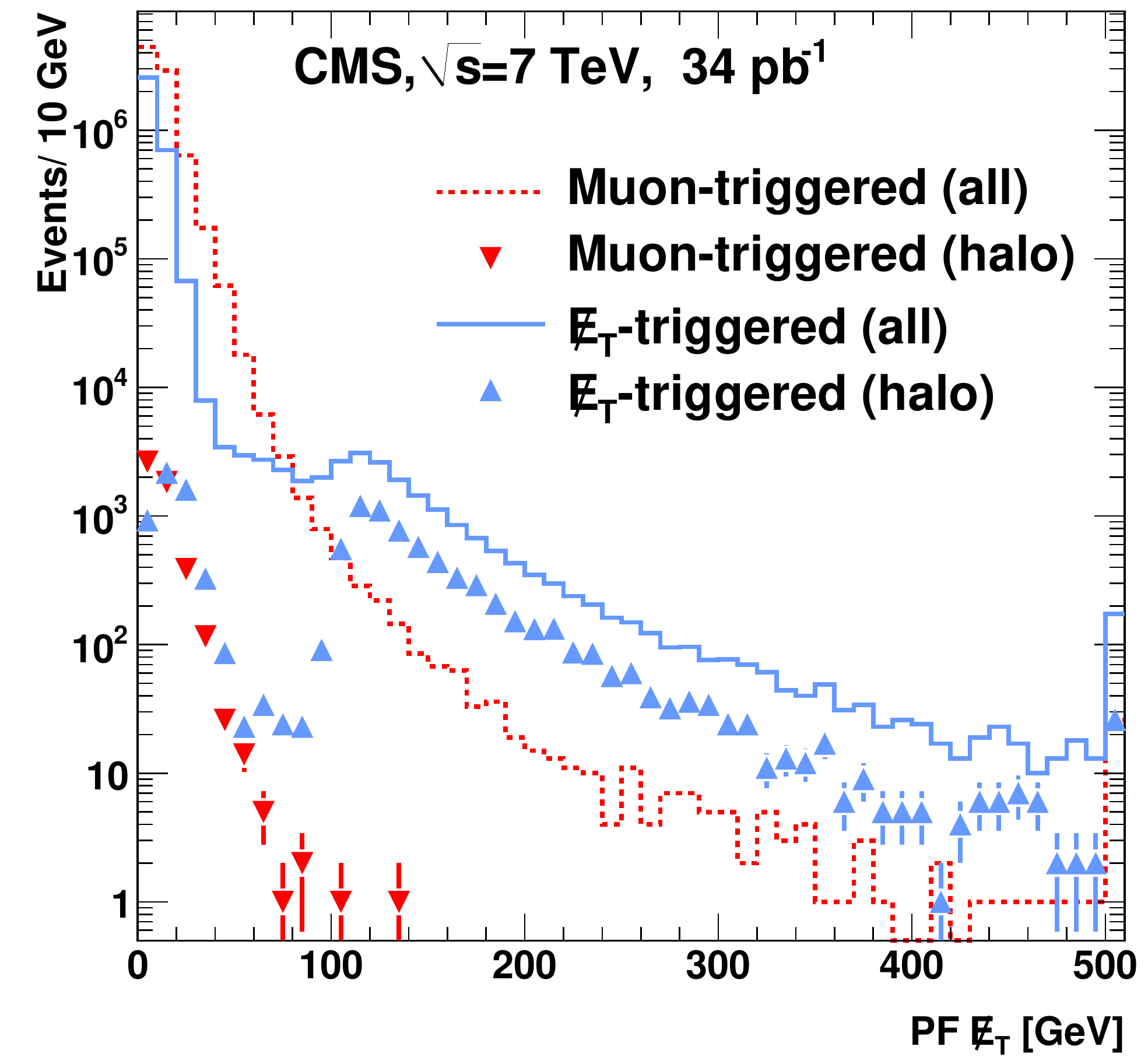}}
\end{center}
\caption{(left) Probability of finding a beam-halo tagged event in muon-triggered events.
 Results are shown as a function
of the beam intensity. (right) \pfmet\ distribution for all the events from muon and \calomet triggers that were analyzed, and for the subset of these events that were identified as beam halo.}
\label{fig:Halo1}
\end{figure}

The CSC-based beam-halo filter was applied to
events passing muon triggers
which had $p_{\rm T}$ thresholds of 9, 11, or 15 GeV, depending on the running period.
Beam-halo muons, because their tracks do not point towards the nominal interaction point
in the centre of the detector, in general do not fire the triggers
for muons from pp interactions.
Thus the beam halo muons in this sample are overlaid on events triggered otherwise.
This sample therefore provides an unbiased comparison of \met in events with and without
a beam-halo muon in coincidence.
Minimum bias events could have been used as well, but, because the minimum bias
trigger was prescaled, the number of available events was small.
The fraction of halo-tagged events for each running period is shown versus
the average beam intensity, with an uncertainty of approximately 10\%,
in Fig.~\ref{fig:Halo1}(left).
The fraction of tagged events increases with the beam intensity, as might be expected.

Figure \ref{fig:Halo1}(right) shows the \pfmet\ distribution for two trigger streams.
The distribution from events recorded
by collision muon triggers is shown by the dashed curve while that of the subset of these events
which met the
requirements of the tight halo filter is shown by the red inverted triangles.  As can be
seen, the halo muons that overlapped with these events did not disproportionately
produce events with large \pfmet,
which indicates that the probability that a halo muon produces large \met\
in events taken from triggers that are uncorrelated with \met is small.
However, events from a trigger on \calomet with a
minimum trigger threshold of 100 GeV (solid curve), show a substantial fraction
identified as halo (blue triangles), since the trigger preferentially selects
events in which the beam-halo muon has deposited large amounts of energy in the calorimeter.  A beam-halo filter is therefore necessary for analyses that make use
of samples based on this trigger.

\subsection{Contributions of non-instrumented or non-functioning detector regions}
\label{sec:cracks}

\begin{figure}[t!]
\begin{center}
{ \includegraphics[scale=0.3]{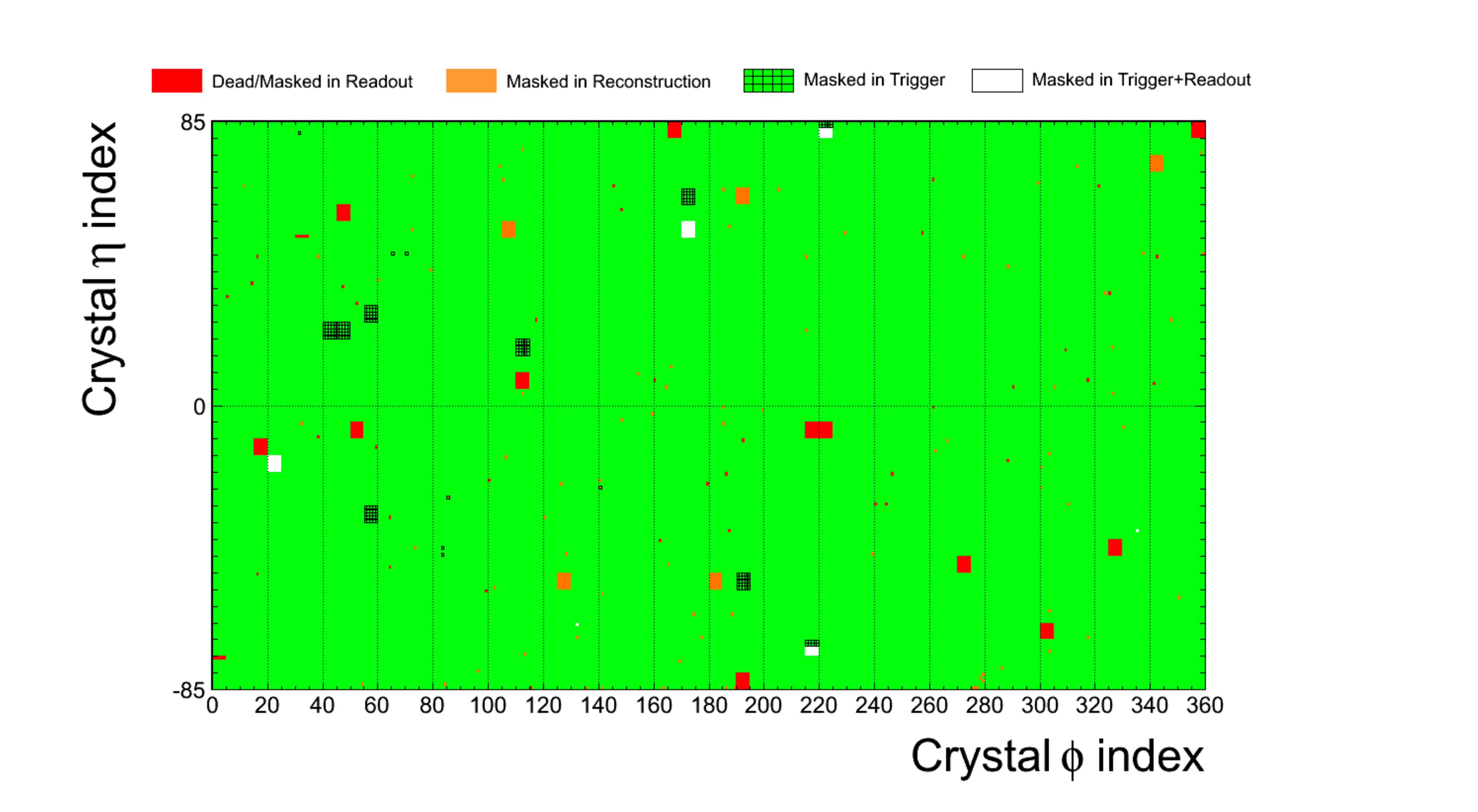} }\\
{ \includegraphics[scale=0.3]{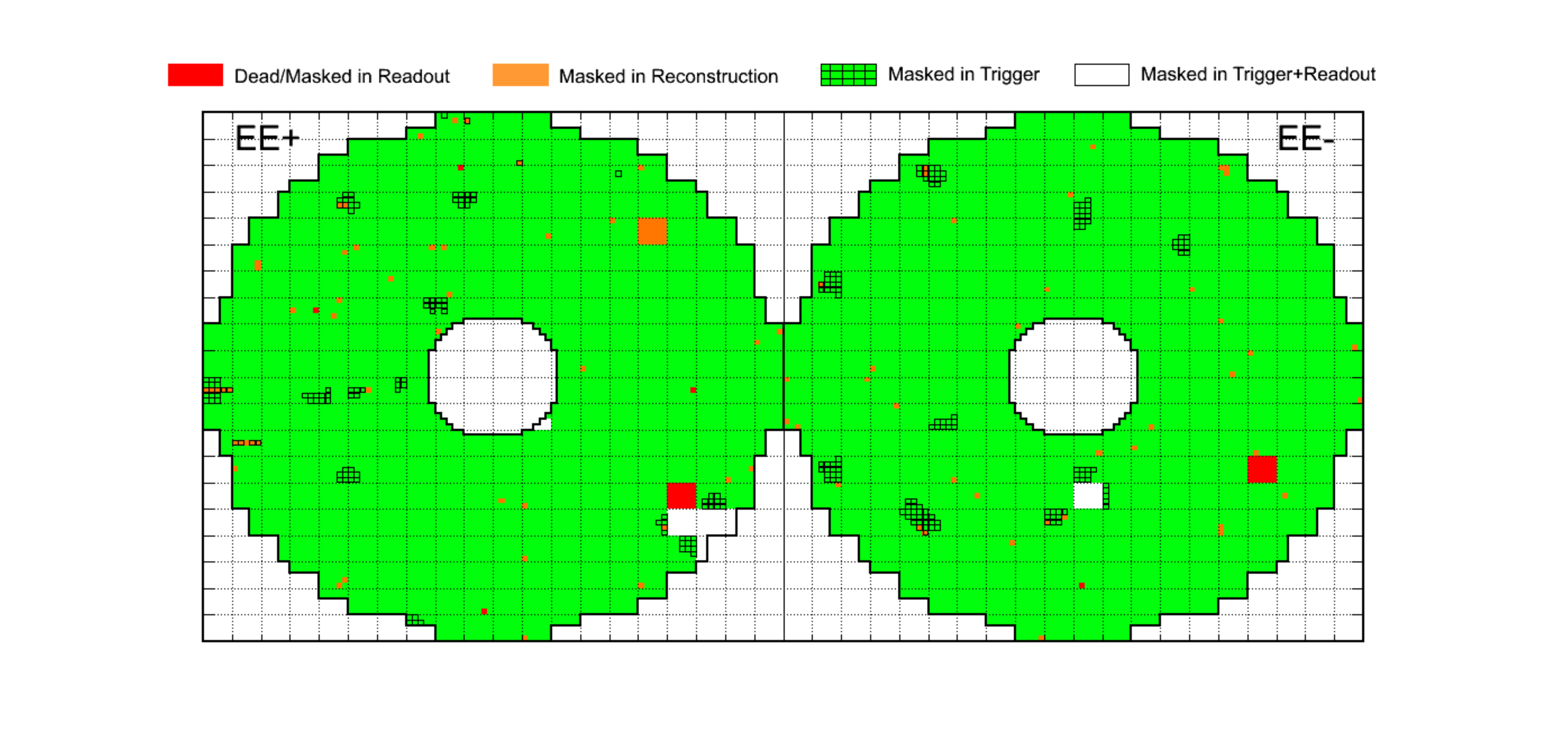} }
\end{center}
\caption{Distribution of masked ECAL channels in (top) barrel: $\eta$-$\phi$ view of 170x360 individual crystals and (bottom) endcaps: $y$-$z$ view of 2x7400 individual crystals.}
\label{fig:deadFEDS}
\end{figure}

Particles traversing
poorly instrumented regions of the detector can be a cause of apparent \vecmet.
While generally hermetic, the CMS calorimeter does have uninstrumented areas
(cracks)
at the boundary between the barrel and endcap sections, and between
the endcap and the forward calorimeters.
The gap between the barrel and endcap sections is about 5 cm and contains
various services, including cooling, power cables, and silicon detector readout.
The crack is not projective to the interaction point.
In addition,
about 1\% of the ECAL crystals are either not operational or
have a high level of electronic noise~\cite{ecalnoise}, and they are masked
in reconstruction.
The $\eta$-$\phi$ distribution of these crystals for the barrel, and $y$-$z$
distribution for the
endcaps,
is shown in Fig.~\ref{fig:deadFEDS}.

In this Section, we illustrate the effect of these features on the \met distribution
and test the reliability of the simulation for events with jets pointing towards
masked ECAL channels or cracks.

\begin{figure}[t!]\centering
{\includegraphics[width=0.32\textwidth]{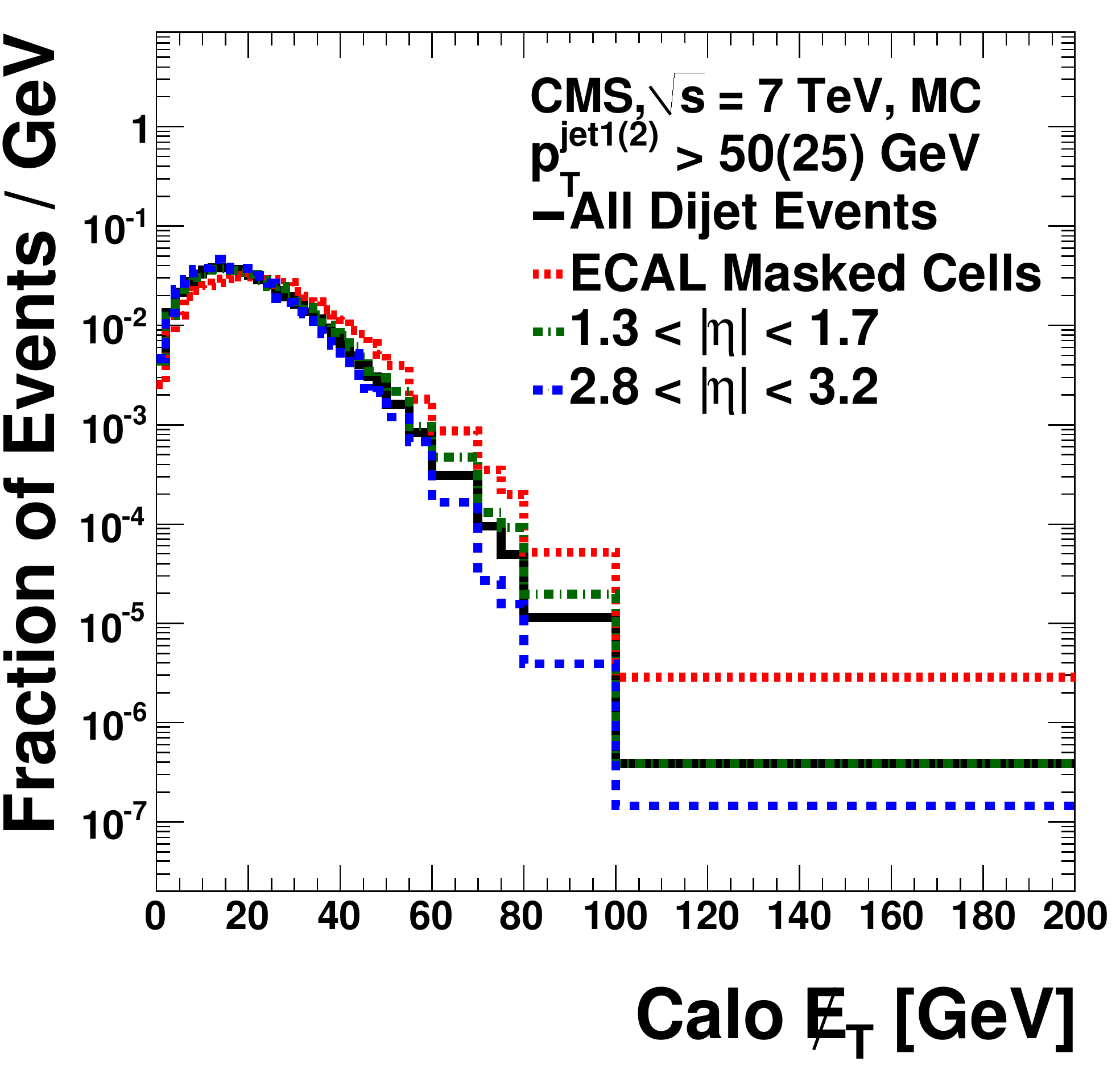}}
{\includegraphics[width=0.32\textwidth]{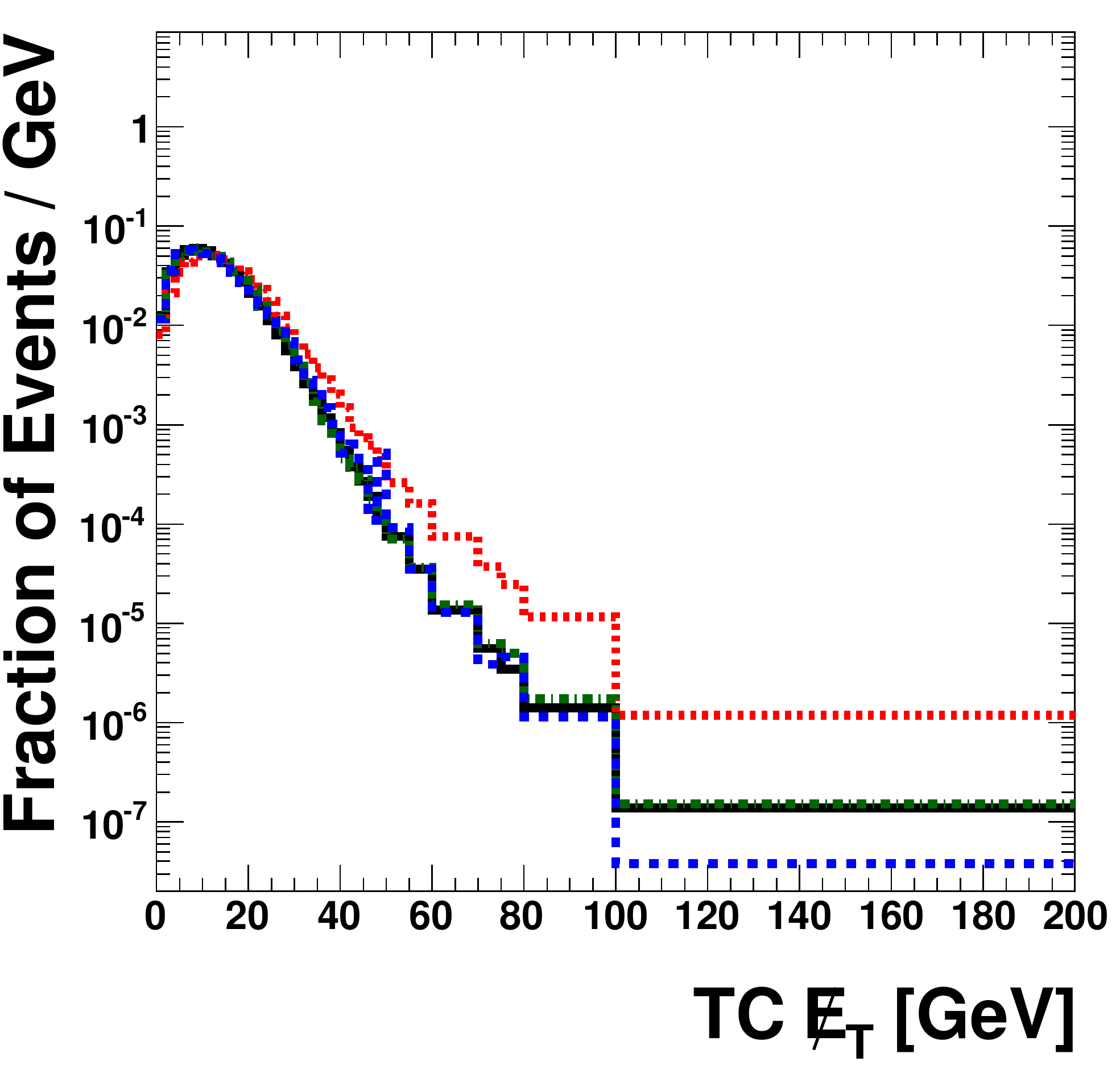}}
{\includegraphics[width=0.32\textwidth]{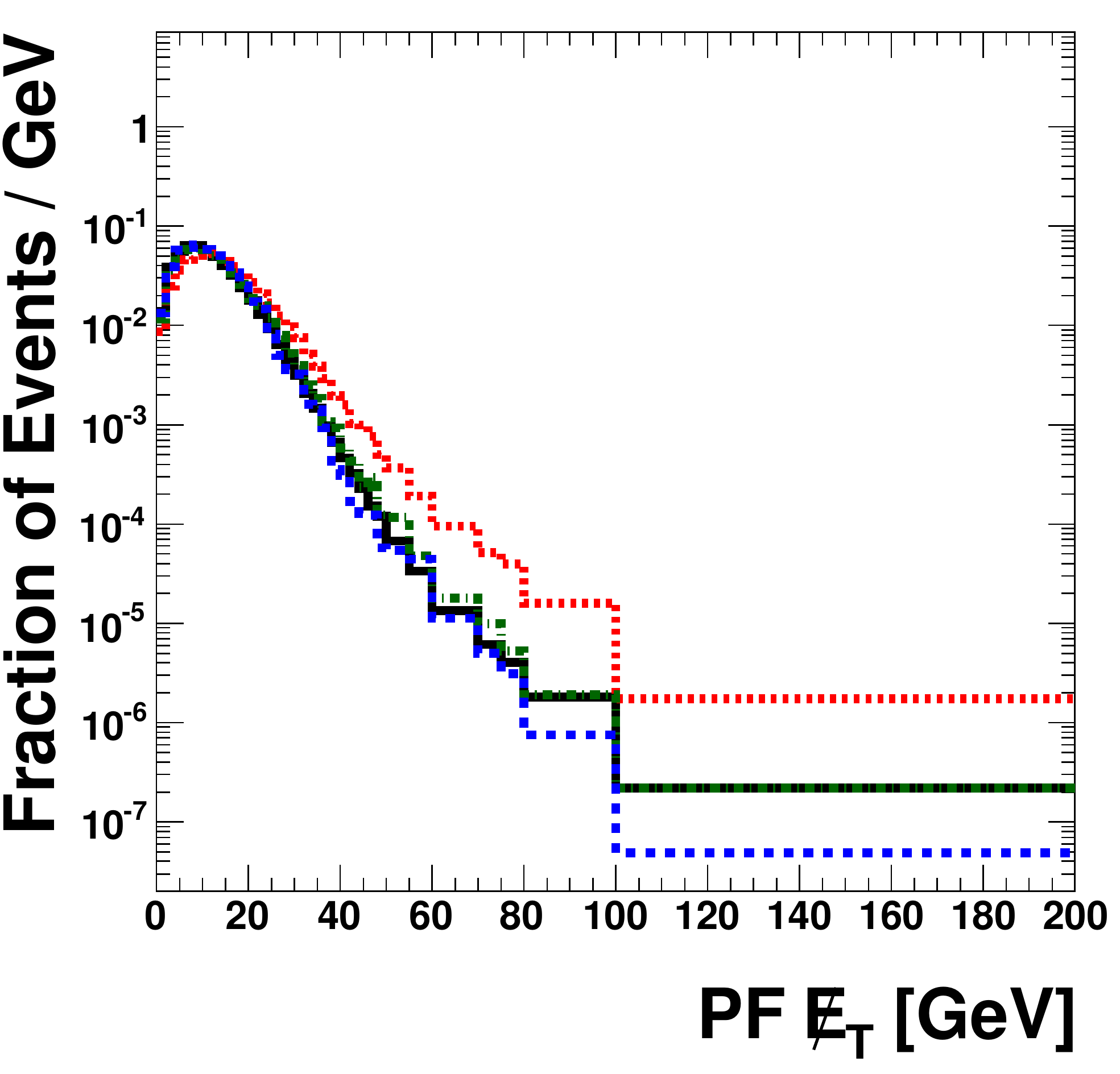}}
\caption{Distribution of (left) \calomet, (middle) \tcmet, and (right) \pfmet, normalized
to unit area,
for events containing at least 2 jets with $p_{\rm T}^{\rm jet1(2)}>$~50 (25) GeV (black solid), and
for the subsets of these events with
a jet aligned with \vecmet within $\Delta\phi(\vecmet,\mbox{jet})<0.2$ and pointing towards a masked ECAL cell (red dotted),
the barrel-endcap boundary (green dot-dashed), and the endcap-forward boundary (blue dashed)
in simulation.
}
\label{fig:MET_Boundary_ECALDeadCell_MC}
\end{figure}

Figure \ref{fig:MET_Boundary_ECALDeadCell_MC} shows
the \met distributions from simulated samples
of events containing at least 2 jets, with
the leading jet satisfying \pt$>50$ GeV and the
second jet satisfying  \pt$>25$ GeV, for \calomet, \tcmet, and \pfmet.
For events with $100<\met<200$  GeV,
the contribution from QCD multijet production is 24--42\%,
depending on the \vecmet\ reconstruction algorithm;
the rest is from W/Z/${\rm t}\bar{\rm t}$ production.
In order to illustrate the effect of the cracks, distributions are also shown
for those subsets of these samples that have
at least one jet that is aligned with the \vecmet to within  $\Delta\phi(\vecmet,\mbox{jet})<0.2$
and that is
pointing towards masked ECAL channels,
the barrel-endcap boundary ($1.3<|\eta|<1.7$), or
the endcap-forward boundary ($2.8<|\eta|<3.2$).
The masked ECAL channels considered here are those that are part of a group of
$5\times 5$ or $5\times 1$ masked channels that are adjacent in $\eta$-$\phi$,
as they
have larger impact on the \met distributions than isolated
masked channels.
A jet is considered to be pointing to one of the masked ECAL channels when
its jet centroid is within $\Delta R<0.2$ of a masked ECAL channel, where
$\Delta R = \sqrt{(\Delta \eta)^2+(\Delta \phi)^2}$.
We can see effects on the \met distribution
from masked ECAL channels, while
the calorimeter boundaries do not appear to have an enhanced contribution to
the events with large \met.

\begin{figure}[t!]\centering
{\includegraphics[width=0.32\textwidth]{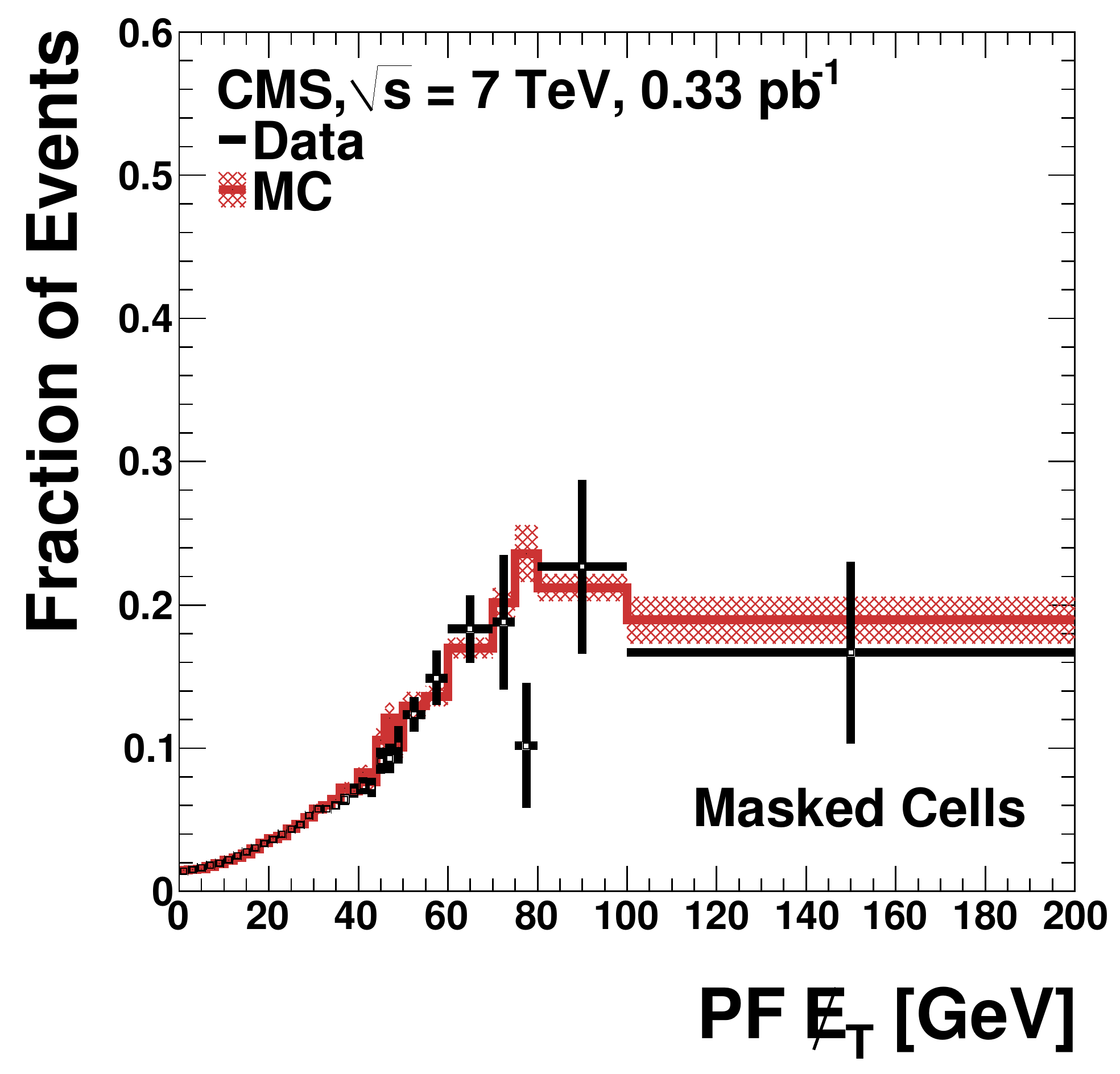}}
{\includegraphics[width=0.32\textwidth]{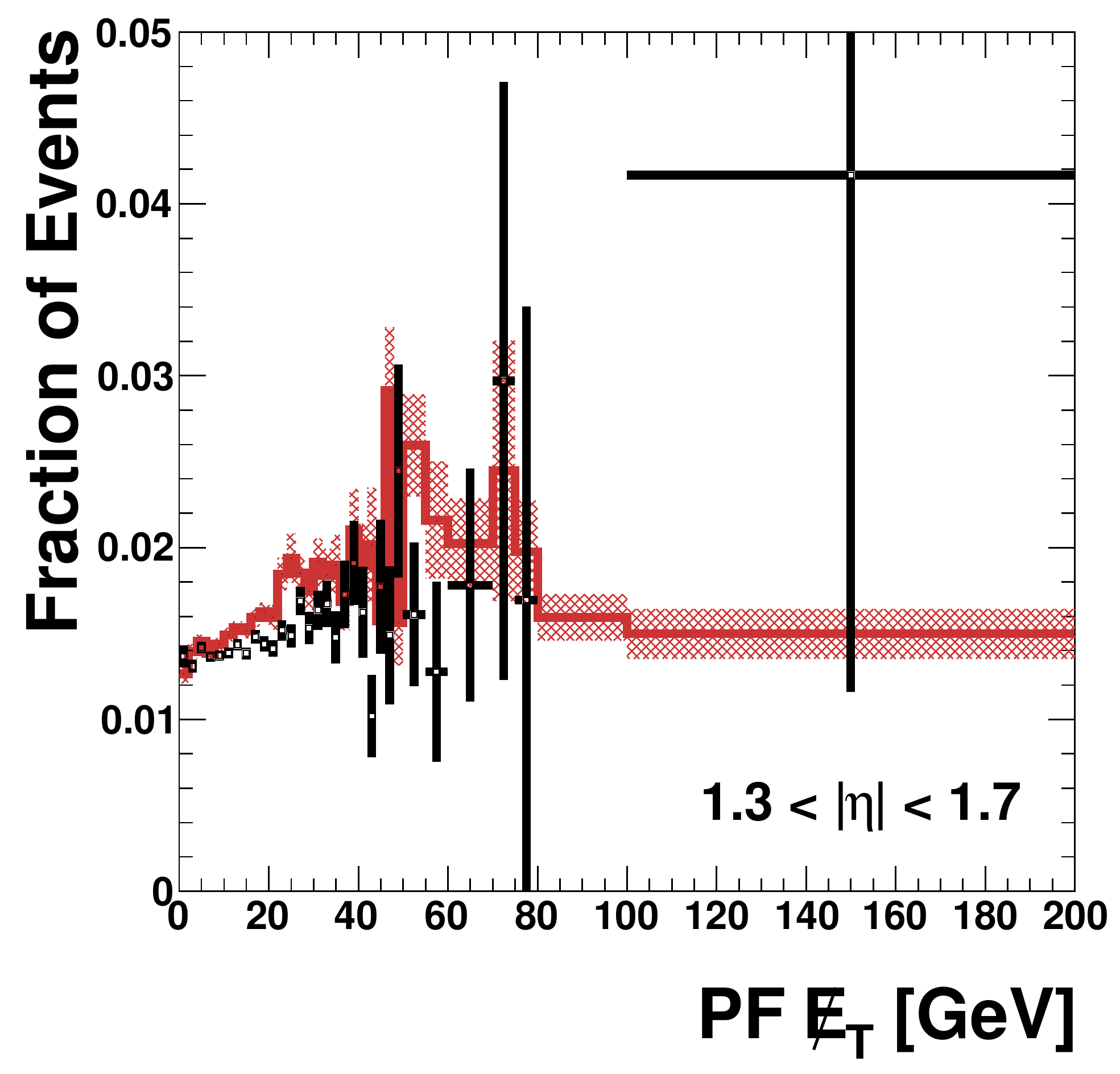}}
{\includegraphics[width=0.32\textwidth]{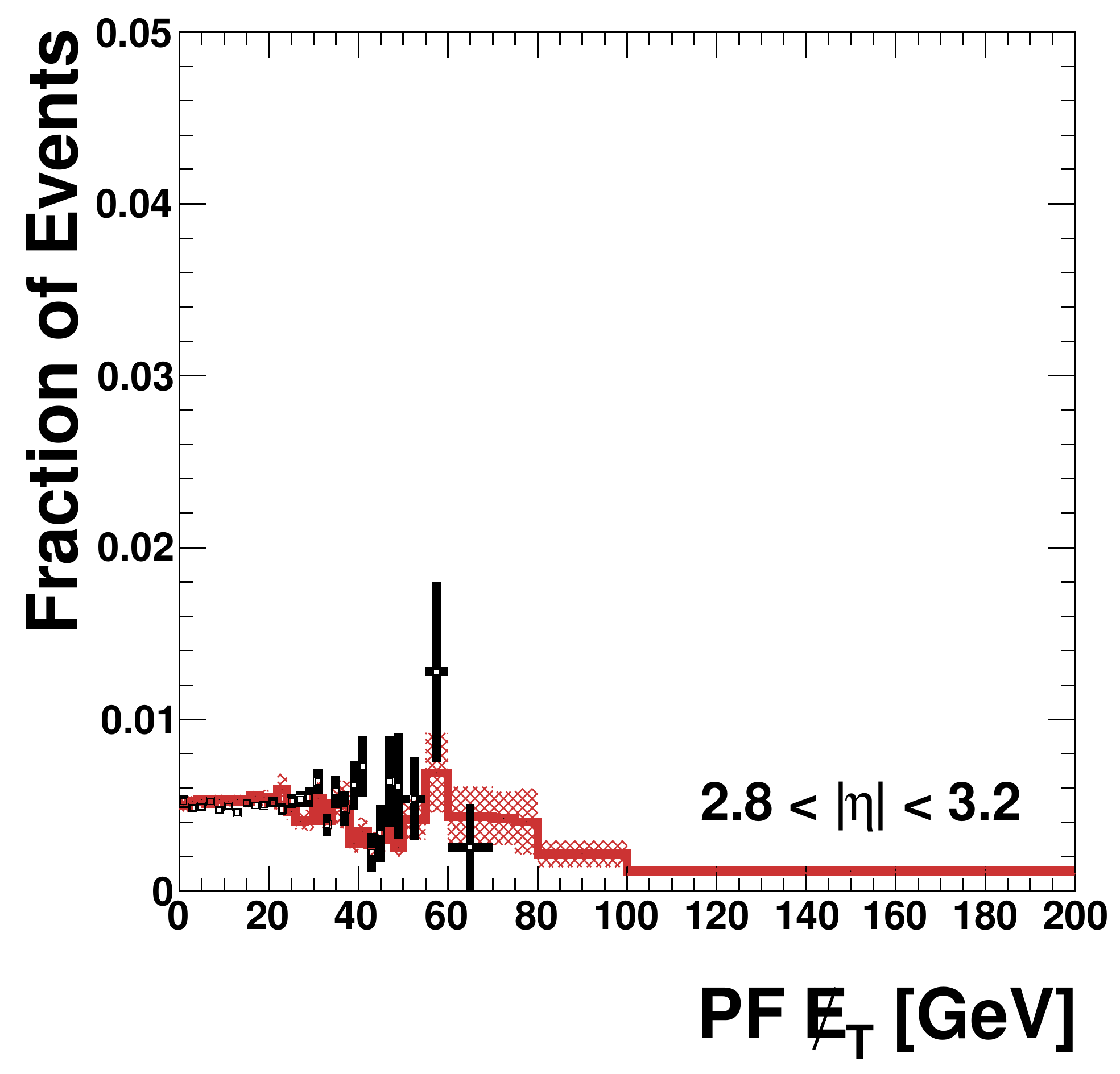}}
\caption{Fraction of dijet events in data (points) and simulation
(red band)
with
a jet aligned to \vecmet within $\Delta\phi(\vecmet,\mbox{jet})<0.2$ and pointing towards
(left) a masked ECAL channel,
(middle) the barrel-endcap boundary,
and (right) the endcap-forward boundary,
in data and in simulation.
}
\label{fig:MC_Data_MET_cracks}
\end{figure}

Figure~\ref{fig:MC_Data_MET_cracks} shows the fraction of dijet events
with
at least one jet aligned with the \vecmet and also pointing towards the masked ECAL channels,
the barrel-endcap boundary, or endcap-forward boundary for data and for simulation.
Figure~\ref{fig:MC_Data_MET_cracks}(left) shows that the masked ECAL channels
enhance the rate of events with large \met in both data and Monte Carlo simulation.
Approximately 20\% of the events with \met$>80$ GeV have contributions to the measured \vecmet
from mismeasurements due to
masked ECAL channels.
Results from simulations indicate that the fraction of events
with large \met\ due to mismeasurements (excluding the predicted contributions
from sources of genuine \met, such as W/Z/${\rm t}\bar{\rm t}$) is 30\%.
As shown in Figs.~\ref{fig:MC_Data_MET_cracks}(middle)
and \ref{fig:MC_Data_MET_cracks}(right),
the fraction of events which contain a jet that is both aligned with the
\vecmet and pointing towards a calorimeter boundary
does not have a strong dependence on \met,
indicating that the calorimeter boundaries are not major contributors
to events that have large apparent \met due to mismeasurements.
Unlike the masked ECAL channels,
the cracks are not projective to the interaction point,
and therefore energies of particles traversing these cracks are still measured,
albeit with degraded resolution.

While the impact of the cracks is small, analyses sensitive to events
with large \met need to take the ECAL masked channels into account.
About 70\% of the ECAL channels that are masked during offline reconstruction
have a useful measurement of their energy from the separate readout of the L1 trigger.
Although the trigger readout saturates,
it can be used to recover energies smaller than this and
to identify events that had more than this amount of energy
in a masked channel.
This saturation energy has been increased from 64 GeV
to 128 GeV in 2011.
Analysts can veto events with a jet pointing towards an ECAL masked channel
that does not have trigger information or that has trigger-readout
energy at the saturation threshold.

\section{Missing transverse energy scale and resolution}
\label{sec:resolution}

In this section, we study the performance of \vecmet using
events where an identified $Z$ boson or isolated $\gamma$ is present.
Events containing vector bosons may be produced in hard parton-parton 
collisions
such as qg$\to$ q$\gamma$, ${\rm q}\bar{\rm q}\to$ Z,
qg$\to$qZ, and ${\rm q}\bar{\rm q}\to$ gZ. 
While there is no genuine \met in these events, we can induce it by removing
the vector boson.
By comparing the momenta of the well-measured and well-understood vector boson to
the \vecmet induced this way,
we probe the detector response to the global hadronic system
and measure the scale and resolution of \vecmet.
While the
lowest order underlying processes may be simple, many physics and
experimental issues contribute to the measured, induced \vecmet in these events.
Effects
due to jet energy scale corrections and fluctuating jet composition
directly impact the measurement of the hadronic products of the hard
collision.
Underlying event activity,
pile-up, detector noise, and finite detector acceptance contribute as well.

The following notation is used: the vector boson
momentum in the transverse plane is $\vqt$, and the hadronic recoil, defined as the
vector sum of the transverse momenta of all particles except the vector boson
(or its decay products, in the case of Z candidates), is \vut.
Momentum conservation in the transverse plane requires
$\vqt+\vut=0$.
The recoil is the negative of the induced \vecmet.

The presence of a well-measured Z or $\gamma$
provides both a momentum scale,
$\qt\equiv|\vqt|$, and a unique event axis, \hatqt.  The hadronic recoil
can be projected onto this axis, yielding two signed components,
parallel (\upar) and perpendicular (\uperp) to the event axis.
Since
$\upar\equiv \vut\cdot\hatqt$, and the observed hadronic system is
usually in the opposite hemisphere from the boson,
\upar\ is typically negative.

The mean value of the scalar quantity $\langle \upar\rangle /\qt$ is
the scale factor correction required for $\met$ measurements in the classes
of events considered here, and is closely related to jet energy scale
corrections and jet parton flavour. We refer to  $\langle \upar 
\rangle /\qt$ as the ``response'' and denote distributions of this
quantity versus \qt\ as ``response curves''. 
Deviations of
the response curve from unity probe the \met\ response as a function of \qt.

Resolution is assessed by measuring the RMS spread of \upar\ and
\uperp\ about their mean values, after correcting for the response, and is denoted $RMS(\upar)$ and
$RMS(\uperp)$. As with the response, we examine the resolutions as
functions of \qt.

\subsection{Direct photon sample}

Candidate photon events
are selected by requiring each event to contain
exactly one reconstructed photon in the barrel portion
of the ECAL ($|\eta|<1.479$), with $q_{\rm T} > 20$ GeV, and which passes the identification and isolation selection described in Section~\ref{sec:datas}.
The total number of events passing all requirements is
15\,7567, of which 67\,621 have only one
reconstructed primary vertex. The prescale factors for the HLT triggers
used to collect this sample varied over the course of the 2010 LHC running period.
As a result, this sample
is dominated by events
recorded during the earlier period of the data taking,
when the fraction of crossings containing pile-up interactions was smaller.

\begin{figure}[t!]
\begin{center}
{\includegraphics[scale=0.35]{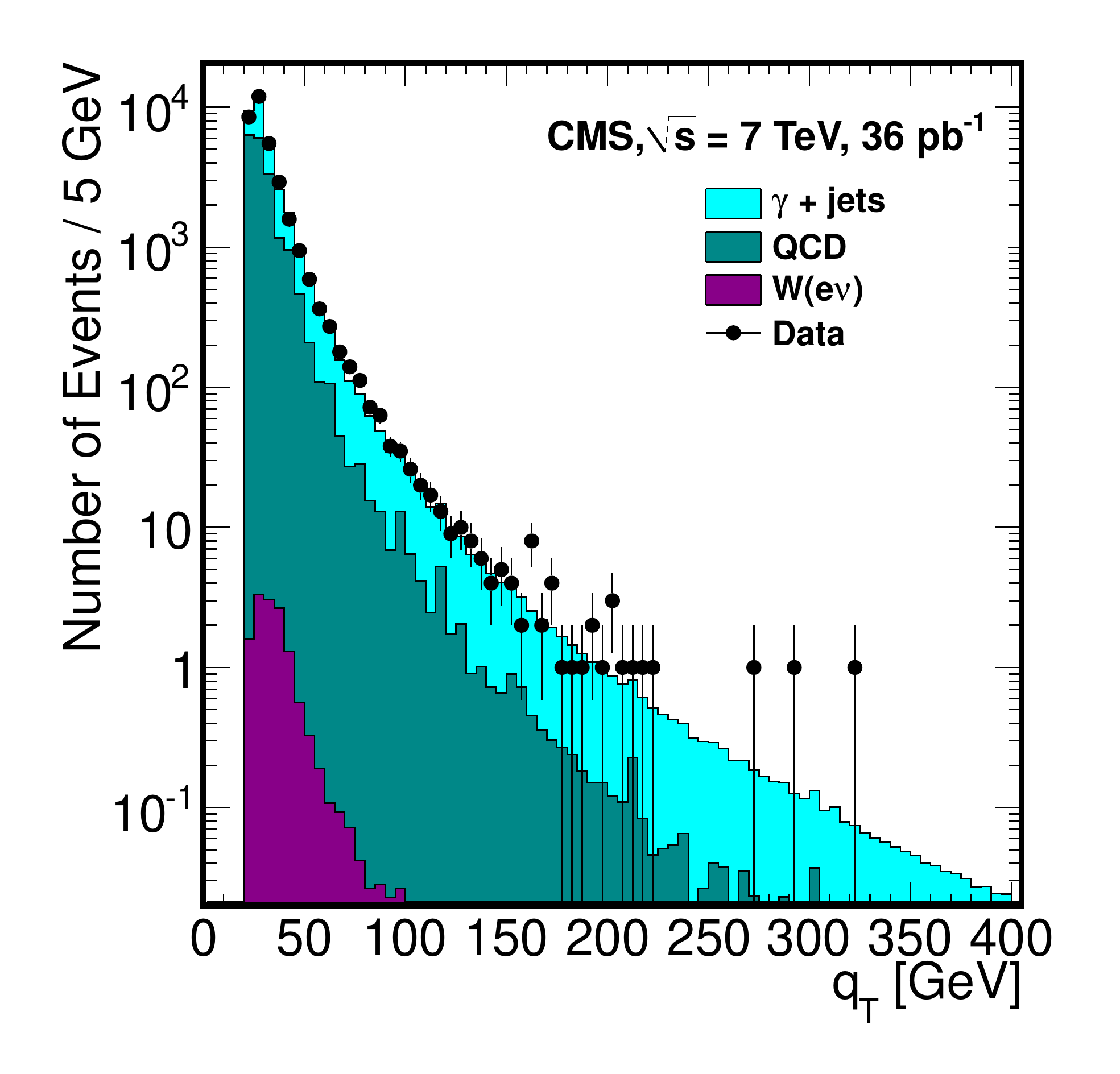} }
\end{center}
      \caption{Distribution of \qt for events selected as photon+jet candidates.
        Predicted rates from simulation for signal and backgrounds are also shown.  QCD refers to multijet events.
}
       \label{f:gammajetbck}
\end{figure}

Figure \ref{f:gammajetbck}
shows the photon \qt\ spectrum for data and for simulation.
About half of the observed rate arises from QCD dijet production
where one jet passes all photon identification requirements. Such
jets are typically highly enriched in $\pi^0\to\gamma\gamma$ and contain
little hadronic activity.
The detector response to these jets is similar to
that of single photons, and studies indicate that
response curves extracted from these QCD background events match the
response of true photon-jet events to within a percent. We therefore make
no further attempt to filter them out.
The detector response to the jet depends on the type of parton
from which it originated.
The leading jet
in photon events is predicted to predominantly be a quark jet.
A prediction for the difference in response for the CMS detector between
quark and gluon jets can be found in \cite{JME-10-011}.
The difference is largest for Calo Jets ($\approx$ 20\% for jets with \pt of 20 GeV), and
decreases with \pt.  The primary reason that the response is lower for gluon jets is that
their particles tend to have lower $p_{\rm T}$s, and the calorimeter response is lower at low \pt.
For PF jets and \pfmet, which use tracker instead of calorimeter momenta for most
charged hadrons,
the difference in response is reduced, and varies from about 5\% at 20 GeV to a percent a high \pt.

\subsection{Z samples}

For the \Zee selection (electron channel),
we require two well-identified and isolated electrons, with \pt$>$
20 GeV, within the fiducial region of the ECAL.
The invariant mass ($M_{\ell\ell}$) of the electron pair is required to be in the
range $70 < M_{\ell\ell} < 120~\text{GeV}$.

For the \Zmm selection (muon channel),
we require two isolated muons with opposite electric
charges, that have \pt $>$
20 GeV, and are within the $|\eta|< 2.1$ region.
The invariant mass $M_{\ell\ell}$ of the muon pair is required to be
at least $60~\text{GeV}$, and no more than $120~\text{GeV}$.

We obtain a total of 12\,635 (12\,383) \Zee (\Zmm) candidates.
The relative contributions of signal and background are estimated from simulation.
By normalizing the invariant mass
distribution from simulation of signal and background
so that it has
the same number of events that is
observed in the data,
a total background of around 143 (35) events is estimated to have a contribution of
97 (2) events from QCD, 28 (9) events from electroweak, and 18 (24)
from final states containing top quarks.

\begin{figure}[t!]
  {
    \begin{minipage}[t]{0.49\linewidth}
      \includegraphics[scale=0.35]{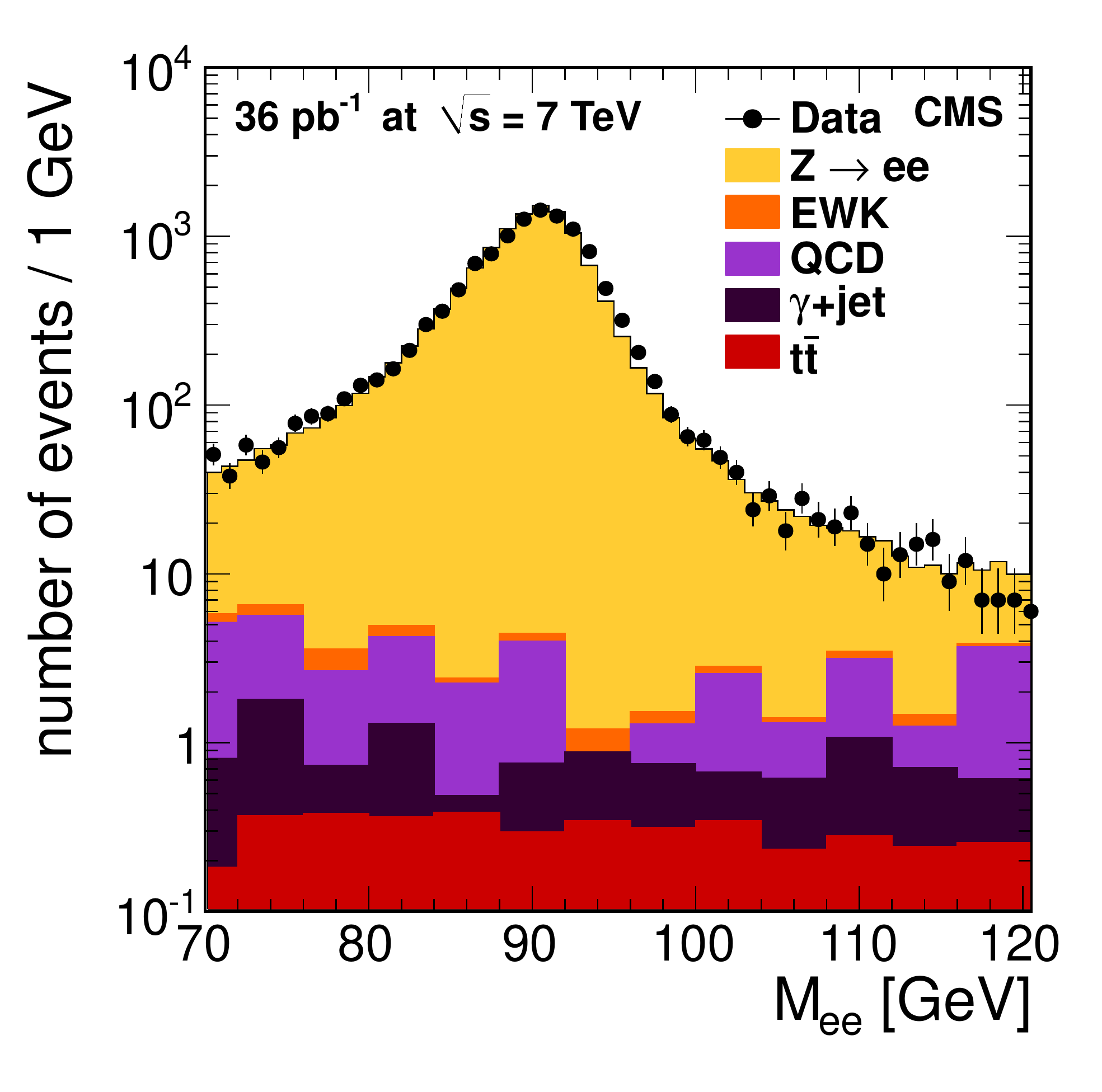}
    \end{minipage}
  }
  {
    \begin{minipage}[t]{0.49\linewidth}
      \includegraphics[scale=0.35]{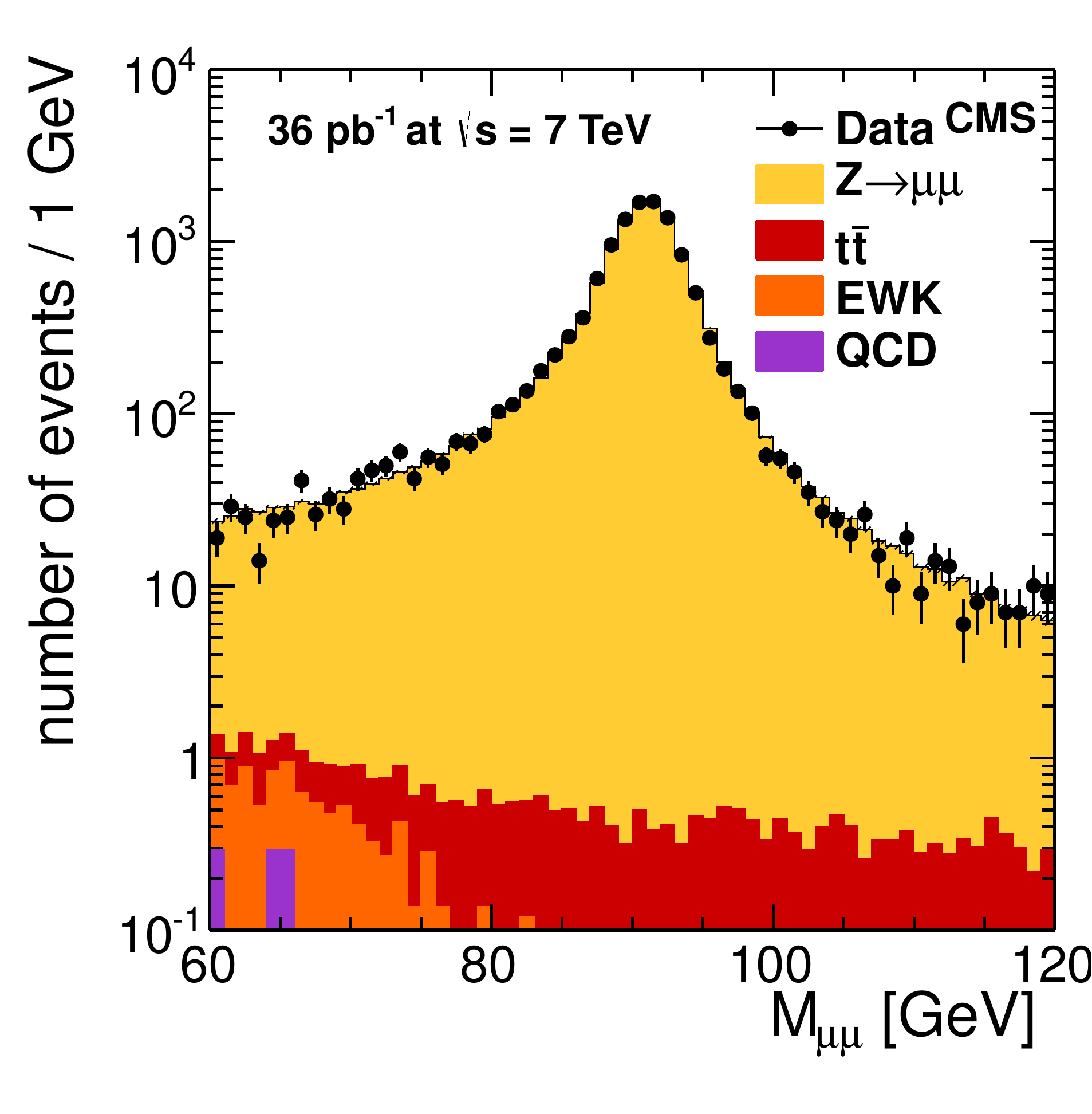}
    \end{minipage}
  }
      \caption{(left) Invariant mass distribution of the two leading electrons and
(right) invariant mass distribution of the two leading muons,
for the Z boson candidates, along with the predicted distribution from simulation.
QCD refers to multijet events.
}
       \label{f:zdata}
\end{figure}

\begin{figure}[t!]
  {
    \begin{minipage}[t]{0.49\linewidth}
      \includegraphics[scale=0.35]{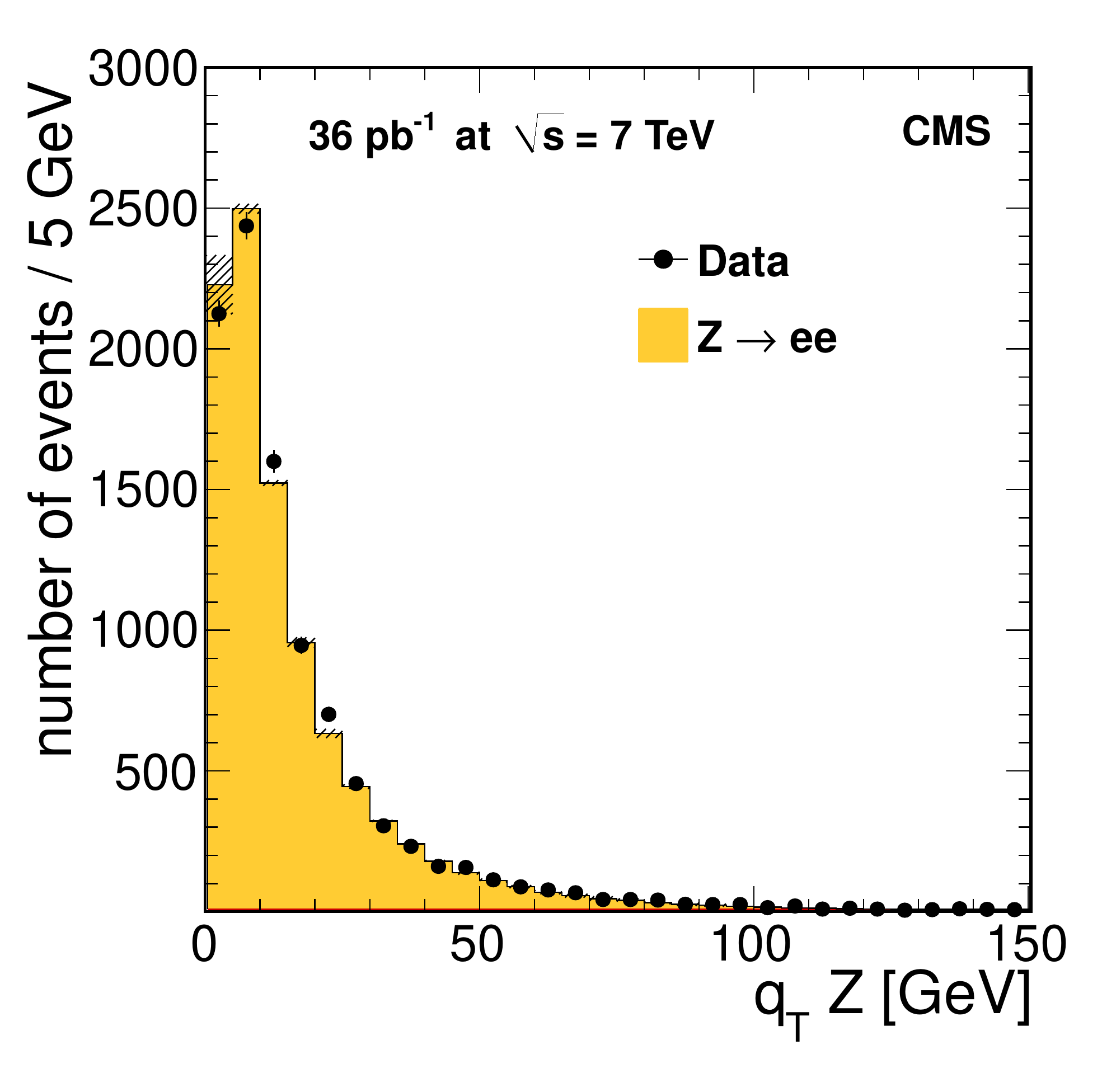}
    \end{minipage}
  }
  {
    \begin{minipage}[t]{0.49\linewidth}
      \vspace{-0.89\linewidth}
      \includegraphics[scale=0.35]{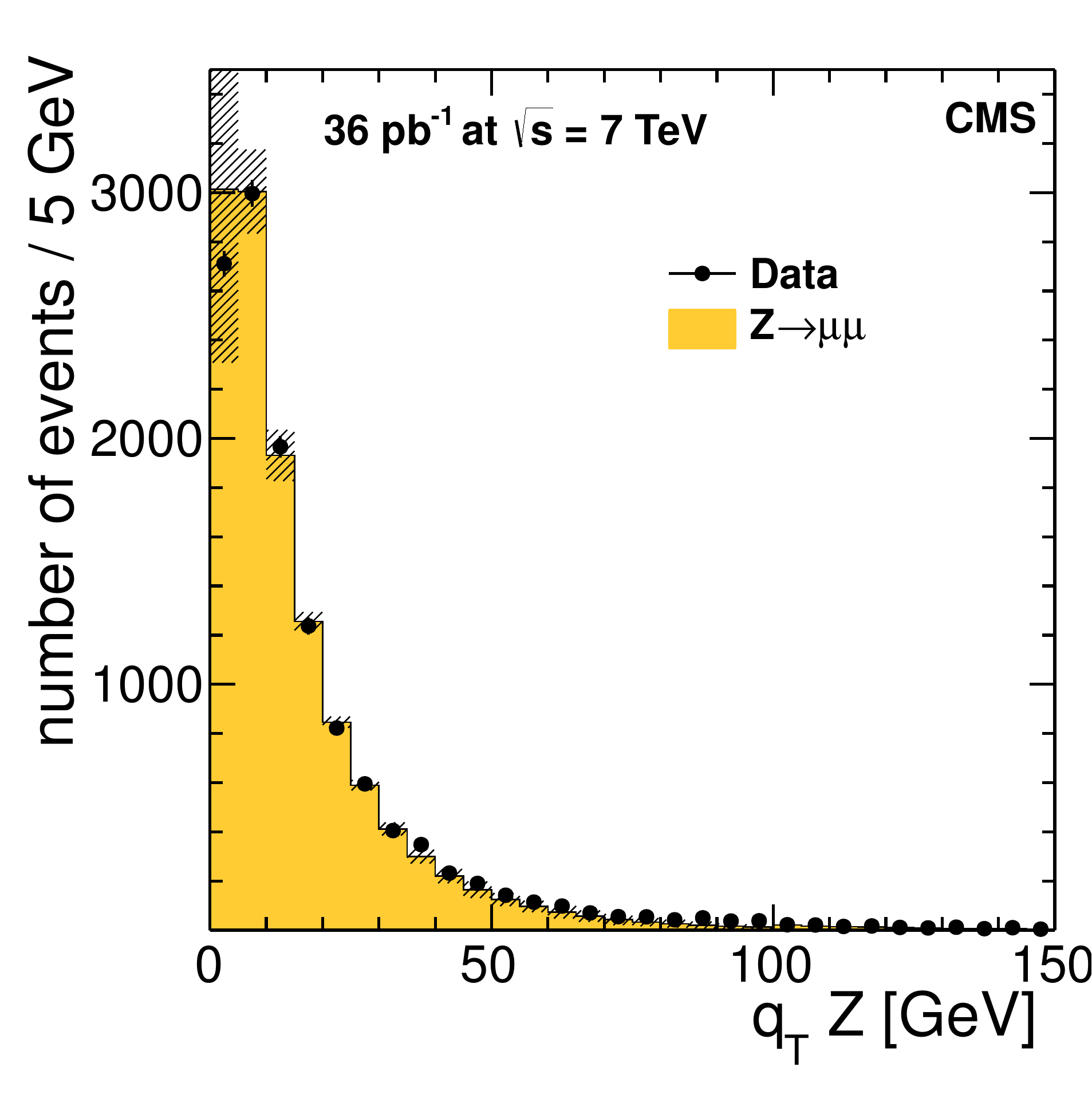}
    \end{minipage}
  }
  \caption{ The $q_{T}$ distribution for Z boson candidates in (left) the electron channel  and (right) the muon channel, along with the prediction from simulation. Systematic uncertainties are
shown as grey bands.}
  \label{f:zqt}
\end{figure}

Figure \ref{f:zdata} shows the $M_{\ell\ell}$ distribution for the electron and muon
samples.
Figure \ref{f:zqt} shows their $q_{\rm T}$ spectrum.
Except at very low \qt, the leading
jet in Z events, as with the $\gamma$ events, should usually be a quark jet.

\subsection{Scale and resolution for events with one primary vertex}

\begin{figure}[!h]
    \begin{center}
        { \includegraphics[height=0.3\textwidth]{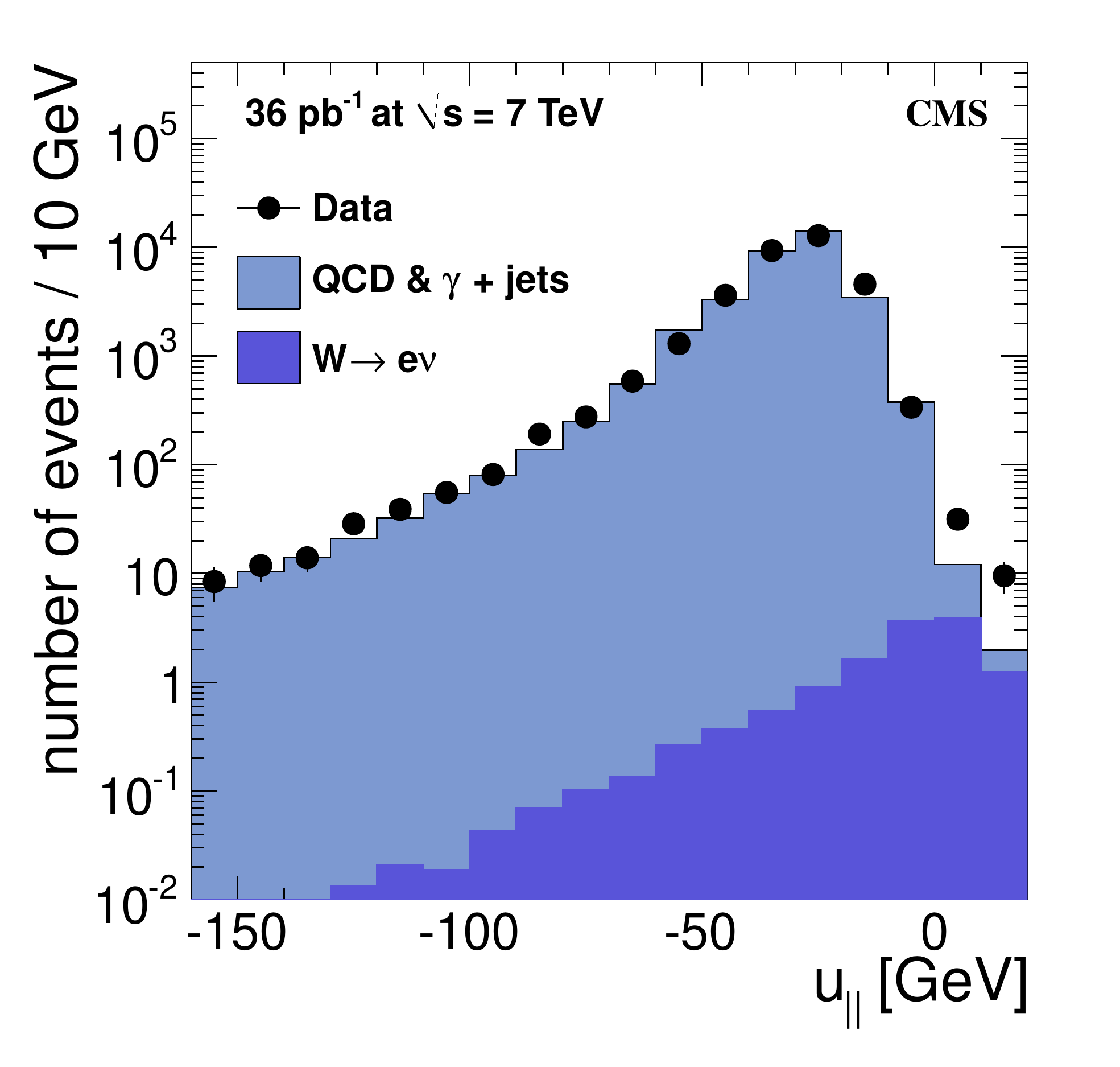} }
        { \includegraphics[height=0.3\textwidth]{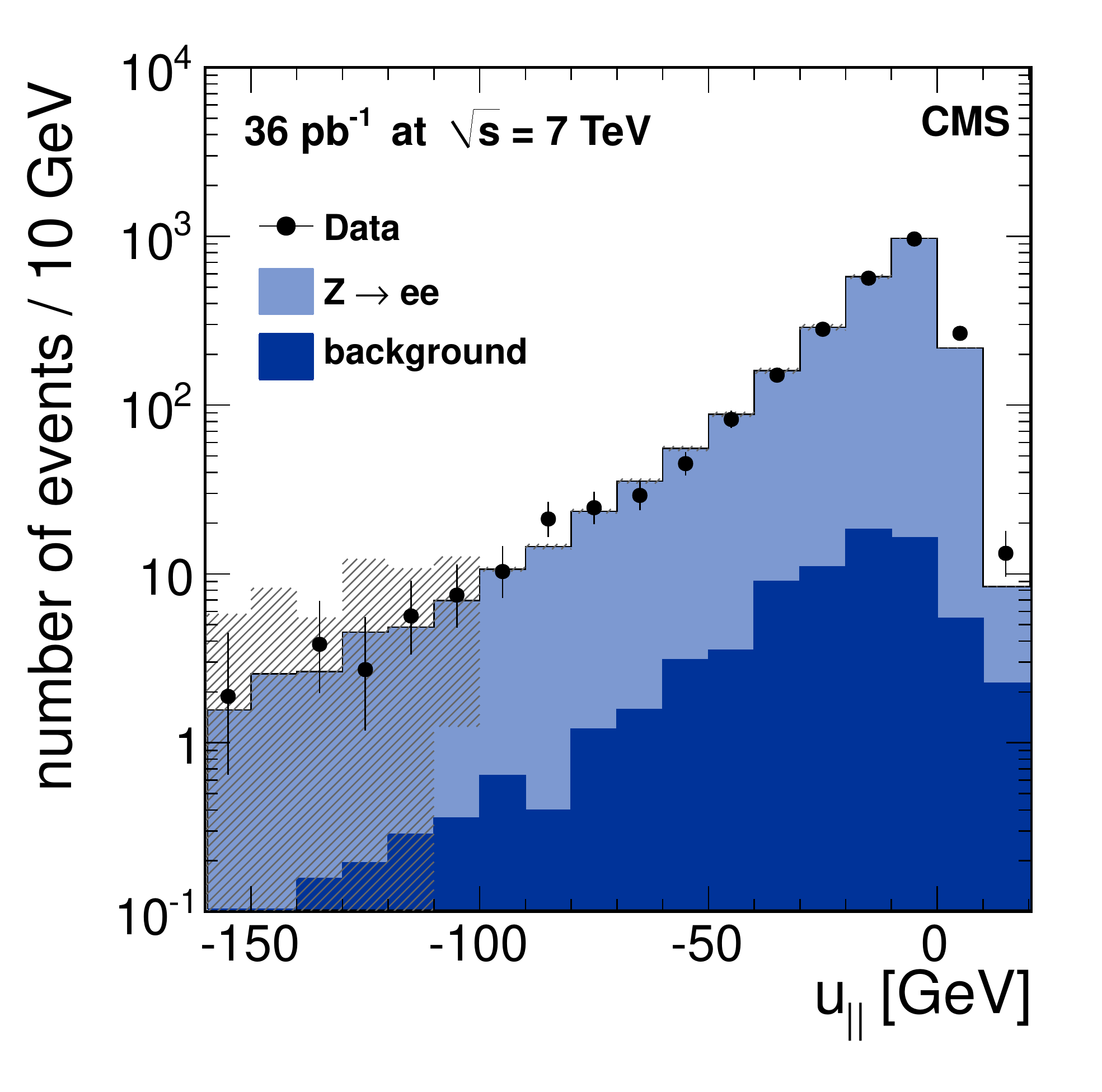} }
        { \includegraphics[height=0.3\textwidth]{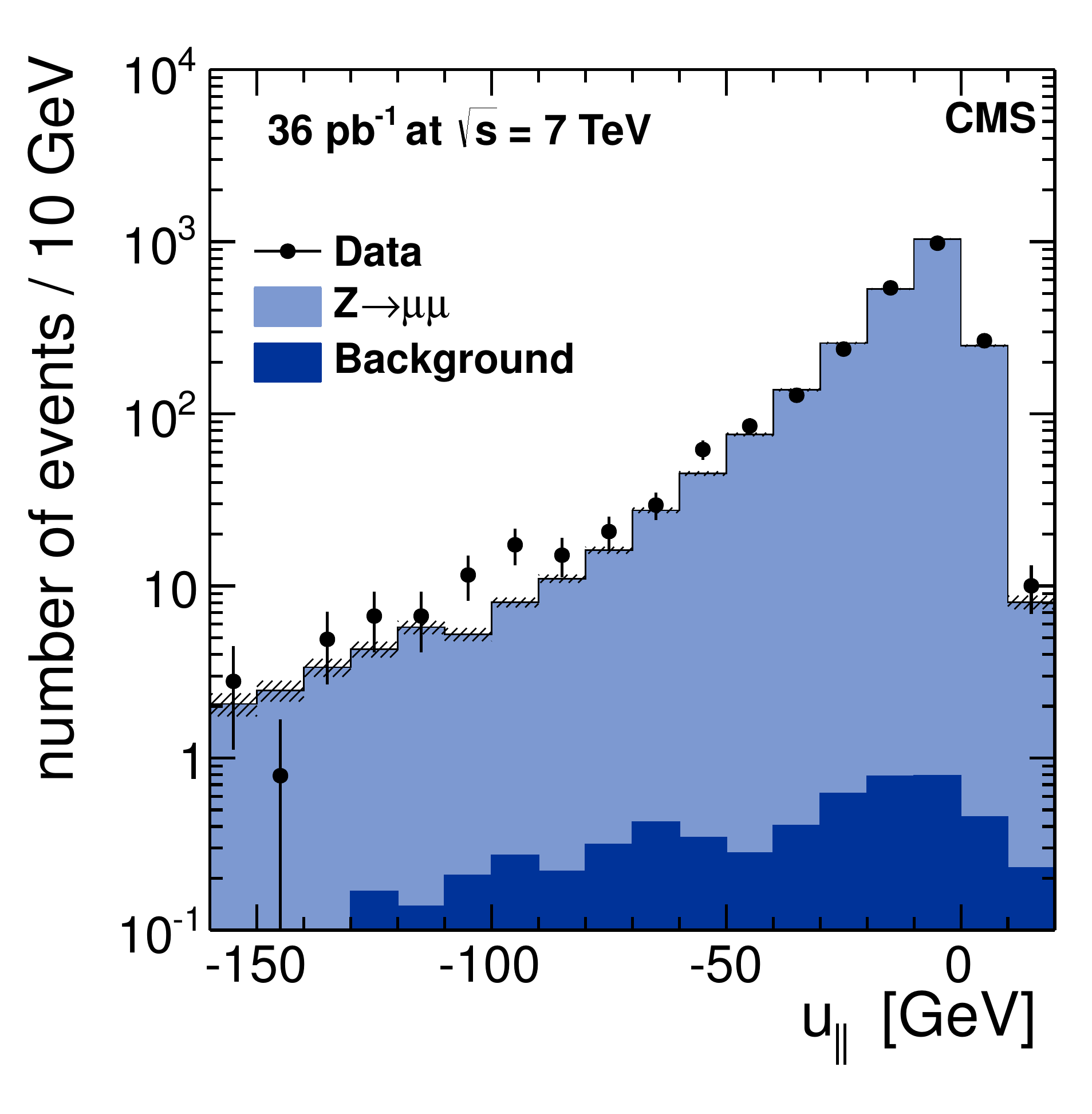} }\\
        { \includegraphics[height=0.3\textwidth]{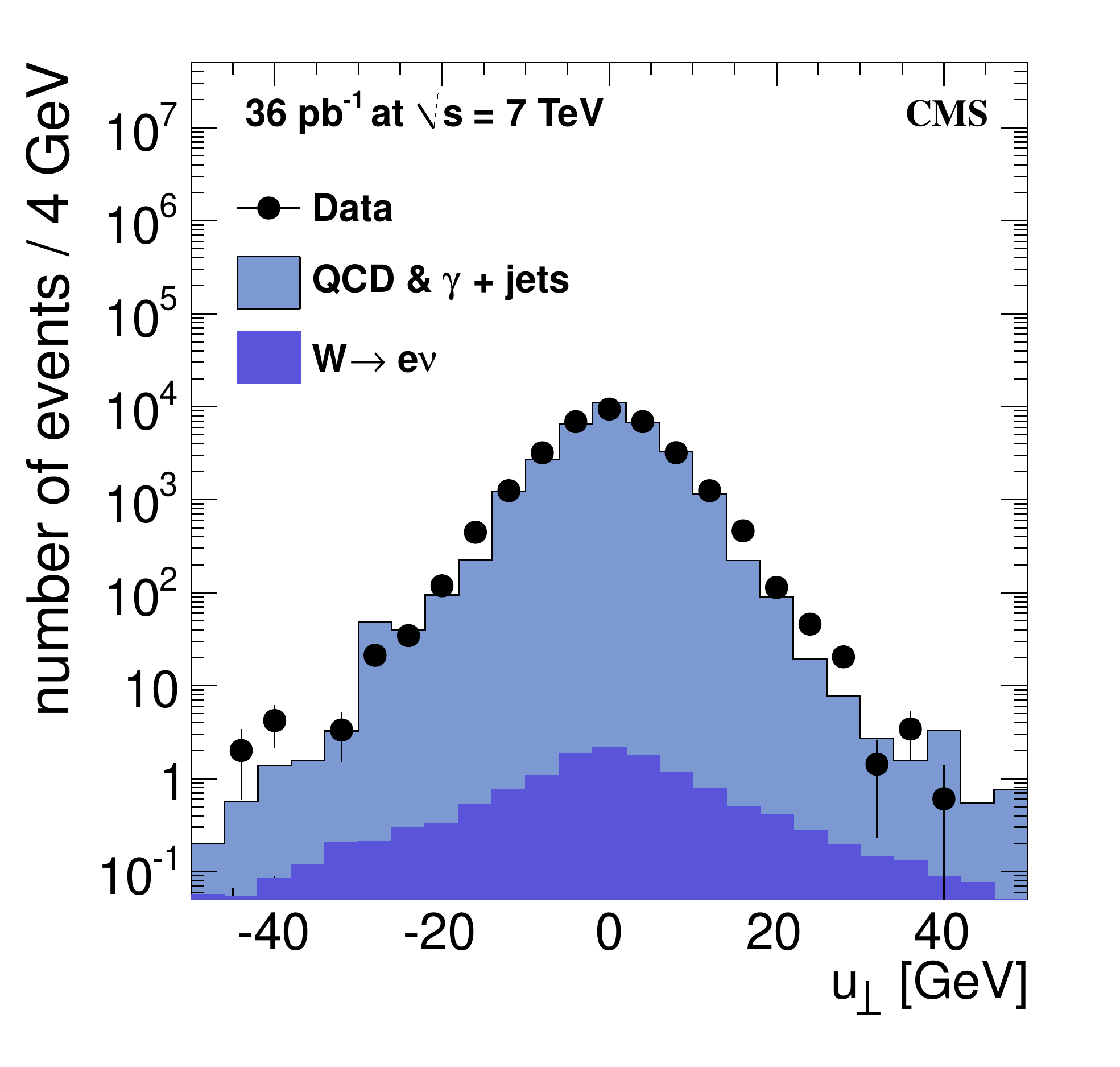} }
        { \includegraphics[height=0.3\textwidth]{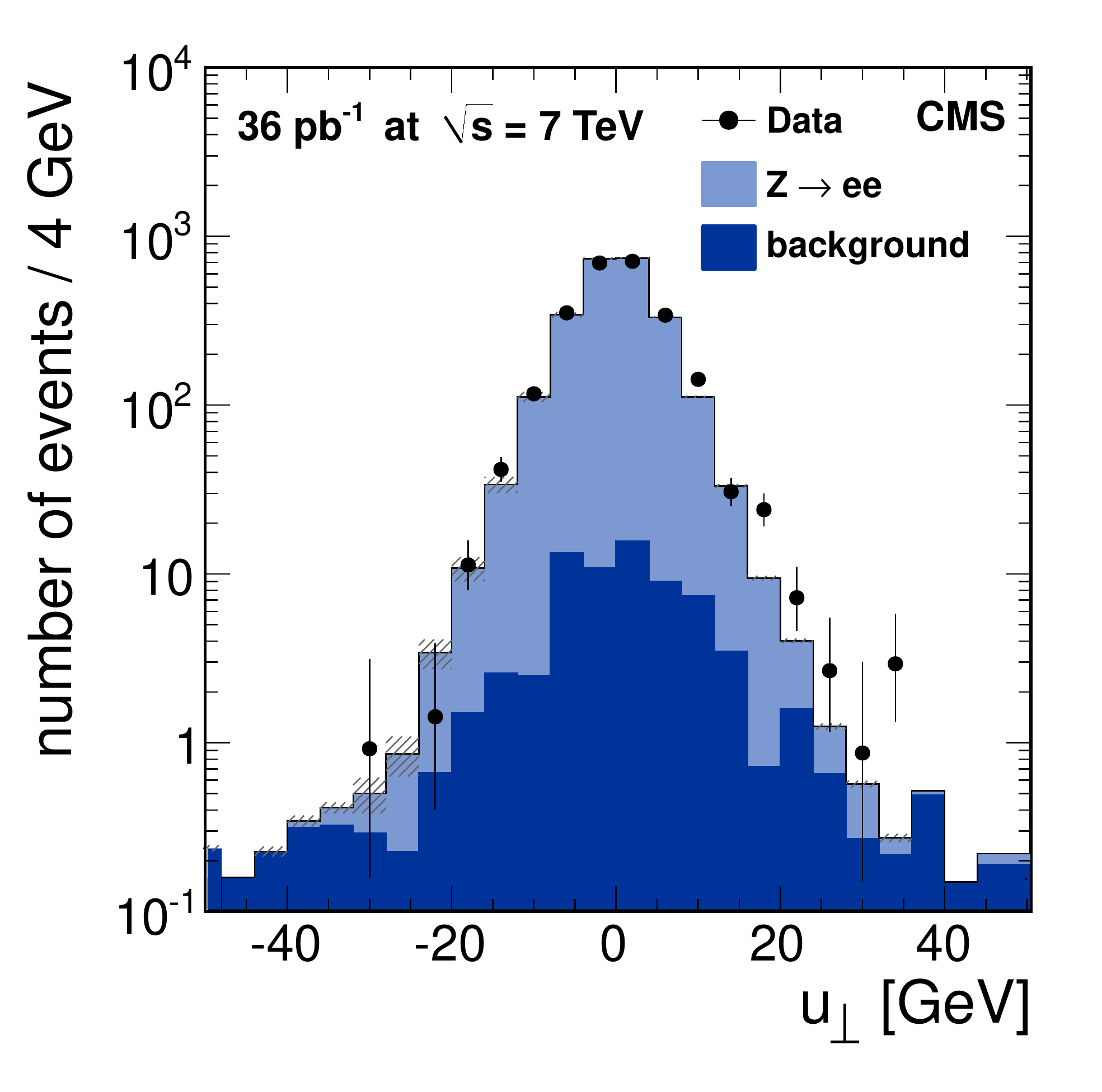} }
        { \includegraphics[height=0.3\textwidth]{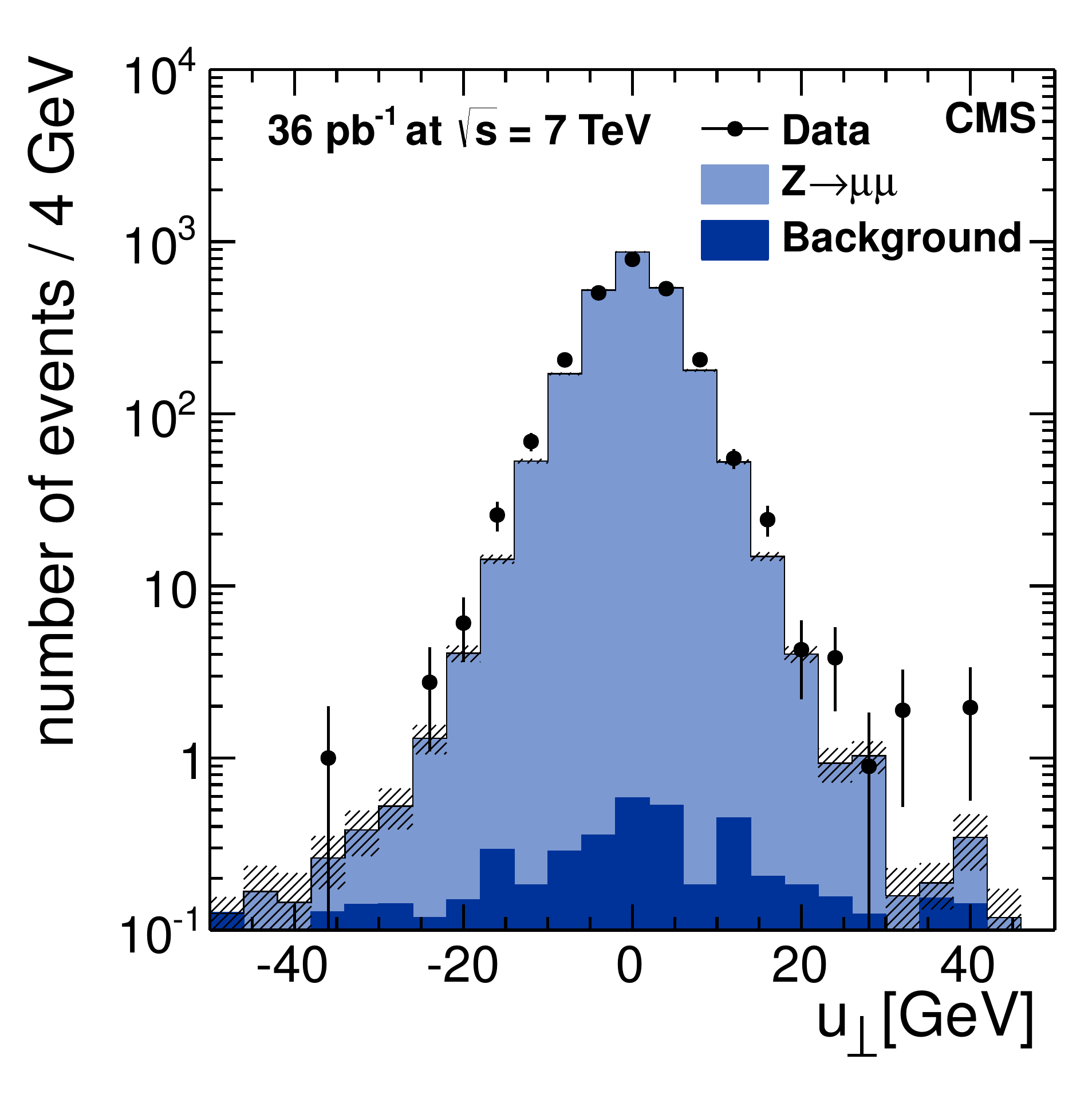} }
      \end{center}
        \caption{
\upar distributions for \pfmet for
        (top, left) $\gamma$, (top, middle) \Zee\, and (top, right) \Zmm\ events;
\uperp distributions for \pfmet for 
(bottom, left) $\gamma$, (bottom, middle) \Zee, and (bottom, right) \Zmm\ events.
Distributions are for events containing one PV, with a correction for residual
pile-up contamination.  Statistical uncertainties and systematic uncertainties on the removal of
events with more-than-one PV and on the boson \qt spectrum are shown as
grey bands on the prediction from simulation. 
QCD refers to multijet production.
}

       \label{f:recoil_perp_gammajet}
    \end{figure}

To study the $\met$ scale and resolution, we decompose the recoil with respect 
to the boson ($\gamma$ or $Z$) direction in the transverse plane. We restrict 
ourselves to events containing one reconstructed primary vertex. $Z$ yields are thus reduced to 
2611 \Zee candidates and 2438 \Zmm candidates.
The effect of pile-up on the scale and the resolution is studied in 
Section~\ref{sec:zpileup}.
Distributions of the components of the recoil calculated from \pfmet that are parallel and 
perpendicular to
the boson axis, $u_{\parallel}$ and $u_{\perp}$, are shown in Fig. \ref{f:recoil_perp_gammajet} 
for direct photon candidates, \Zee\ candidates, and \Zmm\ candidates.
As expected, the parallel component is
mainly negative, consistent with the back-to-back nature of the events, while
the perpendicular component is symmetric.

The distributions are corrected for the residual contamination ($5 \pm 1\%$) from events with more than one 
interaction.  The number of events with more than one scattering that are
reconstructed as a single PV is estimated by convoluting the 
efficiency for reconstructing two vertices as a function of the vertex
separation with the $z$ distribution of vertices.
The distributions are corrected for this contamination 
 by subtracting multi-vertex-event shapes obtained from data, 
rescaled to the estimated contamination, from the
distribution from events with one PV. 
The systematic uncertainty on the residual contamination is obtained by varying
the normalization within its uncertainties.

Events generated with \PYTHIA are reweighted so that the \qt spectrum
matches that predicted by the \RESBOS Monte Carlo program~\cite{resbos}, in order
to take advantage of its resummed calculation
of the boson \qt spectrum. 
The systematic uncertainties due to our imperfect
knowledge of the true \qt distributions for $Z$ bosons are estimated from the difference
between the \qt distributions predicted by \PYTHIA and \RESBOS.
We set the systematic uncertainty, bin-by-bin in \qt, equal to this difference.

In addition, there is a systematic uncertainty on the prediction from the simulation
due to the size of the simulation samples.  In Fig.~\ref{f:recoil_perp_gammajet}, the 
dominant uncertainty on  $u_{\parallel}$ for the part of 
the distribution from the electron channel with $u_{\parallel}<-100$GeV is from
this source and from uncertainties on the removal of the multi-PV contamination.

\begin{figure}[!h]
    \begin{center}
      \includegraphics[height=0.4\textwidth]{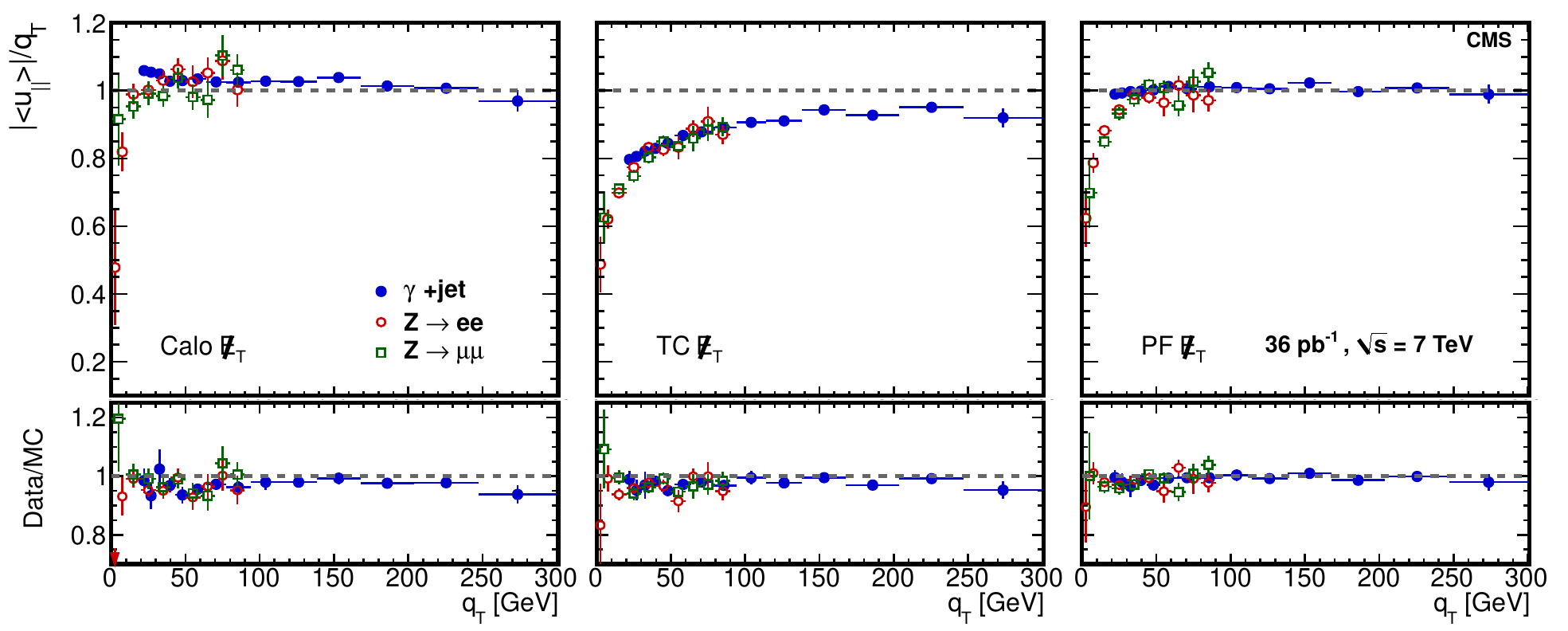}
      \end{center}
	\caption{Response curves for events with one primary vertex,
            for (left) \calomet, (middle) \tcmet, and (right) \pfmet. Results are shown
            for photon events (full blue circles), \Zee events (open red circles) and \Zmm events (open green squares).
            The  upper frame of each figure shows the response in data;
            the lower frame shows the ratio of data to simulation.
            The vertical axis labels at the far left apply to all
            three subfigures.}
       \label{f:PJresponse}
    \end{figure}

\begin{figure}[b!]
    \begin{center}
        { \includegraphics[height=0.45\textwidth]{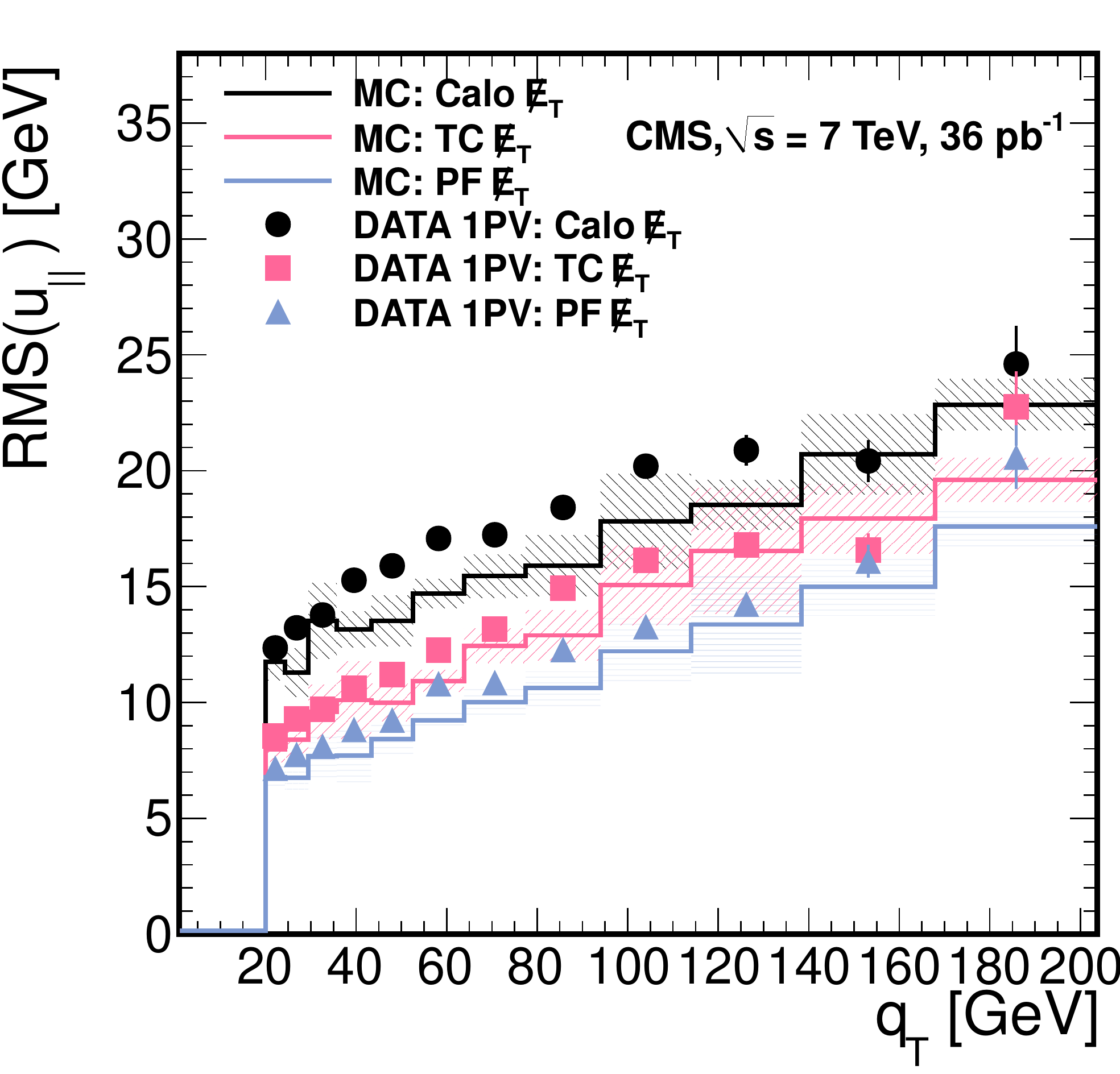} }
        { \includegraphics[height=0.45\textwidth]{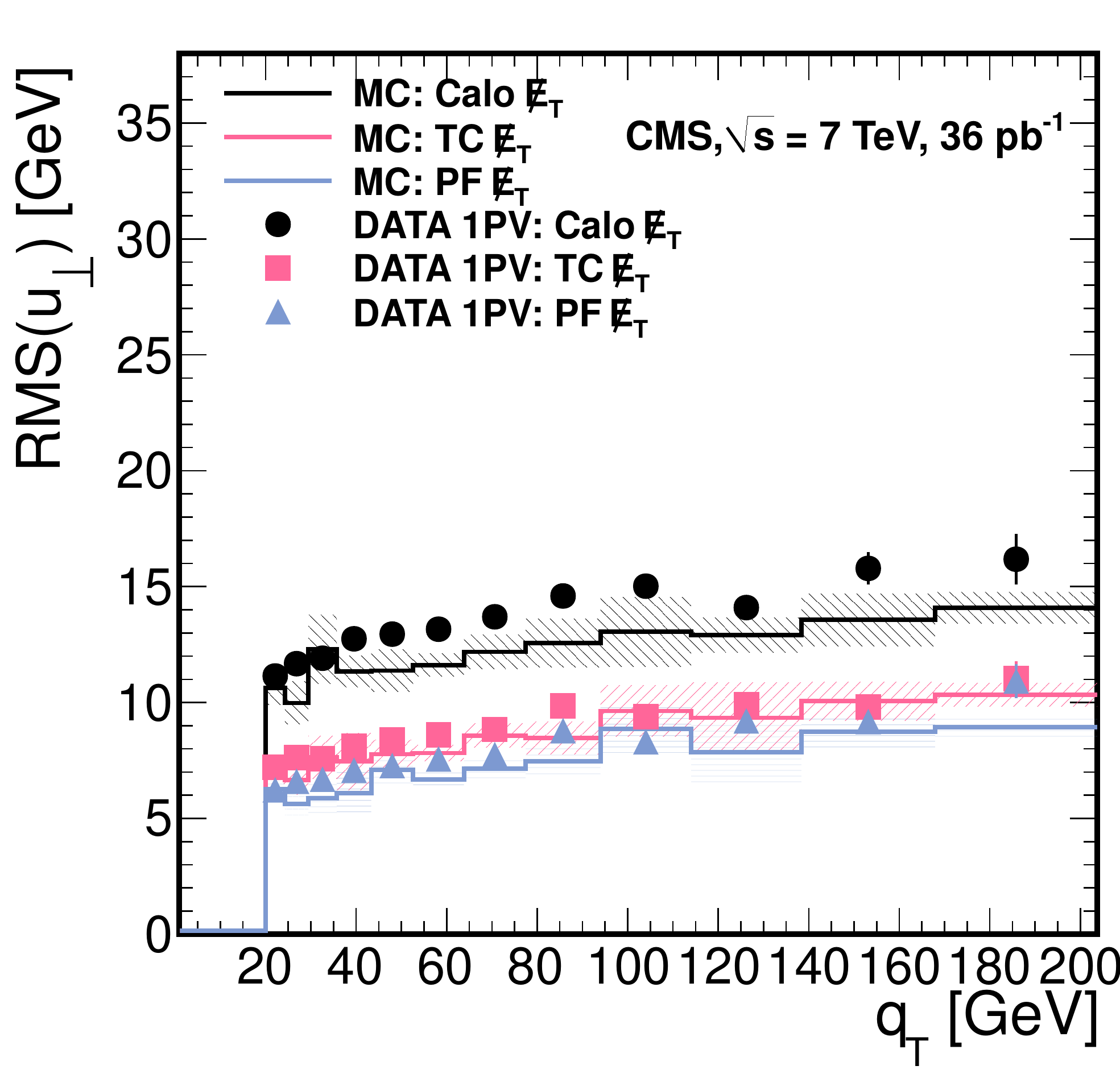} }
      \end{center}
        \caption{Resolution curves for components of hadronic recoil measured in
direct photon candidate
        events with one primary vertex. (left) parallel to boson; (right)
        perpendicular to boson. Data and simulation are indicated by points and histograms, respectively.
        Black circles (upper): \calomet; Pink squares (middle): \tcmet; Blue triangles (bottom): \pfmet. Shaded regions
        indicate statistical uncertainties on the simulation.}
       \label{f:PJresolution}
    \end{figure}

\begin{figure}[!h]
    \begin{center}
        \includegraphics[width=0.90\textwidth]{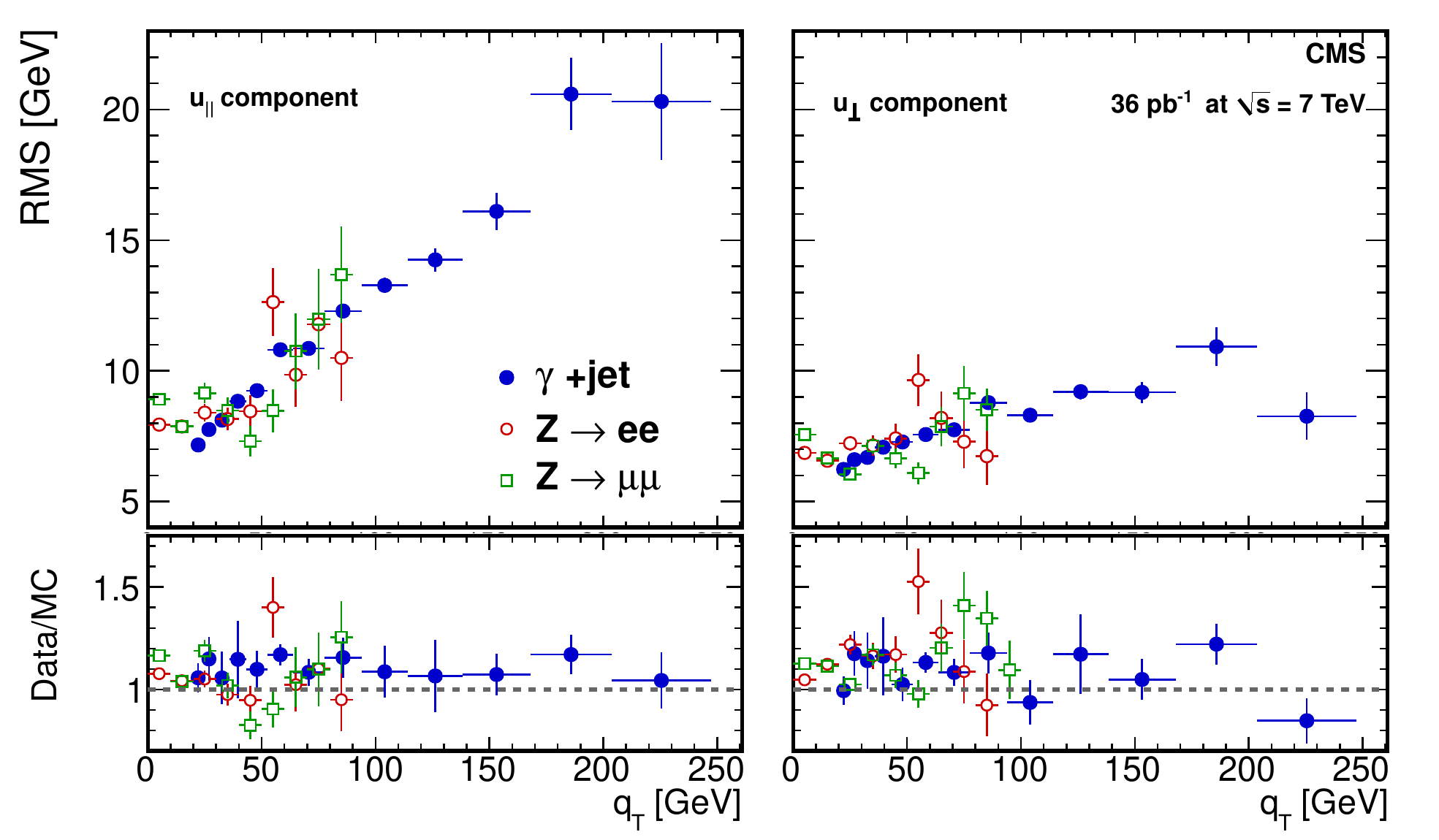}
      \end{center}
        \caption{Resolution curves for components of hadronic recoil 
calculated using \pfmet\, 
measured in  events with one primary vertex. Left: parallel to boson; right:
        perpendicular to boson. \Zmm, \Zee and $\gamma$ events are indicated by open green squares, open red circles and full blue circles, respectively. 
        The lower frame indicates the ratio of data to simulation. 
The vertical axis labels at the far left apply to both subfigures.}
       \label{f:PJresolution2}
    \end{figure}

Figure~\ref{f:PJresponse} shows the response curves, $|\langle \upar  \rangle |/\qt$ versus \qt, extracted from data, for the three
\met reconstruction algorithms, \calomet, \tcmet, and \pfmet
for $\gamma$, \Zee, and \Zmm samples.  The agreement
in response between the different samples is good.
The agreement between data and simulation is good, and
the results indicate that the three reconstruction algorithms are
distinct in their capabilities, performing differently in the recovery
of hadronic activity in the detector. The response for
\calomet is slightly larger than one because the jet energy scale used
in the type-I corrections was determined from a sample with a mixture
of quark and gluon jets, while for these samples the leading jet is primarily 
a quark jet.  
The \tcmet response is lower because it has neither type-I nor type-II corrections.
The \pfmet response is lower than the \calomet response at low values of \qt
because \calomet has type-II corrections while \pfmet has only
type-I corrections.

Figure~\ref{f:PJresolution} shows the resolution curves from
photon candidate events for \upar
and \uperp for data and simulation, for all three reconstruction
algorithms.  Figure~\ref{f:PJresolution2} shows the resolution as
measured in $\gamma$, \Zee, and \Zmm events for \pfmet.  
The measured resolution must be corrected for the scale to avoid a misleading
result; 
e.g., the apparent resolution on \uperp\ is proportional to the scale and therefore
an algorithm with a scale that is smaller than unity could appear to have a
better resolution than one with a scale of unity without such a correction. 
Since only \calomet has been corrected fully for the detector response with
both type-I and type-II corrections, 
the resolution measurements are rescaled, bin by bin, using the corresponding response
curves of Fig.~\ref{f:PJresponse}. 
The data confirm the 
prediction from simulation that tracking information significantly enhances the \vecmet
resolution. The resolutions as measured in the different samples are in
good agreement, but are $\approx$
10\% worse than expected from the simulation.  A similar difference in resolution
for jets for the 2010 run is documented in \cite{JME-10-014}. 
The small discrepancies between data 
and simulation shown in Fig. \ref{f:recoil_perp_gammajet} are due to
this difference.

\subsection{Resolution in multijet events}

The \met resolution can also be evaluated in events with a purely hadronic final state, where
the observed \met arises solely from resolution effects.
Because the \met\ resolution has a strong dependence on the associated
\sumet, it is presented as a function of $\sum E_T$. We characterize
the \vecmet\ resolution using the $\sigma$ of a Gaussian fit to
the distribution of the $x$ and $y$ components of \vecmet ($\eslash_{x,y}$).
In order to make a meaningful
comparison, we calibrate the measured \met\ for the different algorithms to the same scale using
the response from Fig. \ref{f:PJresponse}.
These corrections would not be needed if all types of \vecmet had both
type-I and type-II corrections.

For $\sum E_{\rm T}$, we use the PF \sumet as measured by
the particle-flow algorithm for all types of $\met$, as it gives
the best estimate of the true \sumet, and hence is an accurate evaluation of the event activity.
We use PF \sumet for all algorithms to ensure their measure is the same.
We calibrate PF \sumet\ to the
particle-level $\sum E_T$, on average,
using the predicted average mean value as a function of the particle-level \sumet from
a simulation of events from
the \PYTHIA 8 event generator~\cite{pythia8}.

\begin{figure}[t!]
  \begin{center}
  \begin{tabular}{cc}
    \includegraphics[angle=90,width=0.60\textwidth]{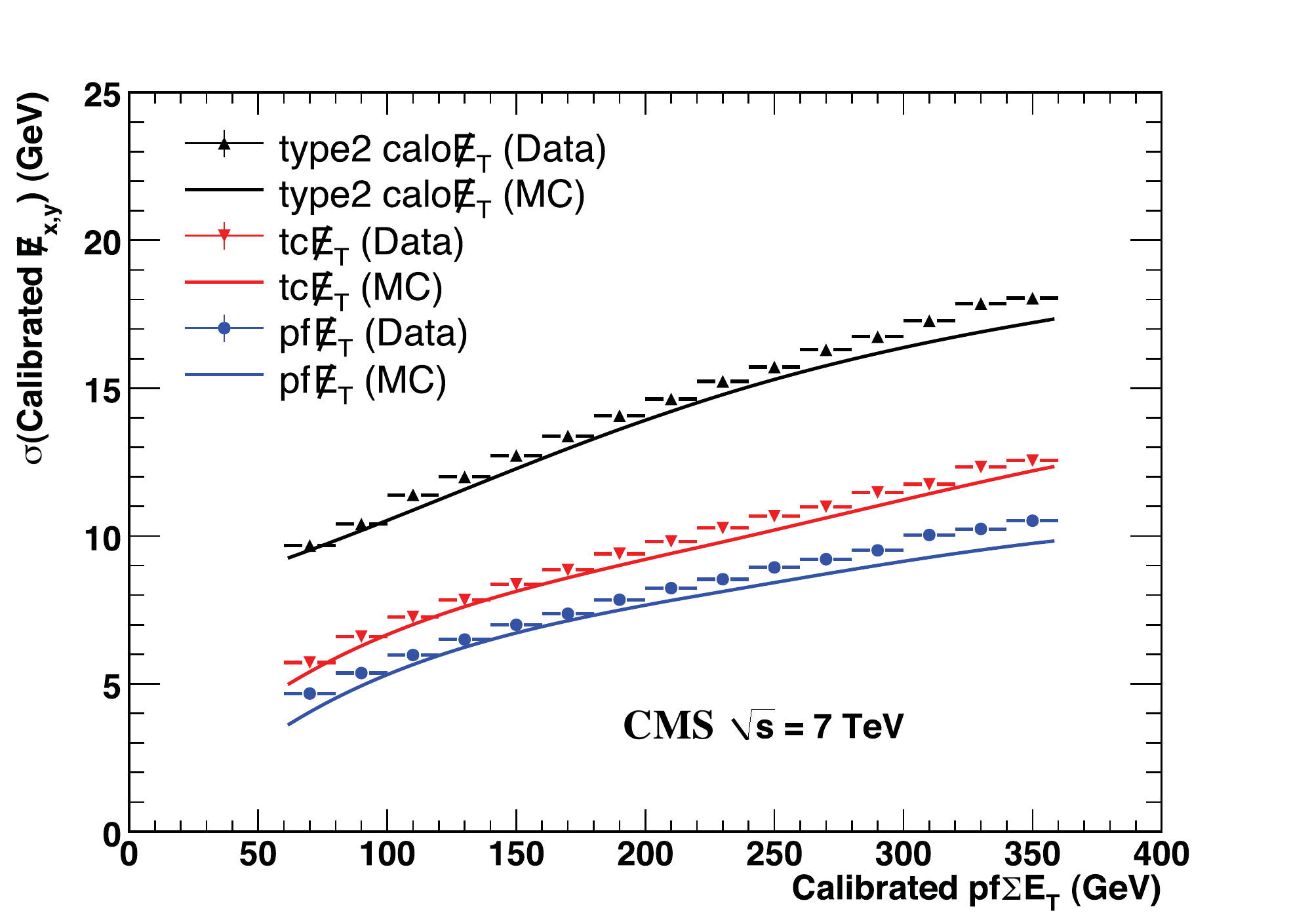}
  \end{tabular}
    \caption{Calibrated $\eslash_{x,y}$ resolution versus calibrated PF \sumet\ for
      \calomet, \tcmet, and \pfmet
      in data and in simulation.}
    \label{fig:CalibratedDistributions}
  \end{center}
\end{figure}

Figure~\ref{fig:CalibratedDistributions} shows the calibrated $\eslash_{x,y}$
Gaussian core resolution versus the calibrated PF \sumet\ for different
\met\ reconstruction algorithms in events containing at least two jets
with $p_{\rm T}>25$ GeV.
Both \tcmet and \pfmet show improvements in the \met\
resolution compared to the \calomet,
and the \pfmet yields the smallest $\met$
resolution.

\begin{figure}[h!]
 \begin{center}                     
  \includegraphics[angle=0,height=0.39\textwidth]{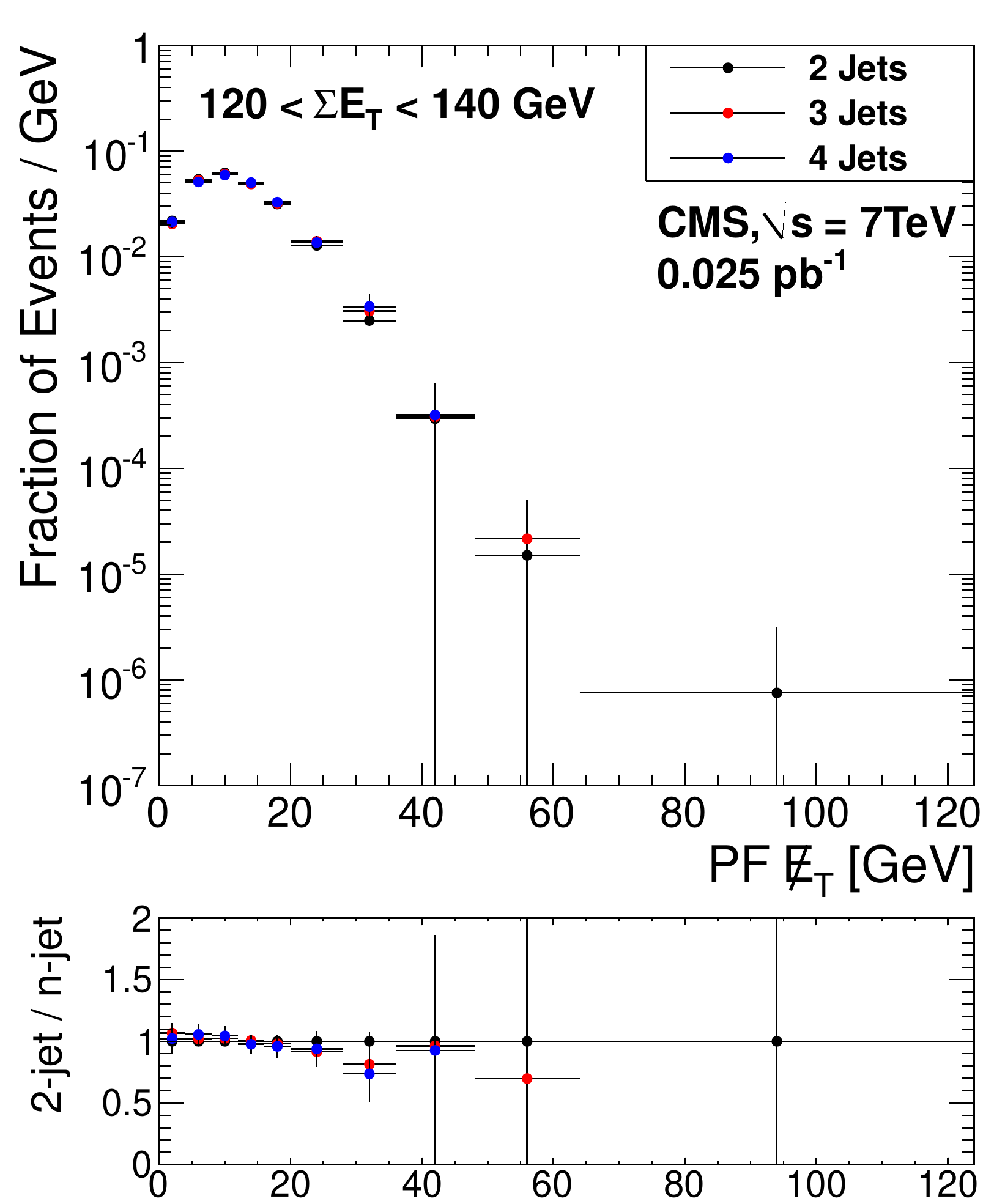}
  \includegraphics[angle=0,height=0.39\textwidth]{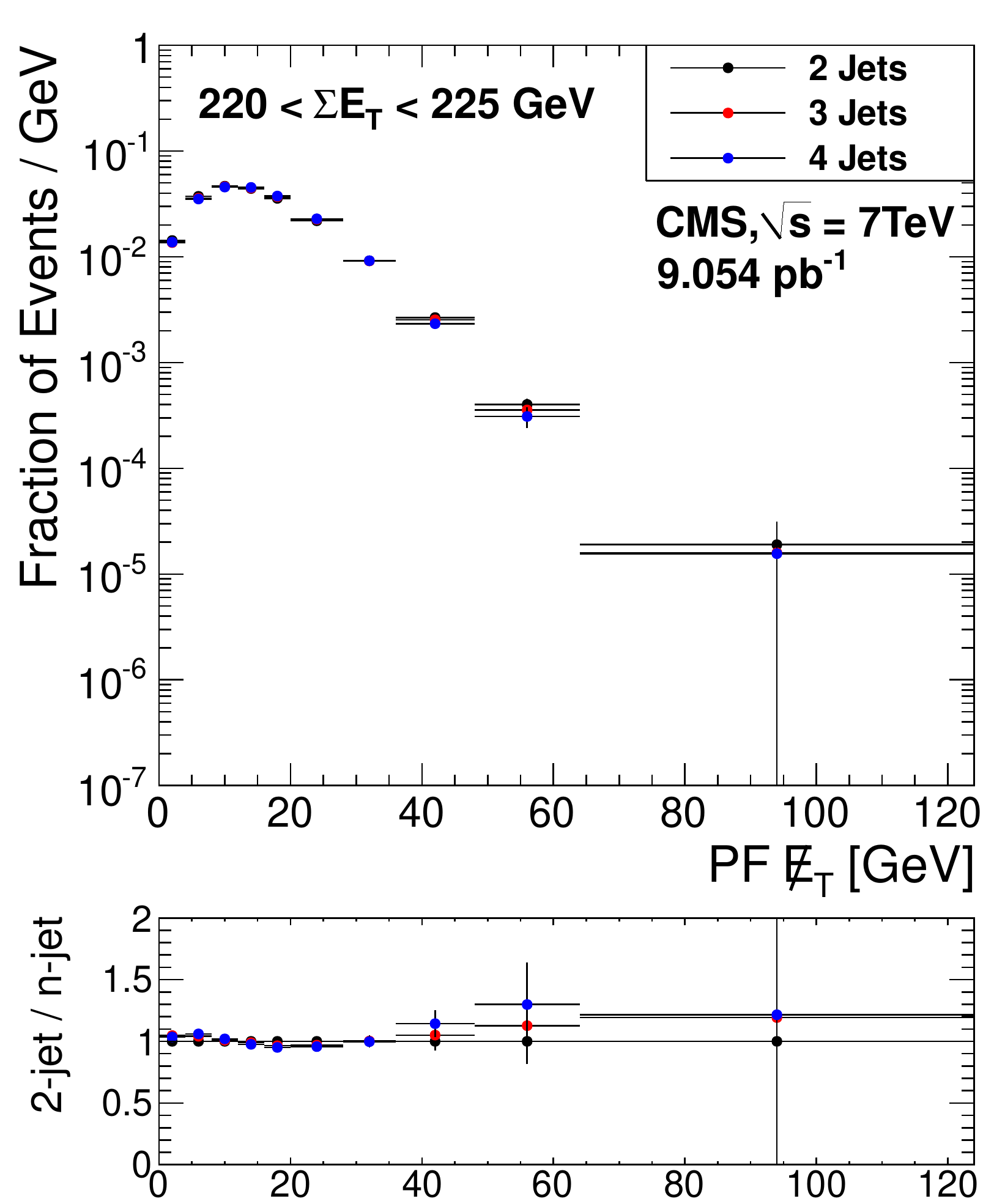}
  \includegraphics[angle=0,height=0.39\textwidth]{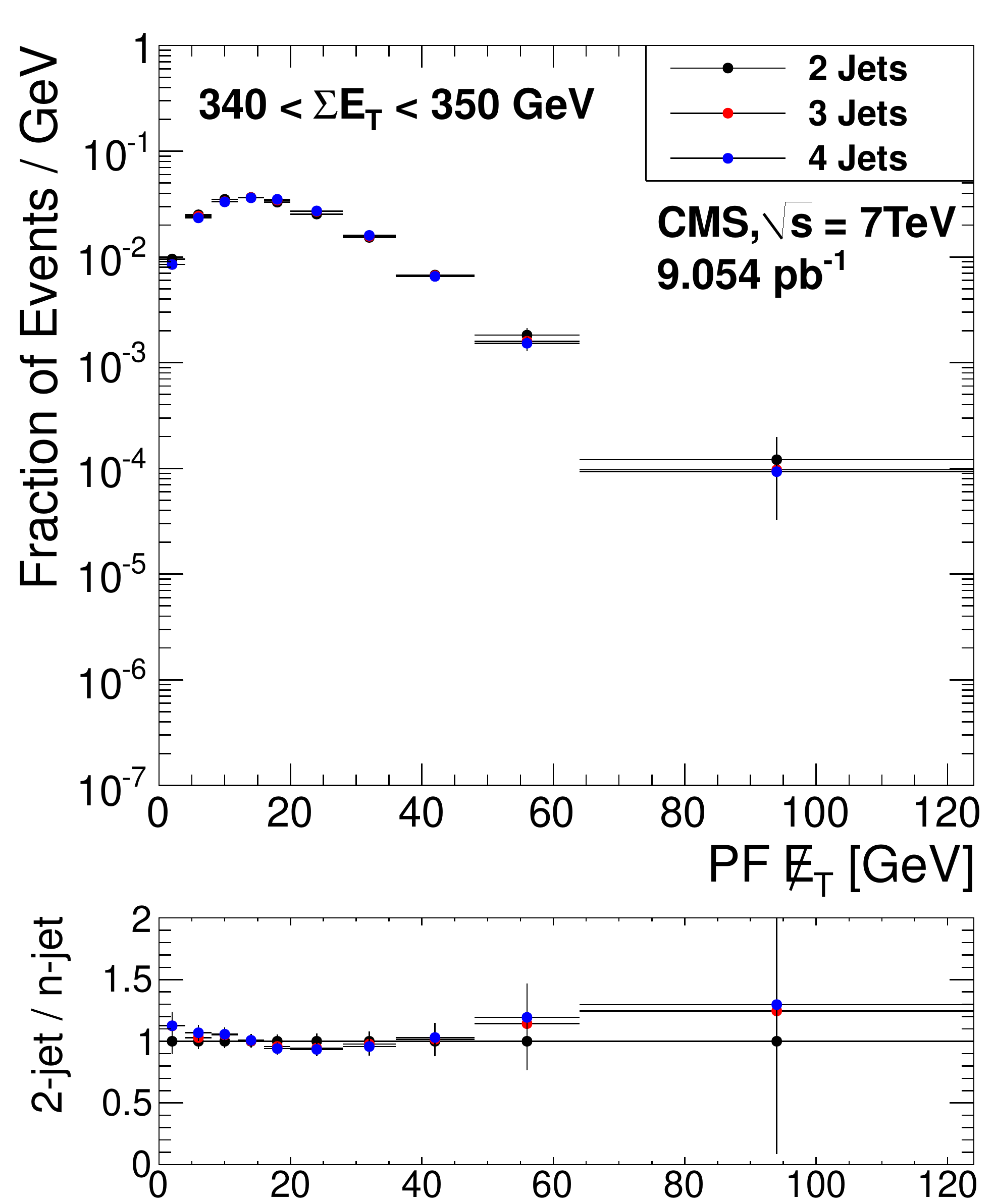}
 \end{center}                                   
 \caption{\pfmet distributions in 2-, 3- and 4-jet events, in selected \calosumet bins.}
 \label{fig:Met_MultiJet}
\end{figure}

Figure~\ref{fig:Met_MultiJet} shows the \pfmet distributions for different 
intervals of \calosumet and for jet multiplicities varying from two to four,
normalized to the same area. 
The jets are required to be above a \pt-threshold of $20\GeV$. 
The good agreement of the normalized shapes in Fig.~\ref{fig:Met_MultiJet} 
indicates that \pfmet-performance in events without genuine \met
is driven by the total amount of 
calorimetric activity (parametrized by \calosumet) and no residual non-linear 
contribution from jets to \pfmet is visible.
Similar behaviour is also observed for \calomet and \tcmet.

\subsection{Effect of multiple interactions}
\label{sec:multint}
Pile-up, namely multiple proton collisions within the same bunch crossing, occurs
because of high LHC bunch currents and can play an important role in \vecmet
performance. 

Because there is no true \vecmet in minimum bias events and because the average value for
a component of \vecmet in these events is zero (e.g., the $x$ or $y$ component),
pile-up should have only a small effect on the scale of the component of the measured \vecmet\
projected along the true \vecmet\ direction.
Pile-up, however, will have a considerable effect on the resolution of the parallel
and perpendicular components.

We investigate the effect of pile-up using multijet samples, $\gamma$, and
Z data.

\subsubsection{Studies of pile-up effects using photon and Z events}
\label{sec:zpileup}

\begin{figure}[!h]
    \begin{center}
        { \includegraphics[width=0.31\textwidth]{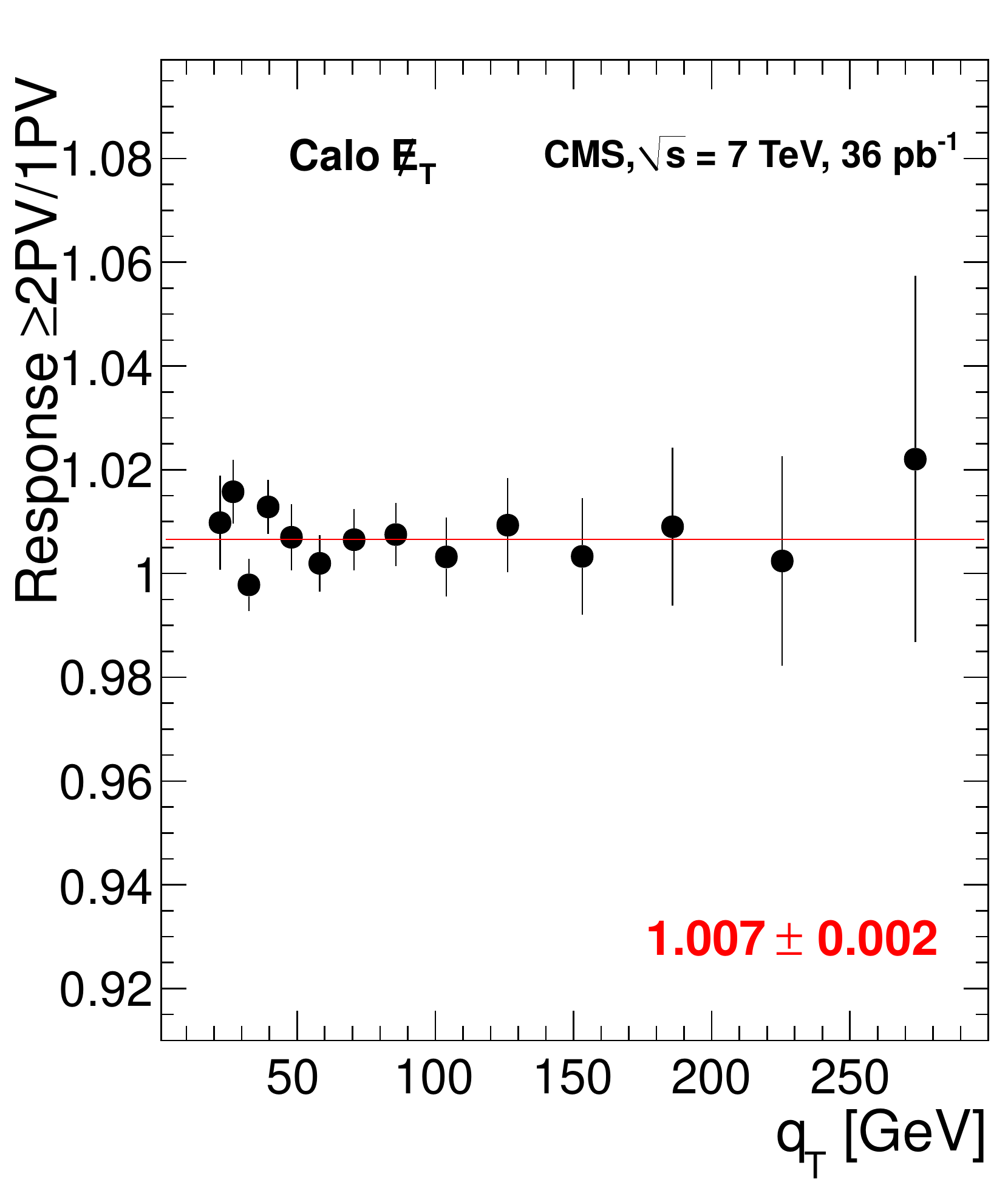} }
        { \includegraphics[width=0.31\textwidth]{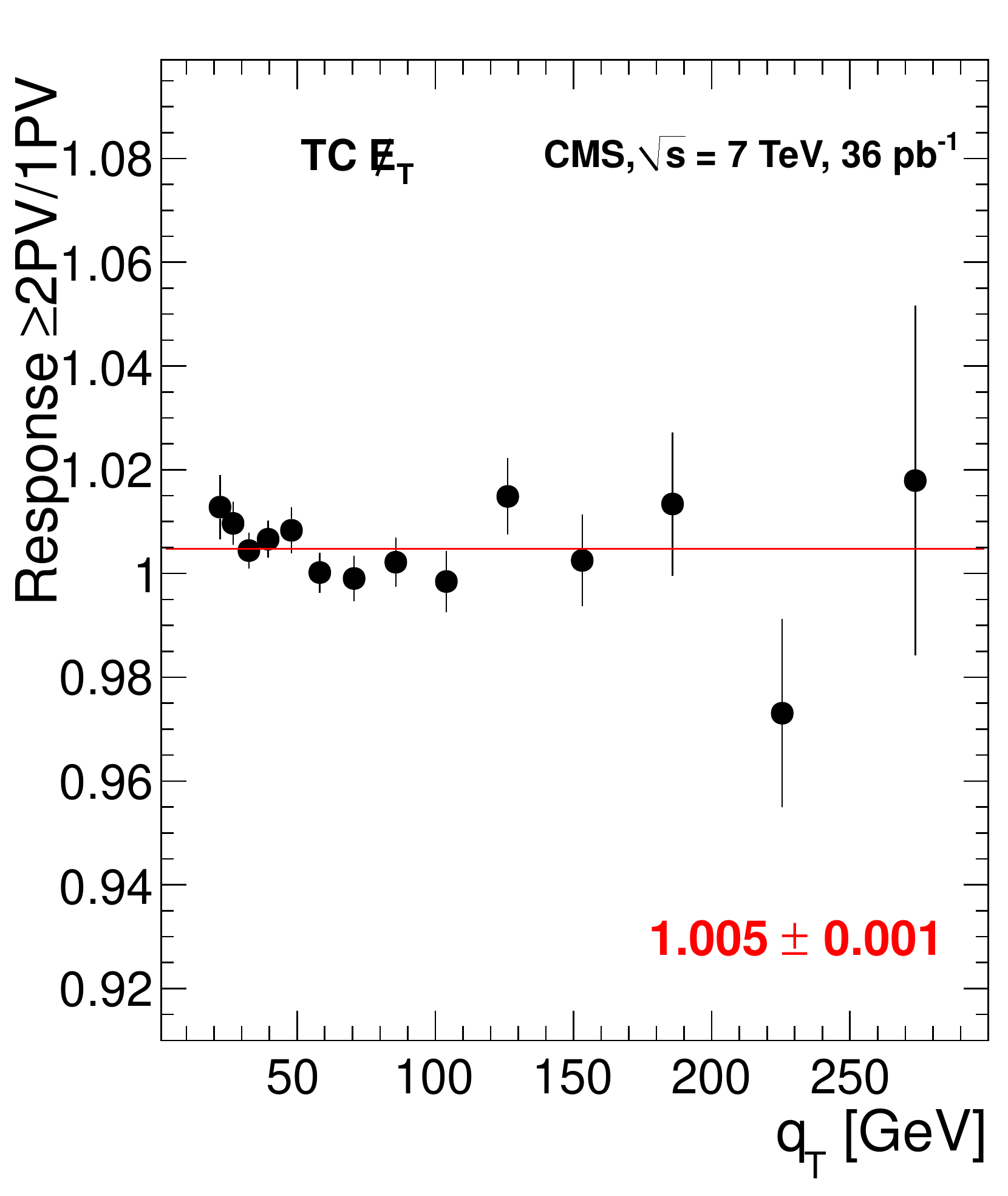} }
        { \includegraphics[width=0.31\textwidth]{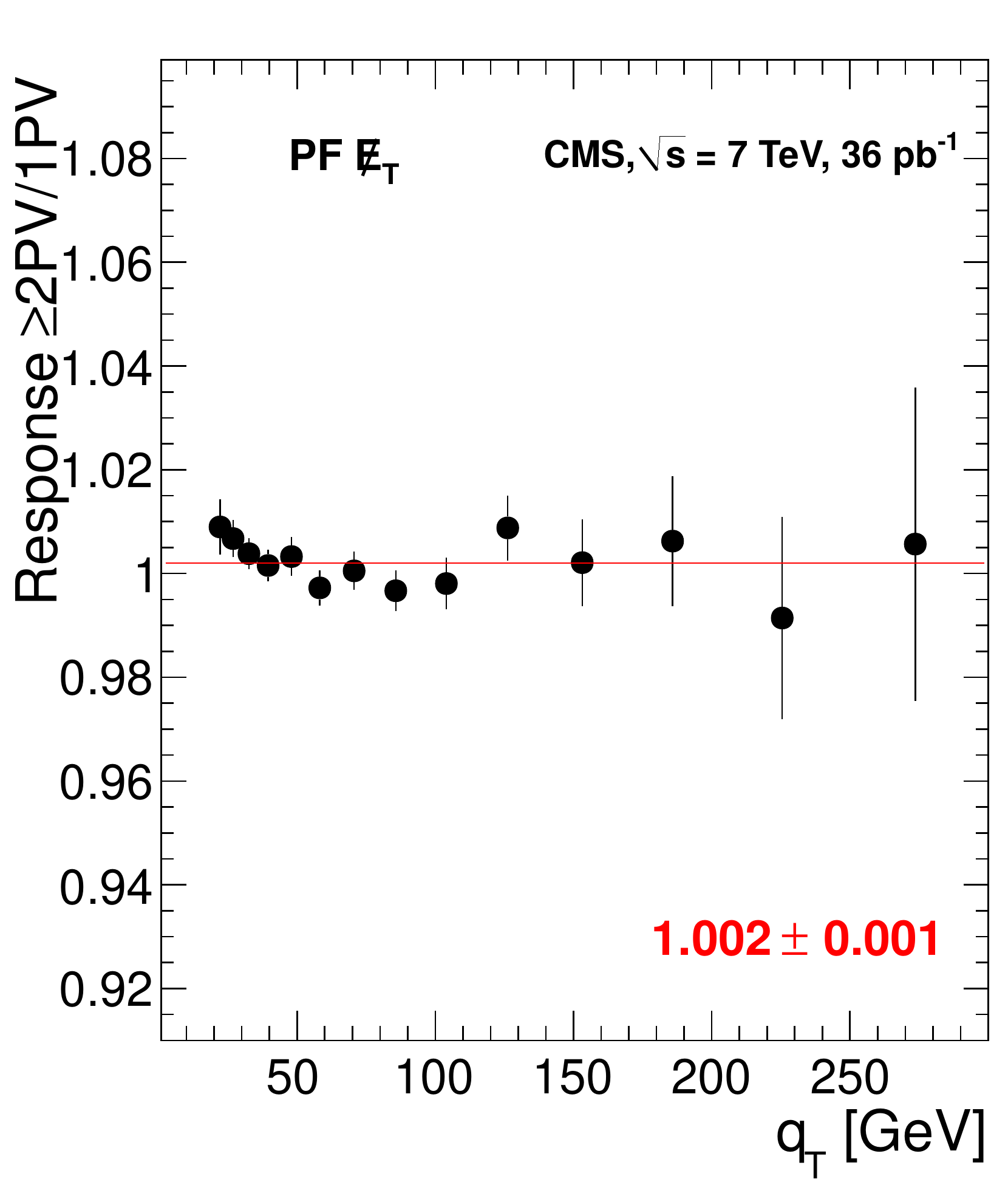} }
    \end{center}
        \caption{The ratio of the response for the component of the induced \vecmet along the boson direction,
measured in $\gamma$ events for events containing 1 PV and 
at least 2 PVs for 
(left) \calomet, (middle) \tcmet, and (right) \pfmet. 
Also given is the best fit value for the average ratio, which corresponds to the solid, red line.}
       \label{f:PJp1}
\end{figure}

In this section, we use samples containing a vector boson to measure the effect of pile-up
on the scale and resolution of a component of \vecmet.
Figure \ref{f:PJp1} shows the ratio of the response as measured in $\gamma$ events
for events containing 1 PV and at least 2 PVs.  The ratio of the responses
of the component of the measured $\vecmet$ along the boson direction is close to one, as
expected.  It is slightly larger at low
\qt when pile-up is present.  This is expected, as pile-up can reduce energy lost due to
zero suppression in the readout of the calorimeter if energy from a pile-up interaction
and from the hard scattering are both in the same readout channel.
If the sum is larger than the zero suppression thresholds, 
more of the energy from the hard
scattering is recorded.

\begin{figure}[!h]
    \begin{center}
        { \includegraphics[height=0.3\textwidth]{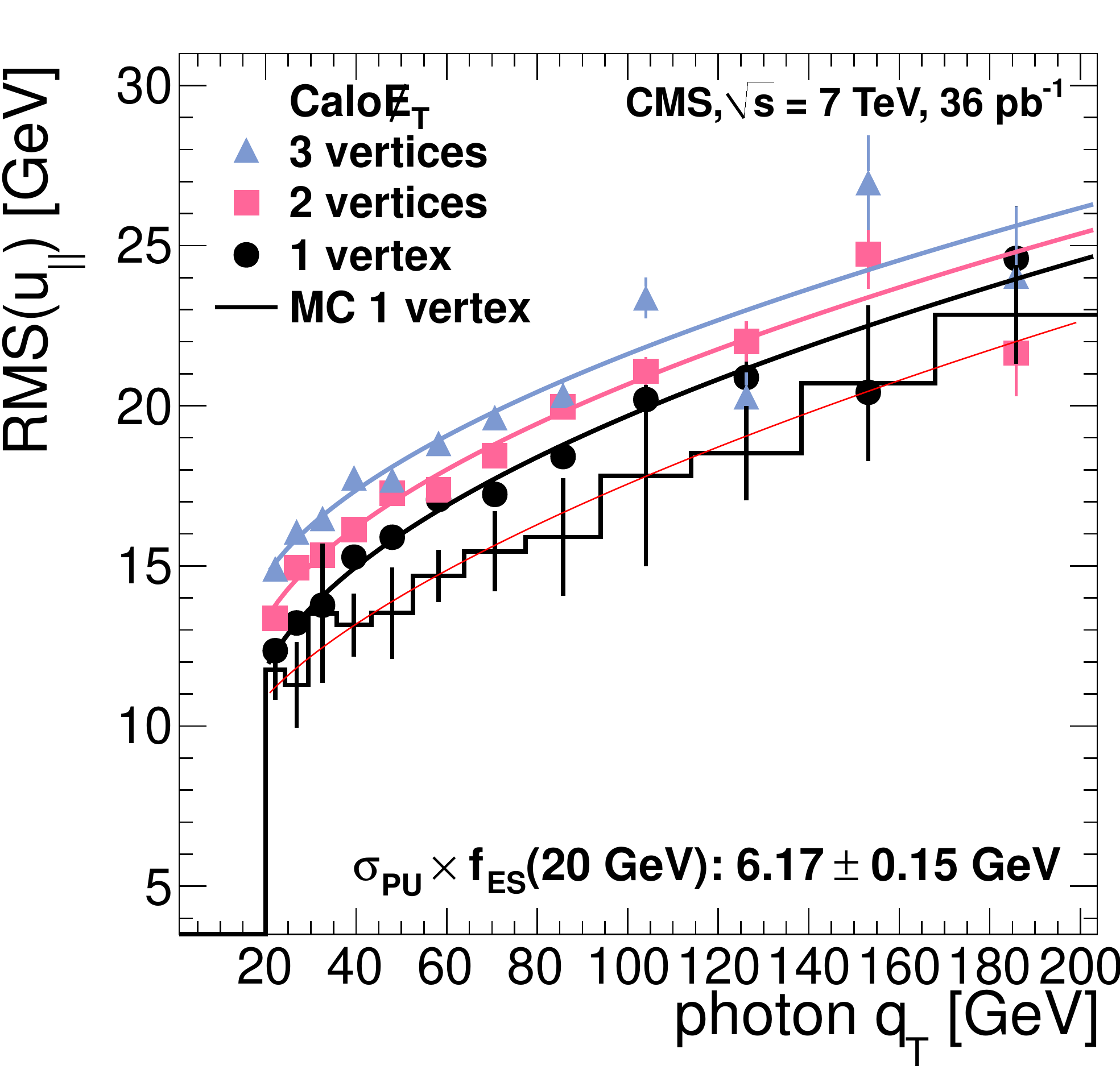} }
        { \includegraphics[height=0.3\textwidth]{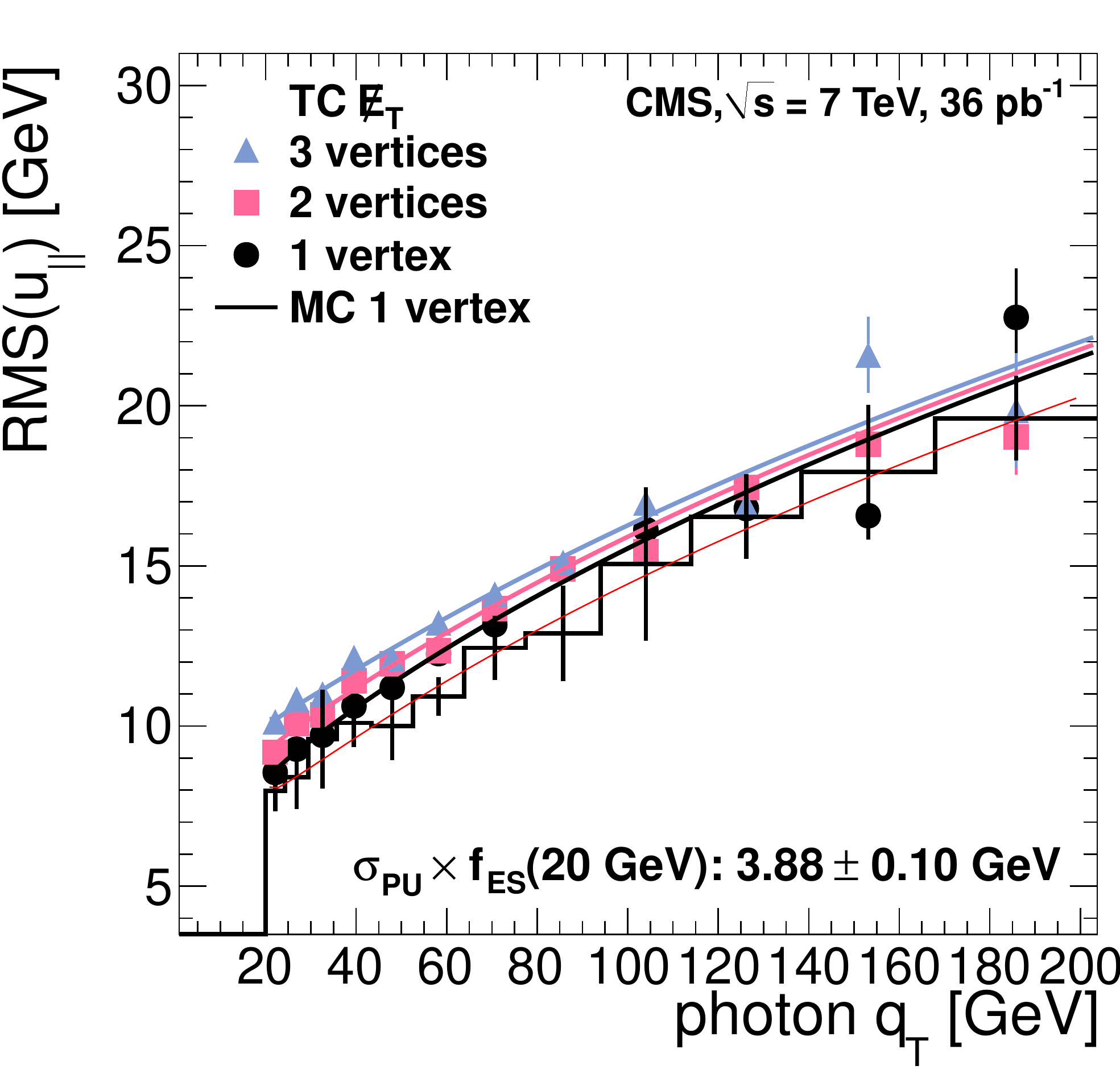} }
        { \includegraphics[height=0.3\textwidth]{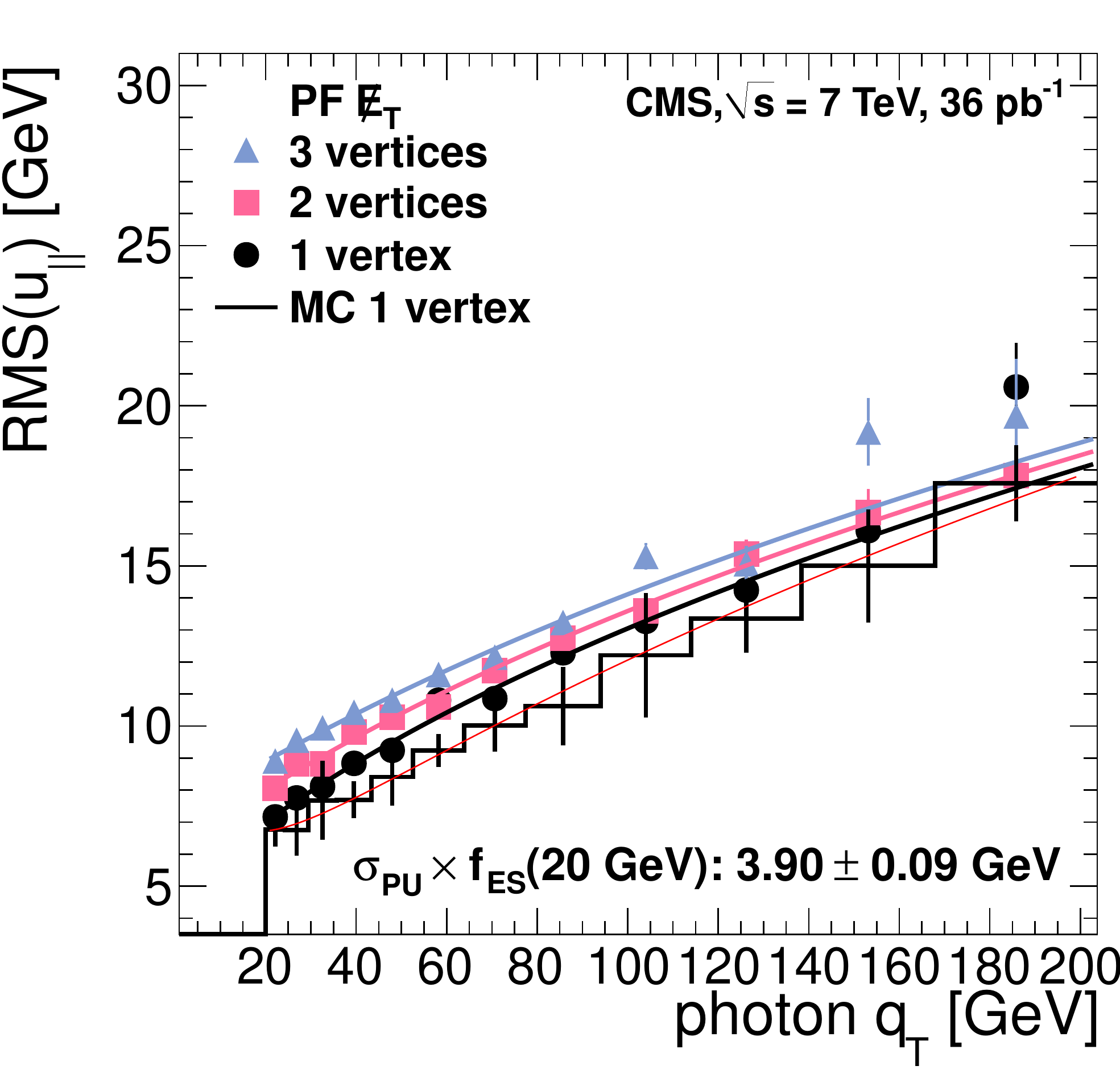} }\\
        {\includegraphics[height=0.3\textwidth]{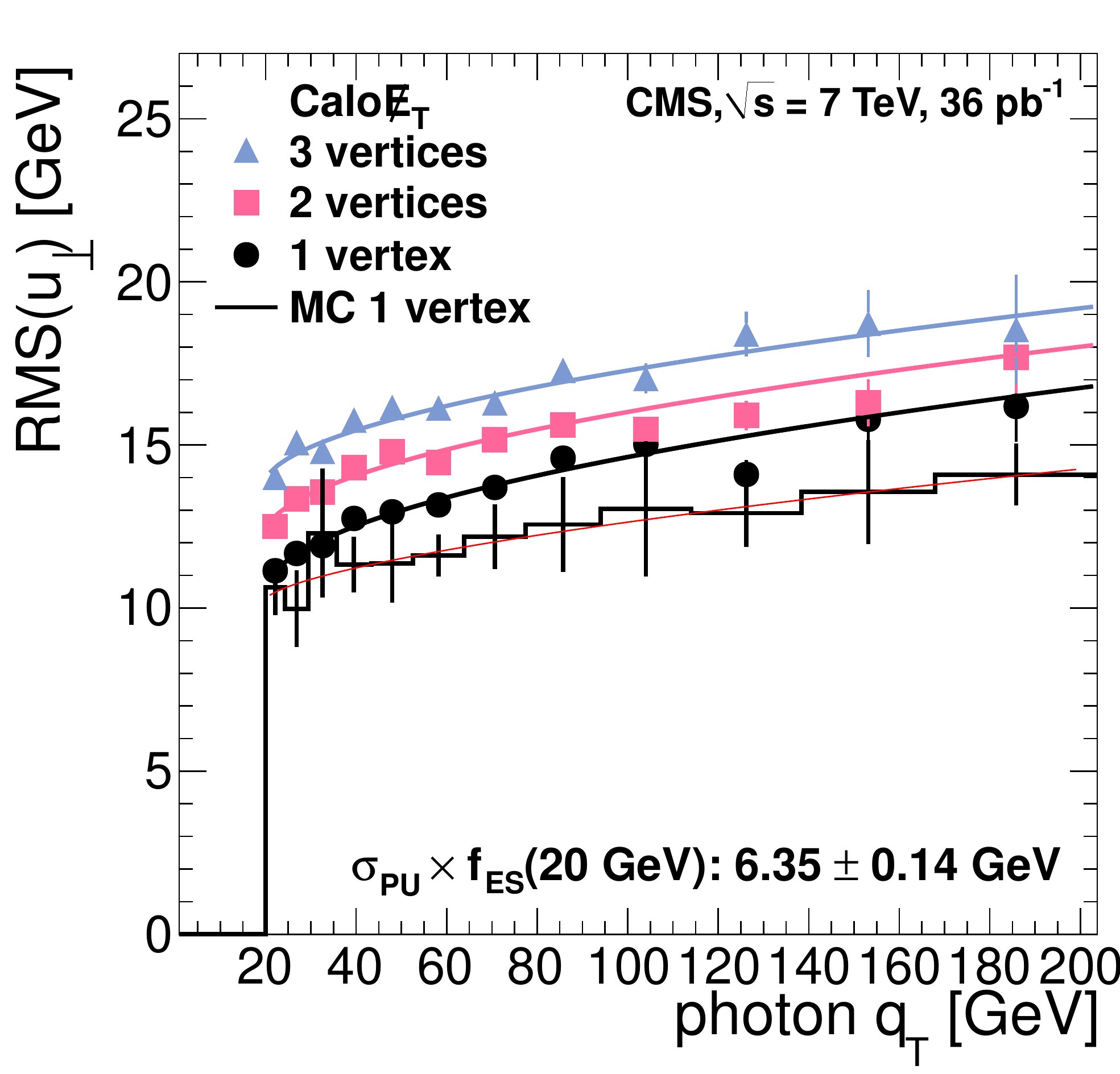}}
        { \includegraphics[height=0.3\textwidth]{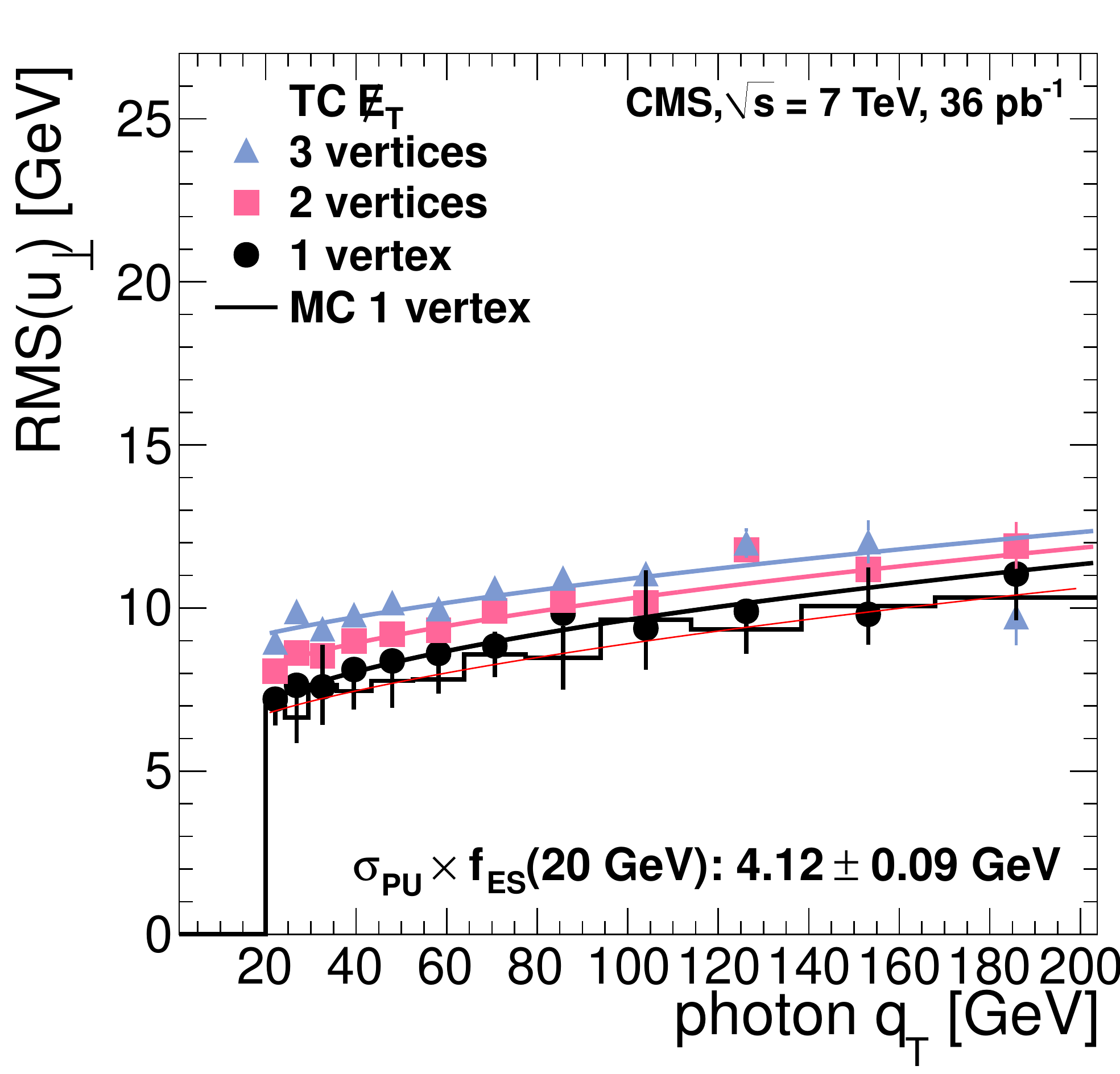} }
        { \includegraphics[height=0.3\textwidth]{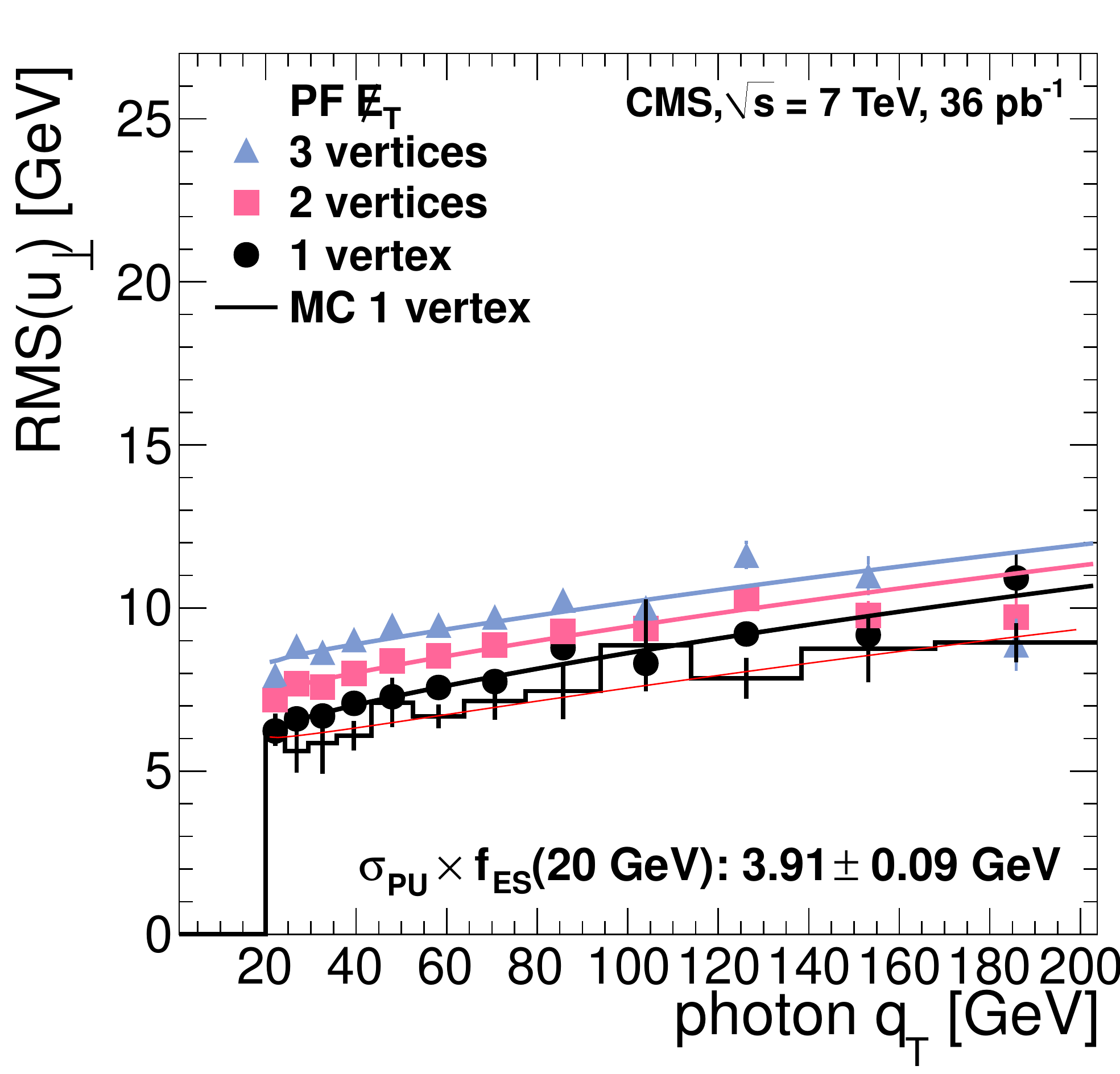} }
      \end{center}
        \caption{
Resolution versus photon $q_{\rm T}$ for the parallel component (top)
and perpendicular component (bottom)
for (left to right) \calomet, \tcmet, and \pfmet, for
events with 1 (circles), 2 (squares), and 3 (triangles) reconstructed
primary vertices. }
       \label{f:PJf2}
\end{figure}

\begin{figure}[!h]
    \begin{center}
        { \includegraphics[height=0.3\textwidth]{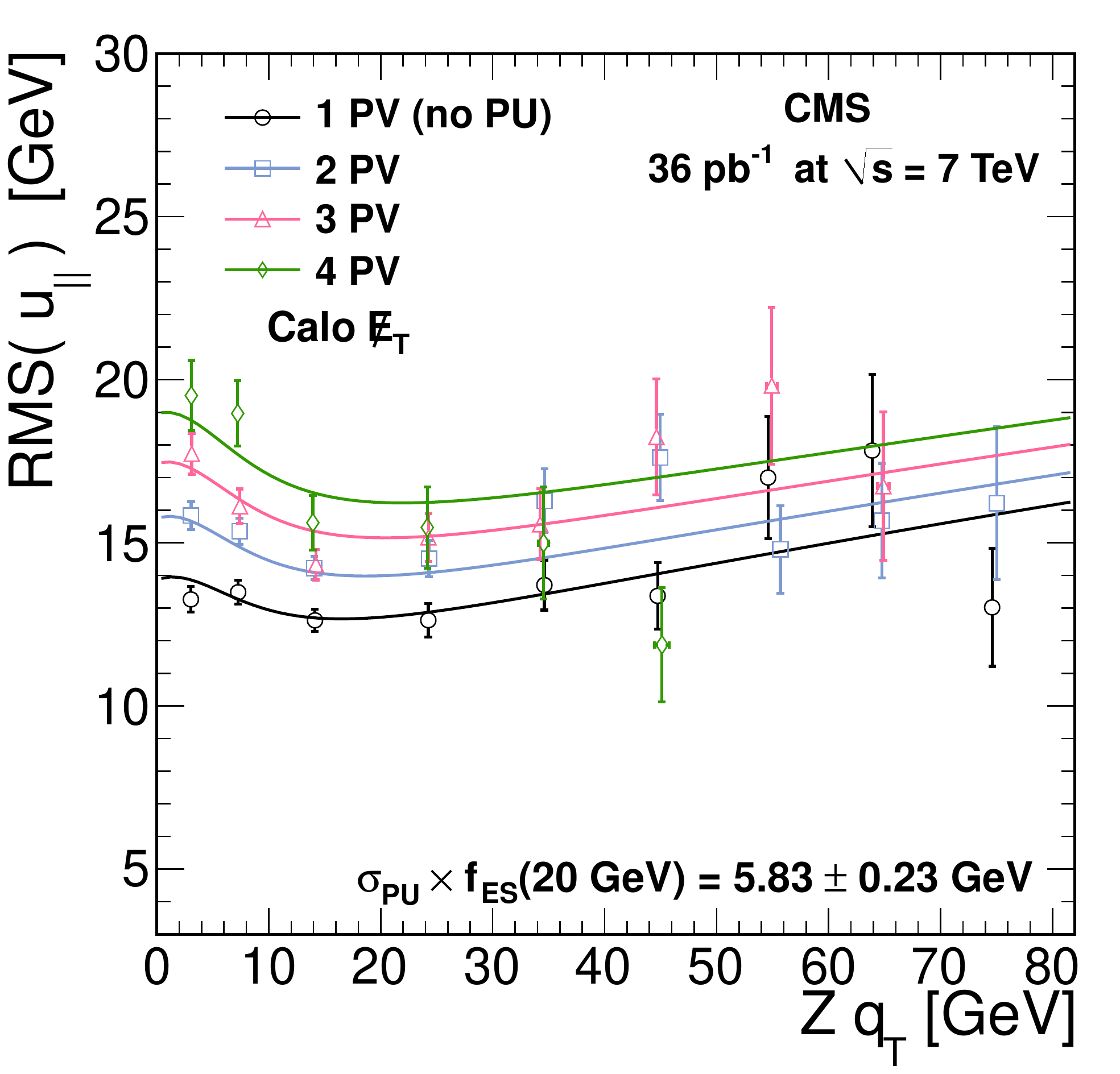} }
        { \includegraphics[height=0.3\textwidth]{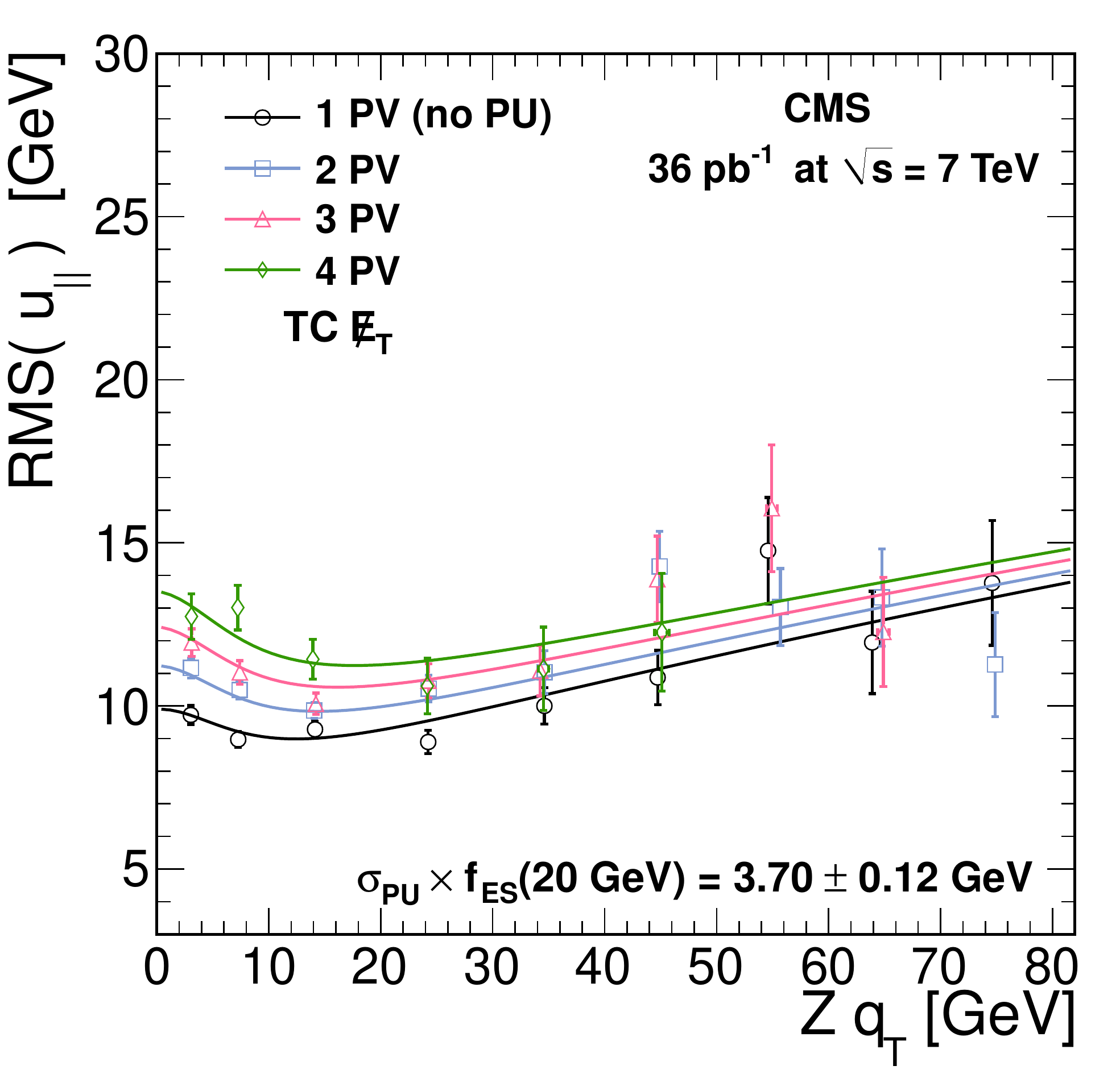} }
        { \includegraphics[height=0.3\textwidth]{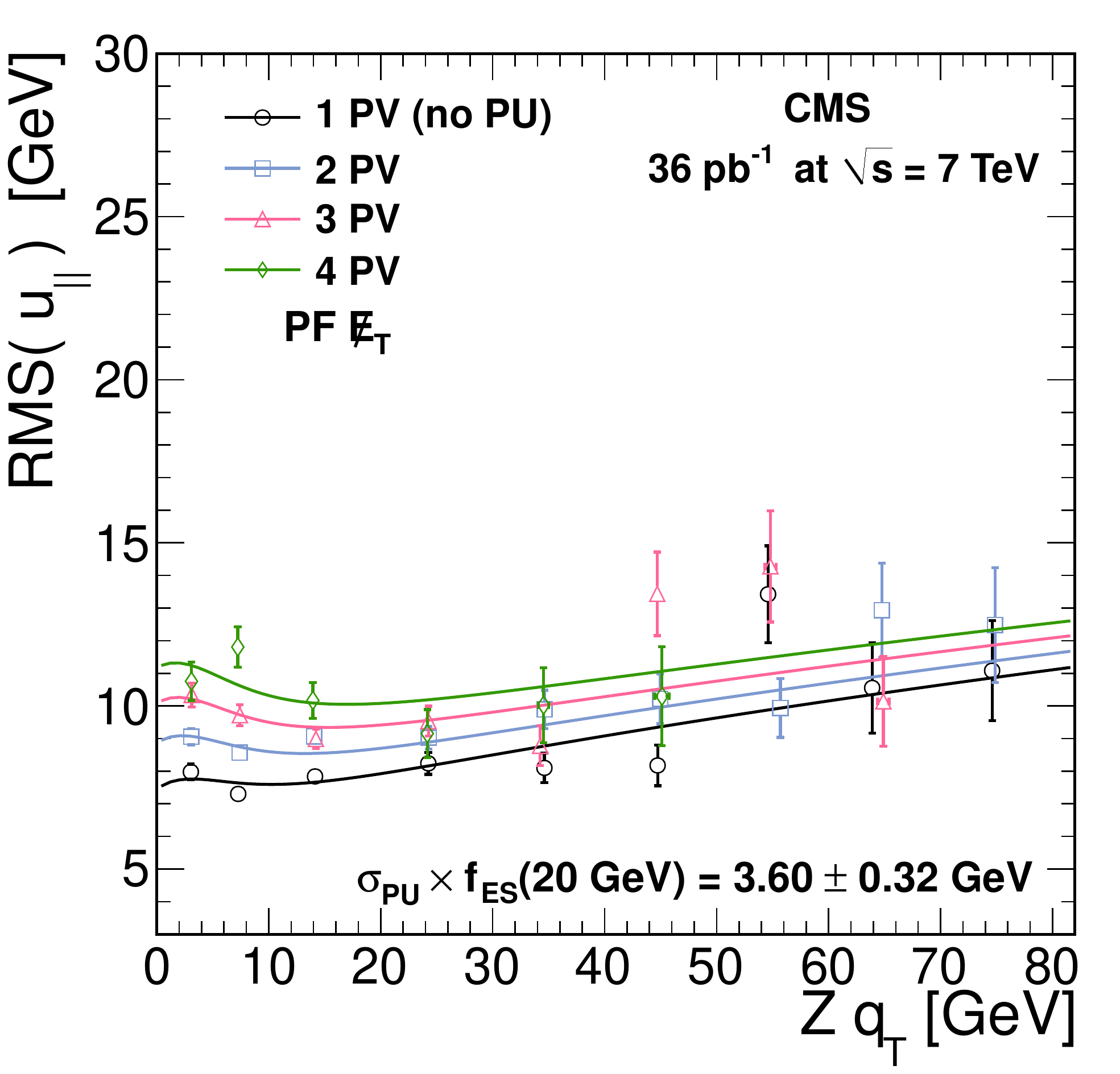} }\\
        {\includegraphics[height=0.3\textwidth]{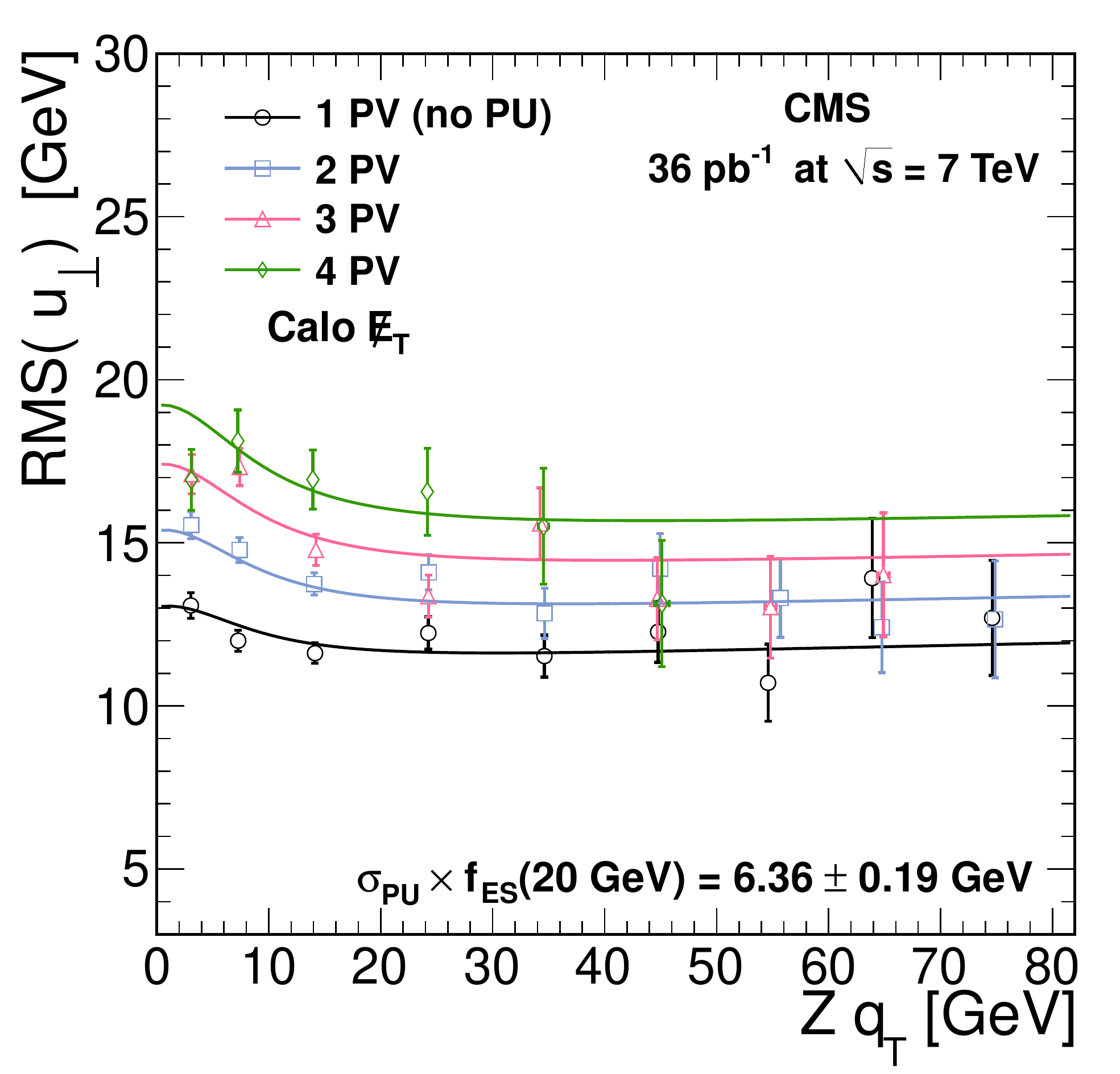}}
        { \includegraphics[height=0.3\textwidth]{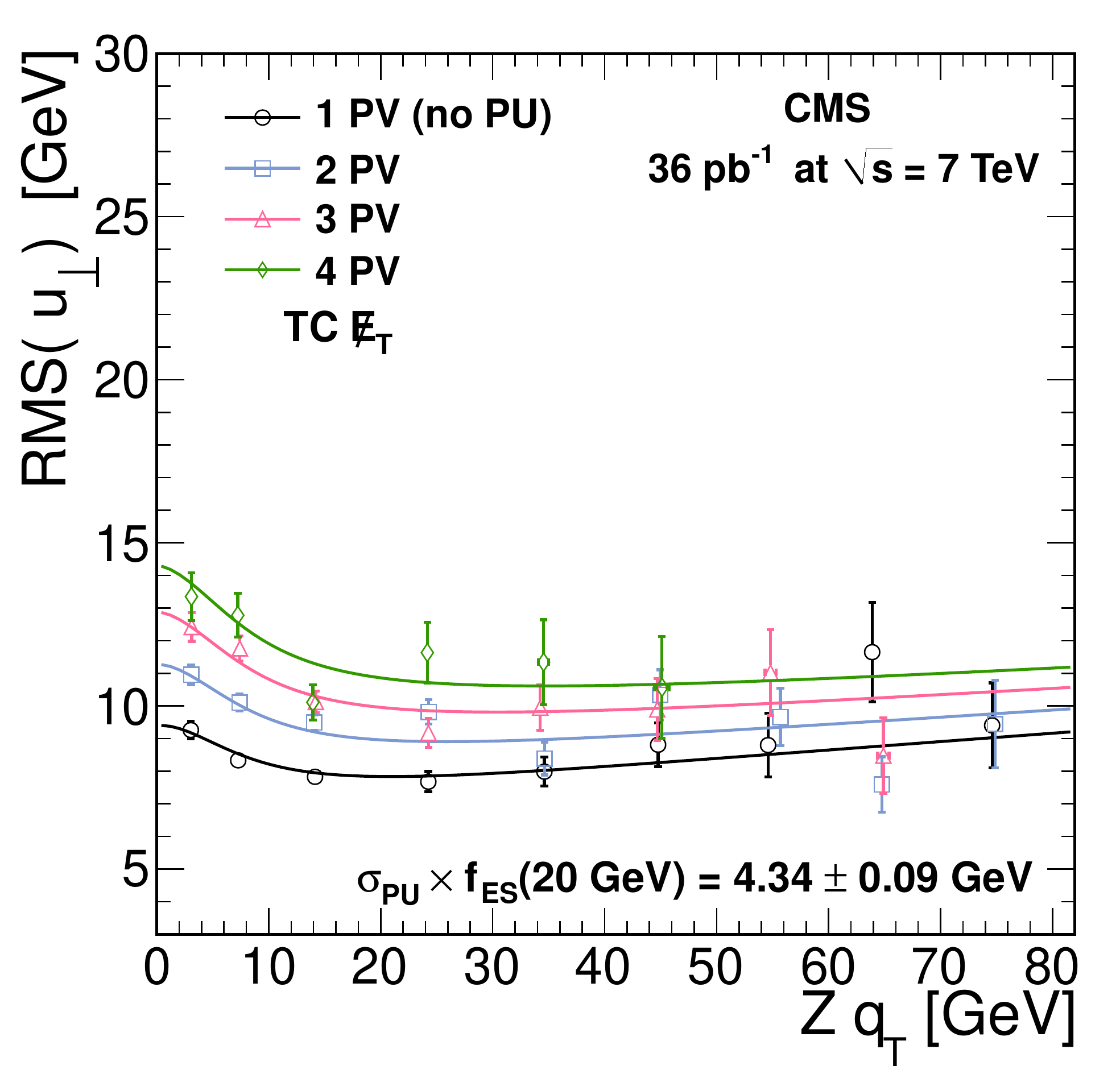} }
        { \includegraphics[height=0.3\textwidth]{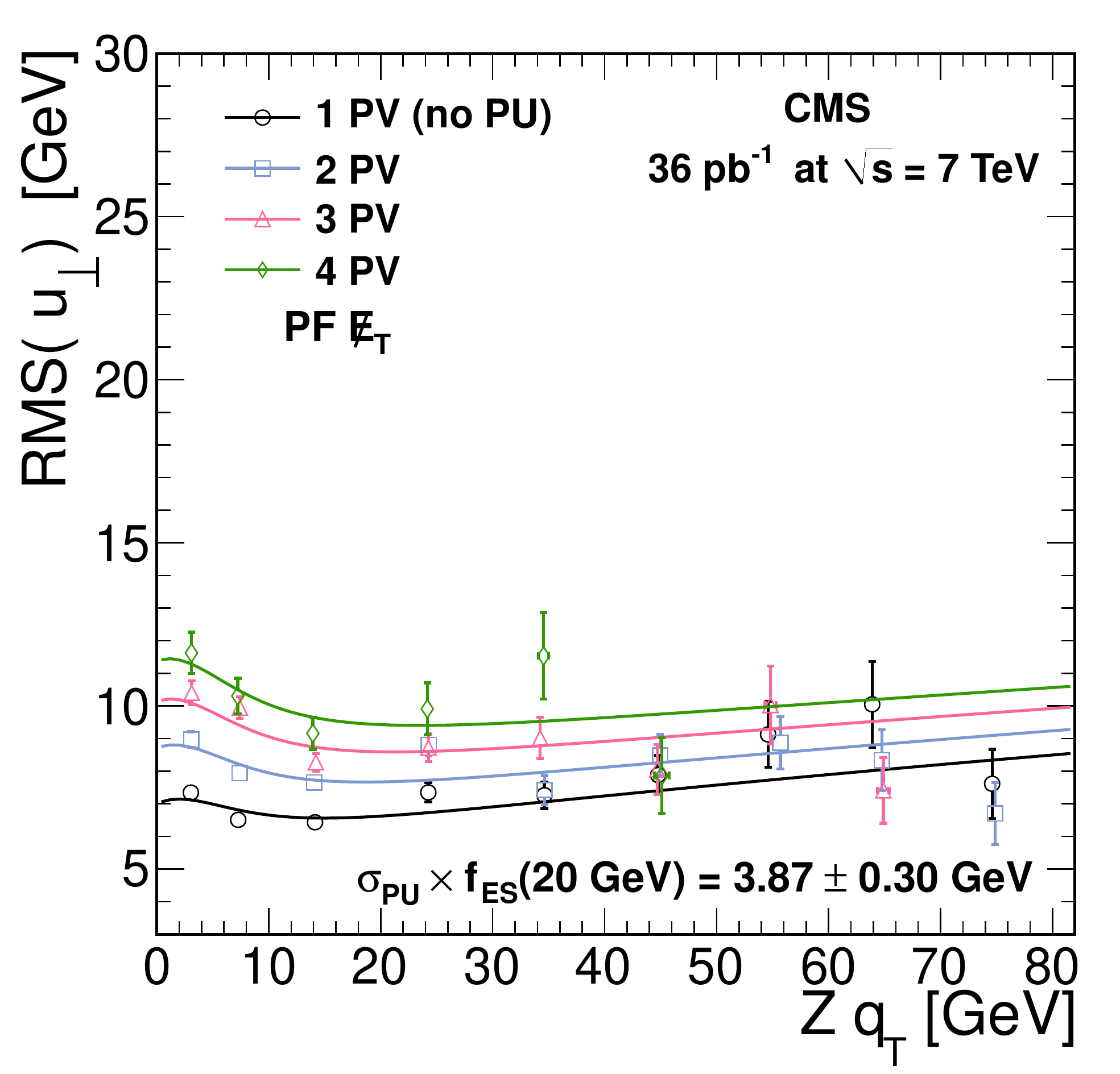} }
      \end{center}
        \caption{
Resolution versus the \qt of the Z for the parallel component (top)
and perpendicular component (bottom)
for (left to right) \calomet, \tcmet, and \pfmet,
for events with 1 (circles), 2 (squares), 3 (triangles), 
and 4 (diamonds) reconstructed primary vertices.}
  \label{f:ZMPVRes}
\end{figure}

Figure \ref{f:PJf2} shows the resolution versus the \qt of the $\gamma$ 
for the components of the hadronic recoil 
parallel and perpendicular to the boson direction 
for 1, 2, and 3 reconstructed PVs.
Also shown is the prediction from simulated $\gamma$ events without pile-up.
Figure~\ref{f:ZMPVRes} shows the resolution versus the \qt of the Z for the parallel
and perpendicular components of the hadronic recoil. 
The parameterization of \met resolution used in Figs.~\ref{f:PJf2} and \ref{f:ZMPVRes}
is given by :

\begin{equation}
 \sigma_{\rm total}^2 = ( a\sqrt{q_{\rm T}} + b )^2 \ + \ \left(\sigma_{\rm noise}f_{\rm ES}(q_{\rm T})\right)^2 + \ (N-1)\left(\sigma_{\rm PU}f_{\rm ES}(q_{\rm T})\right)^2 \label{eq:GZReso}
\end{equation}

where $a$ and $b$ characterize the hard process, $\sigma_{\rm noise}$ is the intrinsic noise resolution, $N$ is the number of reconstructed vertices in the event, $\sigma_{\rm PU}$ is the 
intrinsic pile-up resolution, and $ f_{\rm ES}(q_{\rm T})$ is the energy scale correction applied on each event.
At low  $q_{\rm T}$, the resolution is 
dominated by contributions from the underlying event and detector noise
($\sigma_{\rm noise}$).
Since these contributions can not be distinguished from those due to the particles
from the recoil, and since the recoil measurement needs to be corrected for
the detector response, these contributions are magnified and have a larger
contribution at low boson \qt when energy scale corrections are applied.
As expected, the resolution is degraded with increasing pile-up
interactions. Results from the 
Z and $\gamma$ channels are in agreement and are similar to  
the values 
obtained in Section~\ref{sec:jetpileup} from jet data.

\subsubsection{Studies of pile-up effects in jet data}
\label{sec:jetpileup}

\begin{figure}[h!]
 \begin{center}
 \includegraphics[totalheight=0.9\textheight,width=1\textwidth]{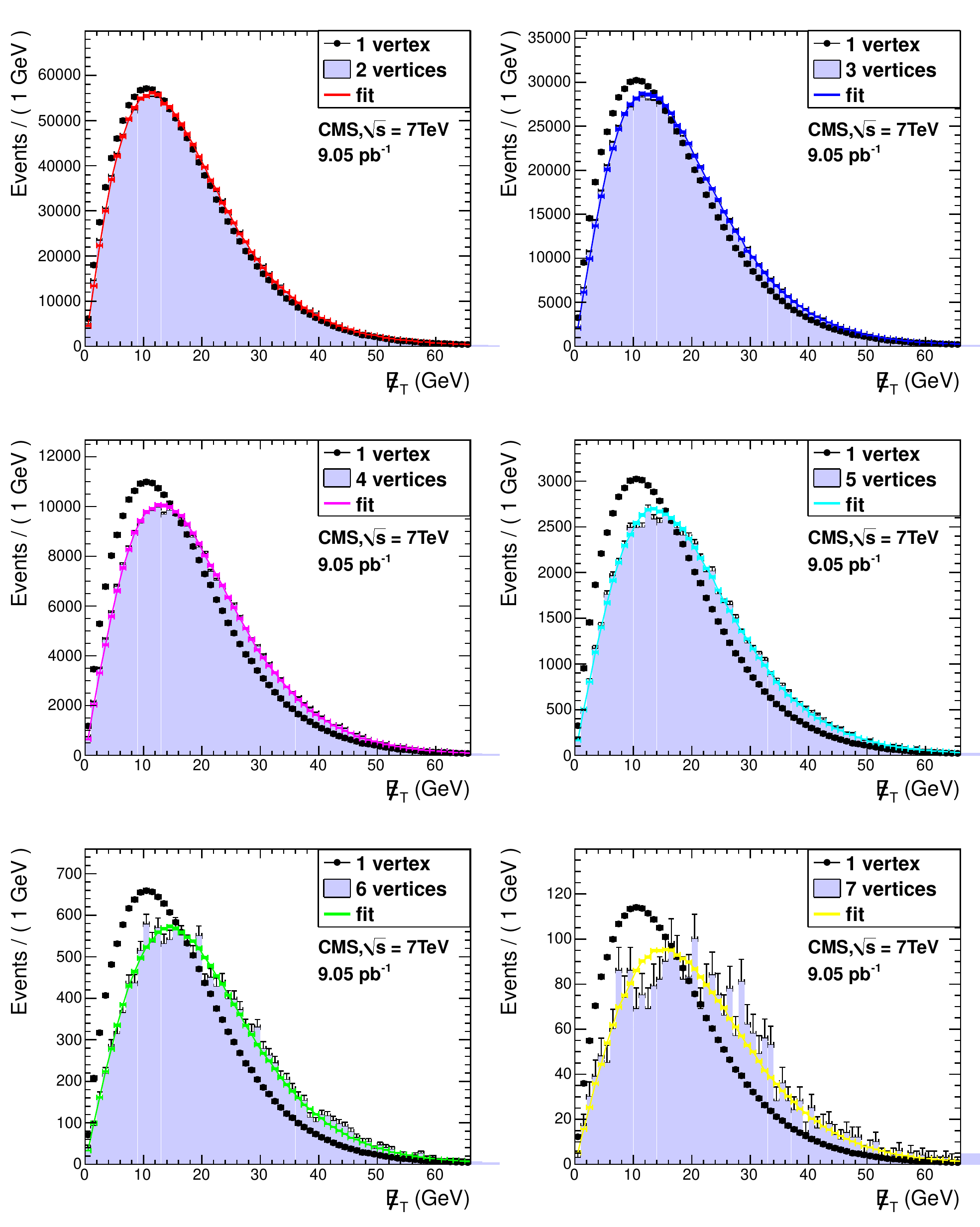}
 \end{center}
 \caption{\pfmet\ distributions in pile-up events. The figures show a comparison between the one- and $n$-vertex shapes ($n$ = 2...7) and the results of a simultaneous fit of Eq. (\ref{eq:vtx1convN}) to the $n$-vertex shapes.
The one-vertex distribution is normalized to the $n$-vertex distribution for each plot.}
 \label{fig:Met_smeared_PU}
\end{figure}

In this Section, we study the behaviour of the \pfmet distributions 
in samples containing high \pt jets when  pile-up is present.
The data are selected using a prescaled $H_{\rm T}$ trigger with a threshold of 100 GeV, where
$H_{\rm T}$ is defined as the scalar sum of the transverse momenta of PF jets ($\pt > 20$ GeV, $|\eta|<$3).
Additionally, in the offline analysis, 
each event is required to have $H_{\rm T}$ (calculated using PF jets)$ > 200$ GeV to avoid bias from the trigger. 
Figure~\ref{fig:Met_smeared_PU} shows that the widening of the \pfmet distribution 
with increasing number of vertices can be modeled by convoluting the $x$- and $y$-components 
of the one-vertex \MET\ shape with a Gaussian (G) whose mean is 
$(n-1)\cdot\Delta\mu_{x}$ and with standard deviation $\sqrt{n-1}\cdot\Delta\sigma_{x}$:

\begin{equation}
  \MET_{,n} = \sqrt{(\eslash_{x1}\otimes G[(n-1)\cdot\Delta\mu_{x},\sqrt{n-1}\cdot\Delta\sigma_{x}])^{2}+(\eslash_{y1}\otimes G[(n-1)\cdot\Delta\mu_{y},\sqrt{n-1}\cdot\Delta\sigma_{y}])^{2}}
  \label{eq:vtx1convN}
\end{equation}  

where $\eslash_{x,y}$ are the $x$ and $y$ components of \vecmet.
Here we assume that each additional vertex contributes with a constant $\Delta\sigma_{x}$ ($\Delta\sigma_{y}$) to the \MET\ resolution such that the resolution with $n$ pile-up interactions is related to that with one primary vertex via:
$\sigma_{xn}^{2} = \sigma_{x1}^{2} + (n-1)\Delta\sigma_{x}^{2}$. 
In addition we also allow for a linear shift of $\eslash_x$ and $\eslash_y$ by $\Delta\mu_{x}$ ($\Delta\mu_{y}$) such that $\mu_{xn} = \mu_{x1} + (n-1)\Delta\mu_{x}$. 
A fit of Eq. (\ref{eq:vtx1convN}) to data results in $\Delta\sigma_{x} = \Delta\sigma_{y} = 3.7$ GeV, consistent with the results from Section~\ref{sec:zpileup}.
This fit is performed simultaneously on the \MET distributions of events containing two to seven vertices.
The shifts of the $x$ and $y$ \MET components are estimated to be $\Delta\mu_{x} = 0.5$ GeV and
$\Delta\mu_{y} = -0.3$ GeV respectively, which are small compared to $\Delta\sigma$ and are consistent with the expected shift seen in simulation due to
non-functional channels.

\clearpage

\section{Studies of physics processes containing genuine \texorpdfstring{\bigmet}{MET}}
\label{sec:gmet}
In this section, we examine distributions relevant to \vecmet in events
containing neutrinos.  Events containing W bosons and b quarks are studied.

\subsection{W events}

The performance of \met is studied in events
that contain large, genuine \met:
${\rm W}\rightarrow \ell \nu$ events, where $\ell$ is a muon or electron.  For most W
events, 
the magnitude of
\vecmet\ is approximately equal to the
\pt of the charged lepton, but its resolution is dominated by the hadronic recoil.
When the W \qt is small compared to the W mass, the \met is approximately

$$\met \approx p_{\rm T}(\ell) - 0.5 u_\ell$$

where $u_\ell$ is the component of the recoil parallel to the lepton 
transverse direction.

In the \Wmn decay channel, events are required to have been collected by a  single-muon high-level
trigger. 
In addition, candidates are selected by requiring a muon with
$|\eta|<2.1$ that has
\pt$>25~$\GeV.                                     
Events with a second muon
with $\pt>25$\GeV are rejected to suppress Z and ${\rm t}\bar{\rm t}$ contamination.
\Wen decays are identified using similar selection criteria.
A single-electron high-level trigger
requirement with a \pt threshold of $15~\text{GeV}$ is applied.
Events are also required to contain an electron with
\pt$>25~$\GeV. 
Events with a second electron
with $p_{\rm T}>20$\GeV are rejected, and
rejection against $\gamma$ conversions is applied. 
A total of 24\,628 (29\,200)
\Wmn (\Wen) events with only one primary vertex are selected.

The main sources of background to the \Wln signal are jet events with one jet 
falsely identified as a
high-$p_{\rm T}$ muon or electron  and \Zll events with one lepton escaping
detection. The jet events usually have  low values of \met.
The apparent \met in these events (that have no genuine \met) 
is amplified by scale corrections to the \met (type-I and type-II)
because, since they are indistinguishable, 
 artificial \met receives the same scale correction factor
as genuine \met, and genuine \met tends to be underestimated without corrections.
Other
backgrounds include W and Z bosons decaying into $\tau$, followed by
$\tau\to\ell\nu\bar\nu$, and ${\rm t}\bar{\rm t}$ events, with one top quark decaying
semi-leptonically. The relative normalization of the different electroweak
(EWK) signal simulation event ensembles, and simulations of those backgrounds that
contain an electroweak boson
(\Wln, \Zll, ${\rm t}\bar{\rm t}$),
are set by the ratios of their
theoretical cross sections computed at next-to-leading order \cite{Gavin:2010az}. 
The normalization of the
composite EWK and the QCD contributions are established through
a one-parameter binned fit 
to the \met distribution from data.

\begin{figure}[!h]
    \begin{center}
      {\includegraphics[height=0.363\textwidth]{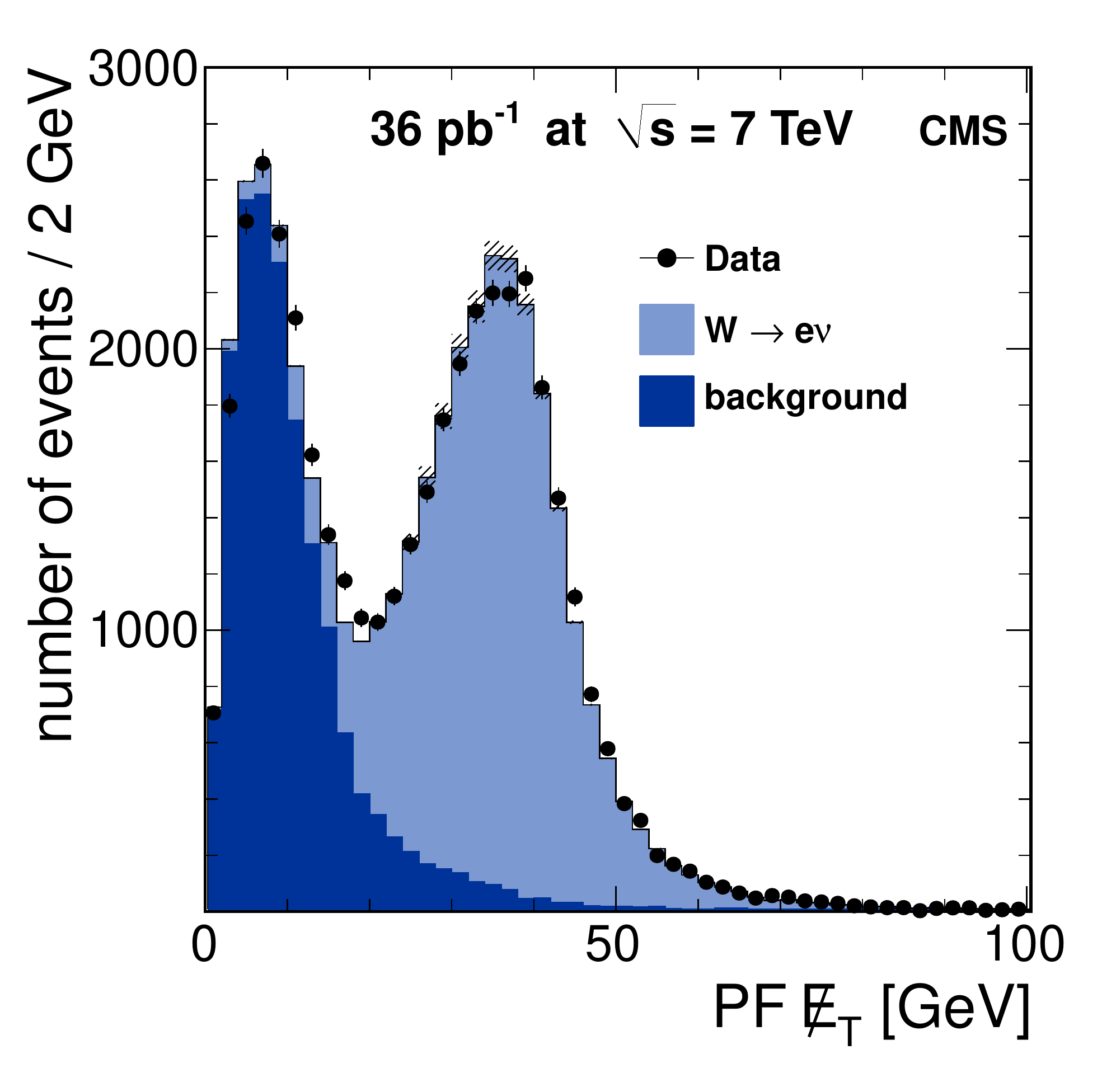}}
      {\includegraphics[height=0.363\textwidth]{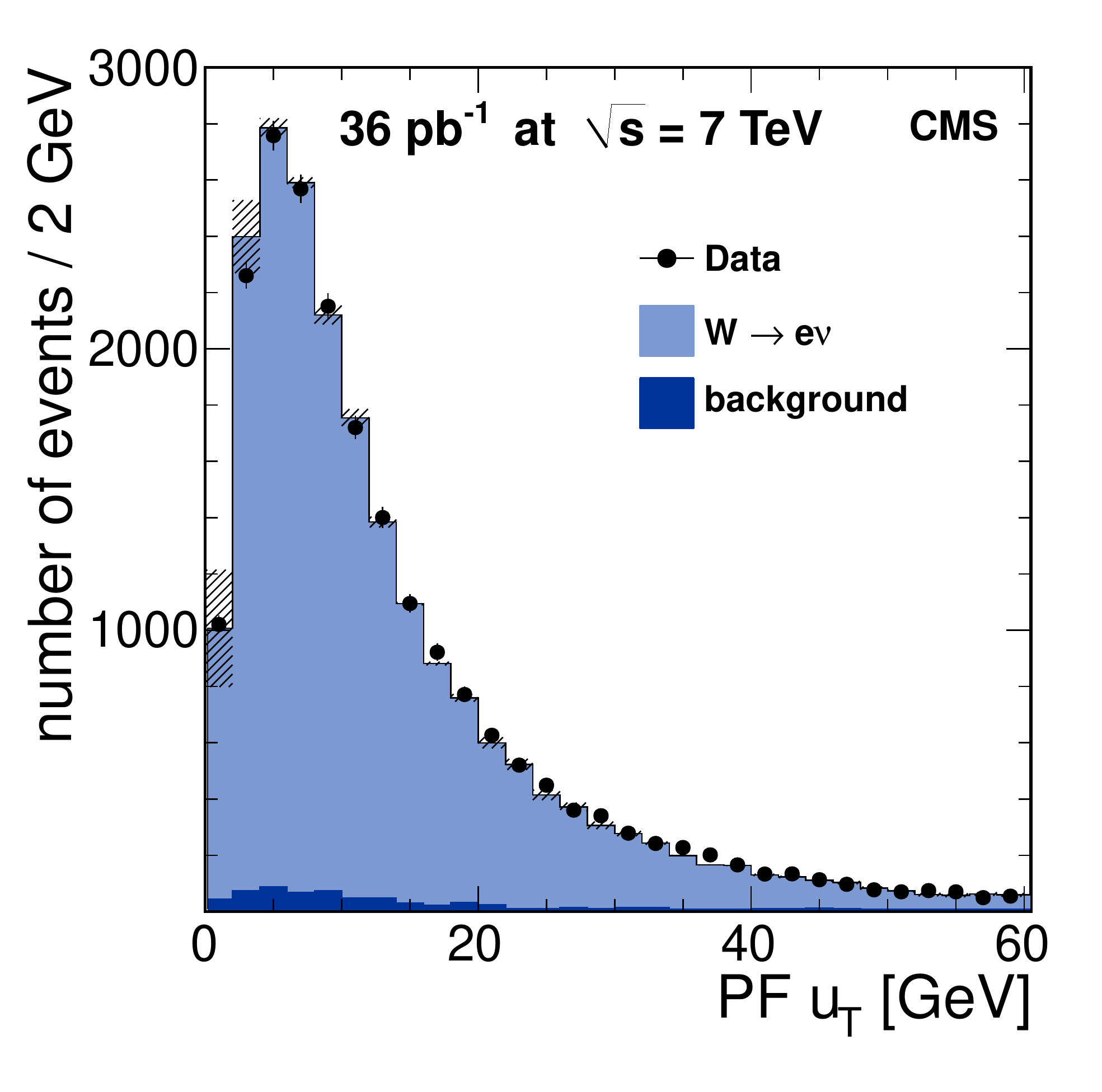}}
      {\includegraphics[height=0.363\textwidth]{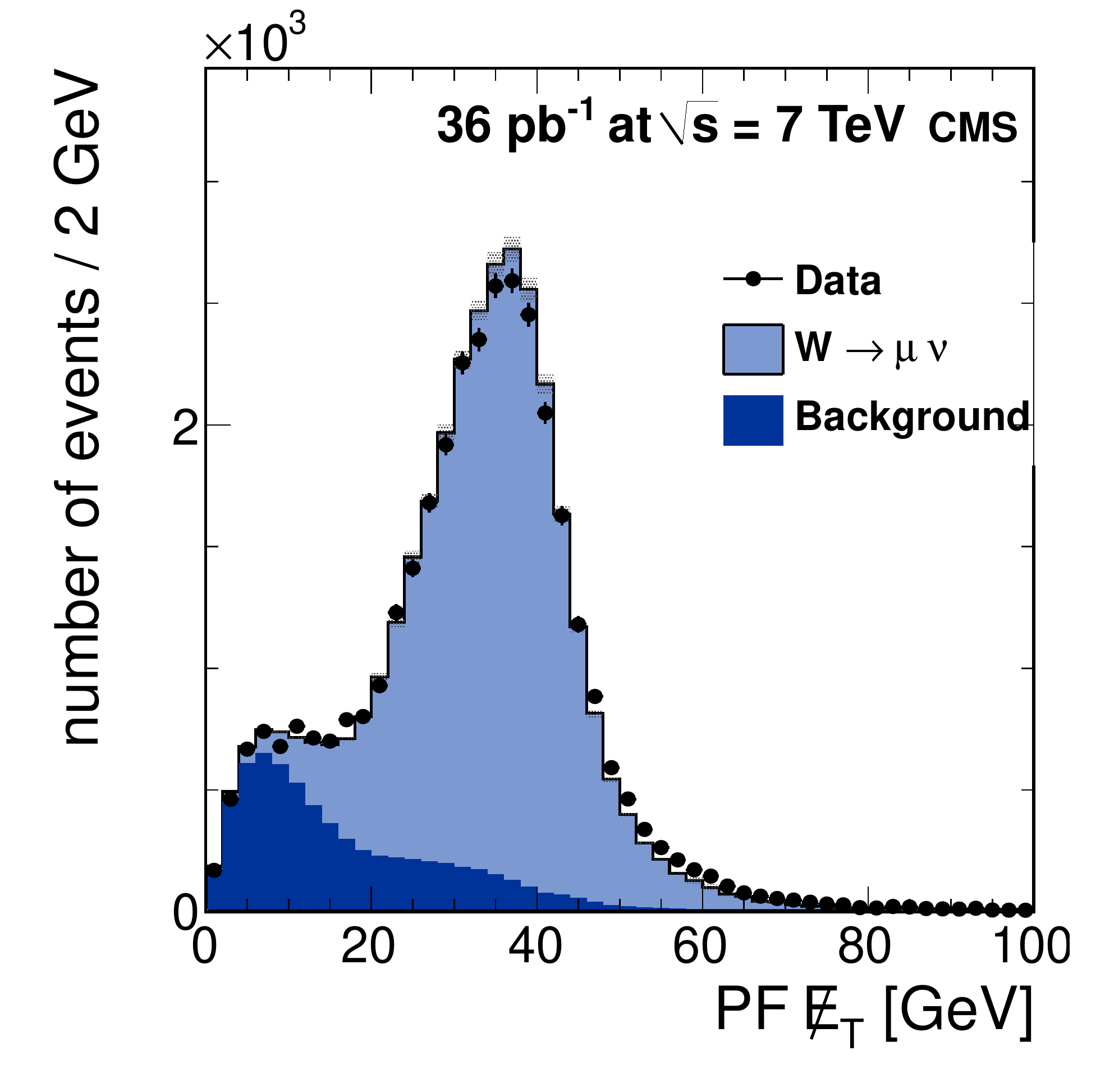}}
      {\includegraphics[height=0.363\textwidth]{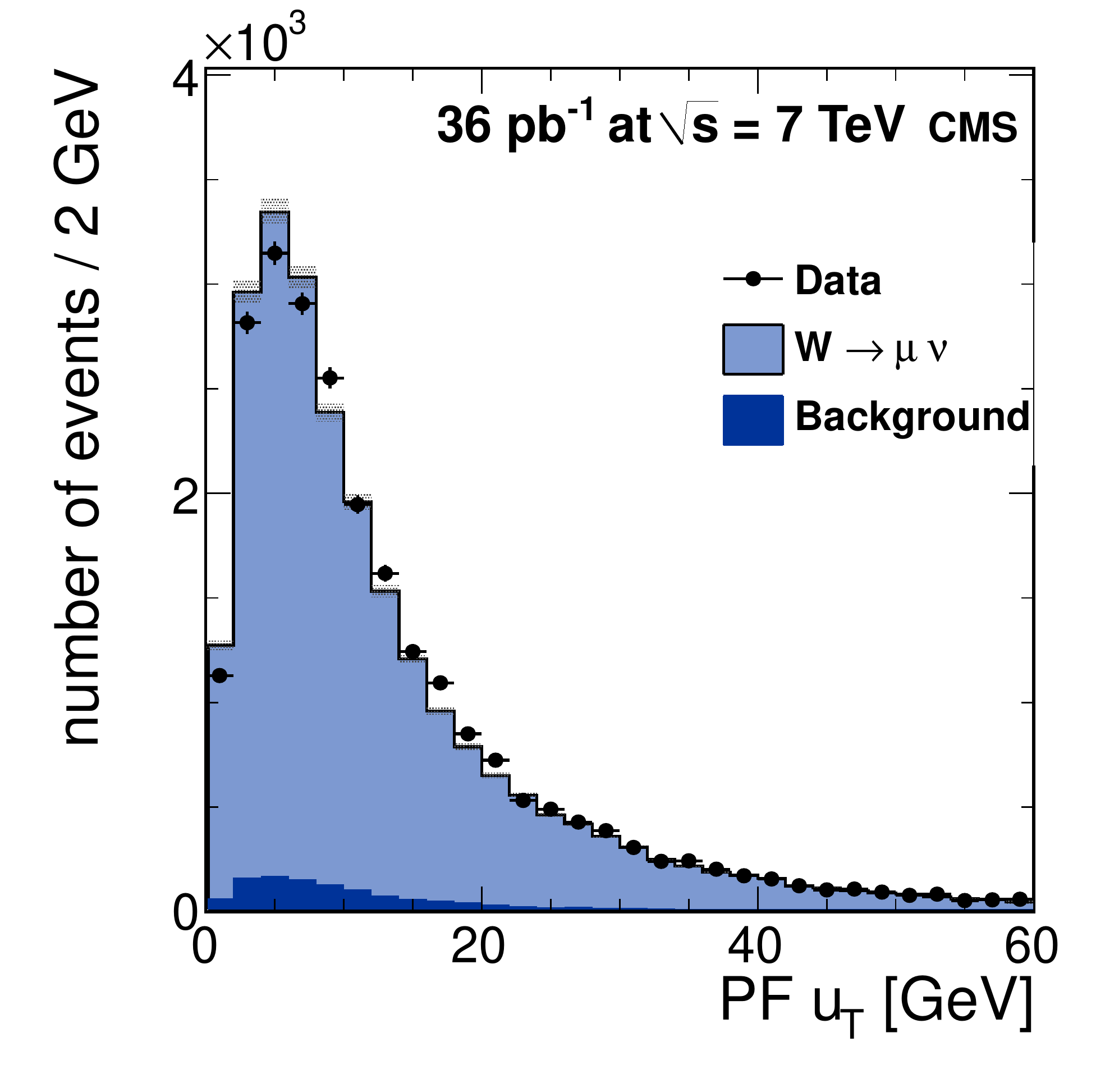}}
        \caption{The \pfmet (left) and
         $|\ut|$ (right) distribution in \Wen (top) and \Wmn (bottom) candidate events.
          Both data (points) and simulation (solid lines) are shown.  The plots on the right
include selection requirements on \met and $M_{\rm T}$, while those on the left do not.
         }
        \label{f:W_metut}
    \end{center}
\end{figure}

Figure \ref{f:W_metut}  shows the \pfmet distribution for the \Wen and \Wmn candidate samples,
along with the expectation from simulation. 
As for the analyses using Z events, the background distributions include a grey-shaded band
indicating the estimated uncertainty due to the size of the simulation 
samples, modeling of the
W \qt\ spectrum, and the pile-up correction procedure. In most cases
this uncertainty is too small to be visible.
Data and simulation agree well, and the W shows up prominently
as expected.

Figure \ref{f:W_metut} also shows the \ut distribution.  
To suppress QCD background, for the \ut studies only, we further require that the W candidate pass
a requirement on the transverse mass, defined as 
$M_{\rm T}=\sqrt{p_{\rm T}(\ell) \cdot \met \cdot (1-\cos{\Delta \phi})}$,
where $\Delta \phi$ is the opening angle in the transverse plane between the lepton candidate
and the \vecmet.
We require $\MT>50~\GeV$, and a
minimum \met threshold, $\met>25~\GeV$. 
The \met\ resolution has
substantial contributions from the mismeasurement of the many particles in the
underlying event.  These contributions can be more clearly seen in the \ut
distribution, since they are not obscured by the contributions
from the charged lepton.  Again we see reasonable agreement between data and simulation.

\subsection{Heavy flavour production and \texorpdfstring{\bigmet}{MET}}

The \met\ distributions from jet
samples containing b quarks can differ from those of inclusive jet samples
because the B hadrons have unique fragmentation properties, and sometimes their
final states contain neutrinos.
Neutrinos from b  jets are one of the
main sources of severe underestimations of jet energies (the other main source is the ECAL
masked channels discussed in Section~\ref{sec:cracks}).  In this section, we study the
induced \met\ in an inclusive b-tagged jet sample.

\begin{figure}[h!]
 \begin{center}
{  \includegraphics[angle=0,height=0.45\textwidth]{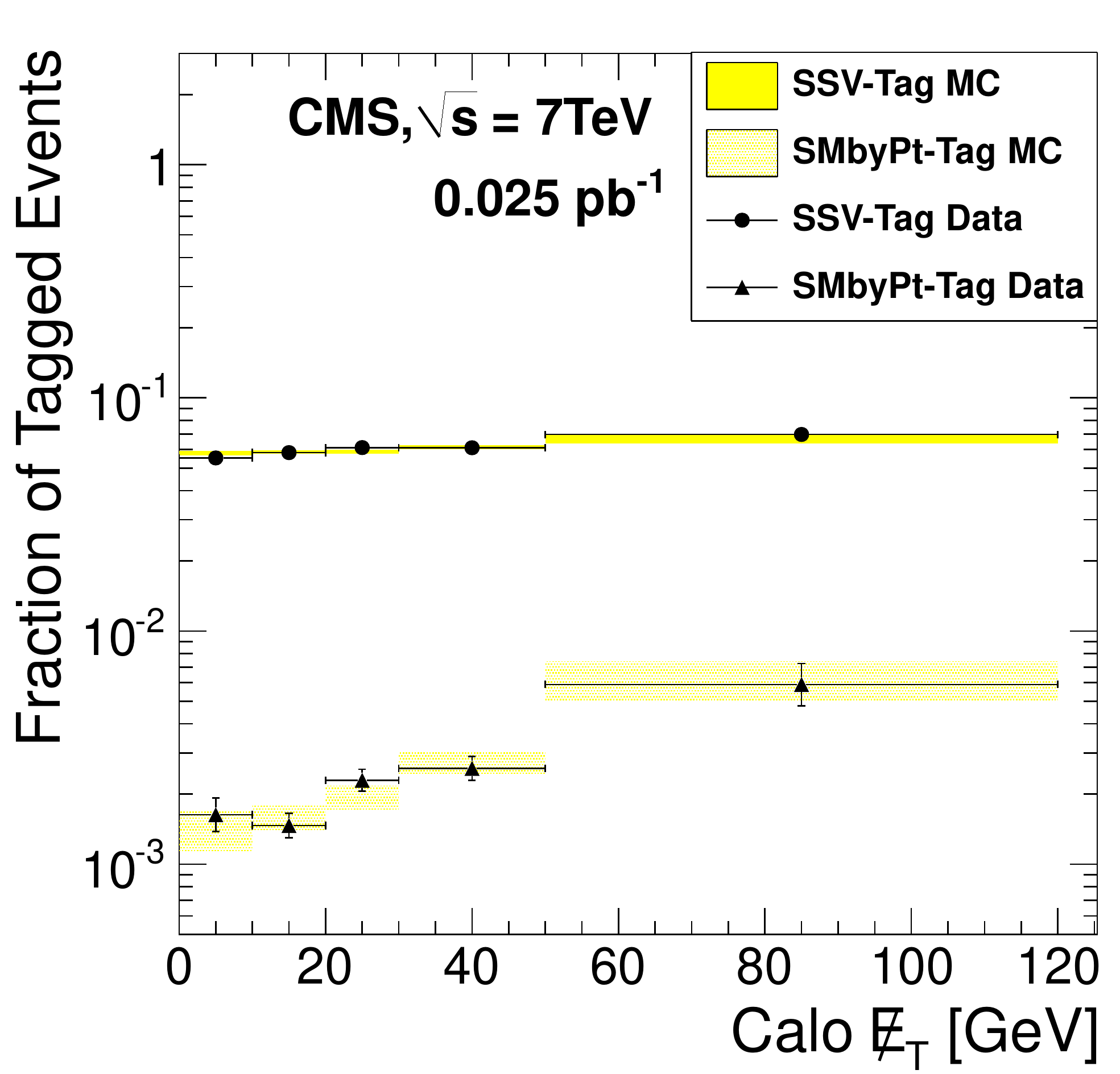}}
{  \includegraphics[angle=0,height=0.45\textwidth]{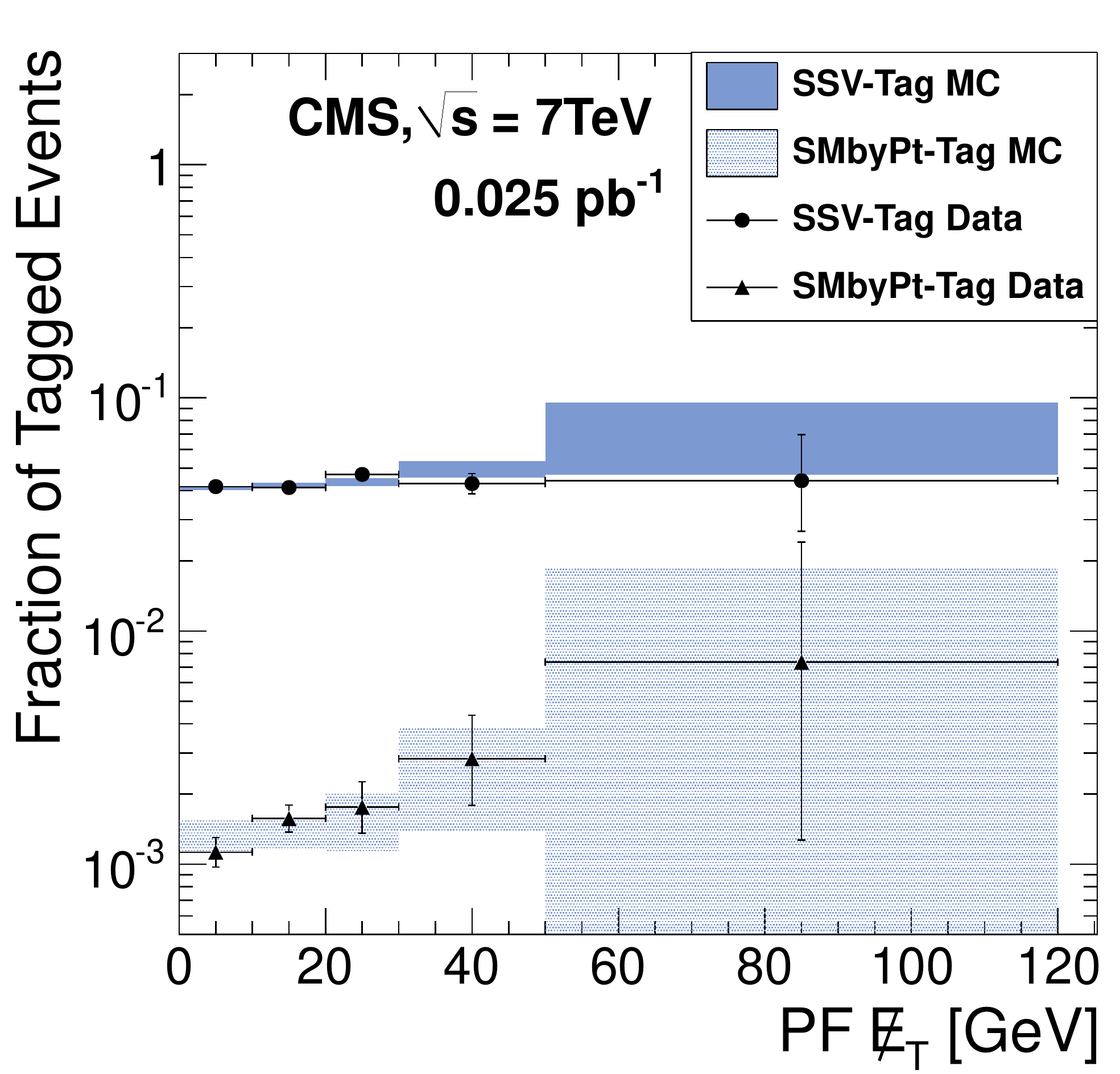}}
 \end{center}
 \caption{Fraction of events in a jet sample that contain a jet tagged as a b jet 
   by the SSV- and SMbyPt-taggers for (left) \calomet and (right) \pfmet. 
   \label{fig:f_MET_HF_Ratios}}
\end{figure}

We compare the \MET distributions in dijet events with and
without a secondary vertex, {\it i.e.}, events where the leading jet has a
positive SSV (displaced vertex) tag
or a SMbyPt (lepton) tag \cite{SSV,BTV10} (Section~\ref{sec:datas} includes a description
of these algorithms).
Since these taggers rely on tracking information, we require that the
leading jet has $|\eta|<2.1$.
Also, we require the leading two jets to have \pt$>40$ GeV.
Below this value, the b-quark tagging purity is significantly reduced.
A prescaled jet trigger with a \pt threshold of 
15 GeV  was used; the resulting sample corresponds to an integrated 
luminosity of $0.025$ $\mbox{pb}^{-1}$.
Figure~\ref{fig:f_MET_HF_Ratios} shows the fraction of events from this sample
with a b-tagged 
jet as a function of \met
for the two tagging algorithms. 
The larger increase in the fraction of b-tagged events at large \met
for SMbyPt than for SSV  is due, in part, to the higher probability of
neutrinos in leptonically tagged events.

\section{\texorpdfstring{\bigmet\ }{MET} significance}
\label{sec:performance}

\newcommand{\ex}				{\ensuremath{\varepsilon_x}\xspace}
\newcommand{\ey}				{\ensuremath{\varepsilon_y}\xspace}
\newcommand{\exi}				{\ensuremath{E_{x_{i}}}\xspace}
\newcommand{\eyi}				{\ensuremath{E_{y_{i}}}\xspace}
\newcommand{\like}				{\ensuremath{{\cal L}}\xspace}
\newcommand{\signif}			{\ensuremath{{\cal S}}\xspace}
\newcommand{\signife}			{\ensuremath{{\cal S}_E}\xspace}
\newcommand{\signifh}			{\ensuremath{{\cal S}_H}\xspace}
\newcommand{\signifpf}			{\ensuremath{{\cal S}_{\rm PF}}\xspace}
\newcommand{\gevc}				{GeV\xspace}
\newcommand{\probchi}			{\ensuremath{{\cal P}(\chi^{2})}\xspace}
\newcommand{\vethat}			{\ensuremath{\vec{e}}_{\rm T}\xspace}
\newcommand{\vethati}[1]		{\ensuremath{\vec{e}}_{\rm T_{#1}}\xspace}
\newcommand{\veceti}[1]        {\ensuremath{\vec{E}_{\rm T_{#1}}}\xspace} 
\newcommand{\lkgHideText}[1]	{}

A spurious nonzero \vecmet in an event can have contributions from many
sources, including measurement resolution,  reconstruction inefficiencies, instrumental defects,
and improper pattern recognition.  Events in which
the reconstructed \vecmet is consistent with contributions solely from particle-measurement
resolutions and efficiencies can be
identified by evaluating the \vecmet significance, $\mathcal S$.  The significance
offers an event-by-event assessment of the likelihood that the observed $\MET$
is consistent with zero given the reconstructed content of the event and known
measurement resolutions.  A similar variable used by the CDF collaboration
is described in \cite{cdfmetsig}.

\subsection{Definition}

The significance requires evaluation of the uncertainty in the total measured transverse energy, which is given by

\begin{equation}
\vecet^{\rm \!\!\!total} = \sum_{i\,\in\, X } \veceti{i} = -\vecmet,
\label{eq:metsig:etsum}
\end{equation}

where $\veceti{i}=(\exi,\eyi)$
is the measured transverse momentum of the $i^{\rm th}$ reconstructed object. $X$ is
the set of reconstructed objects, such as
calorimeter towers (for \calomet) or PF particles (for \pfmet), used to calculate \MET.
In the derivation of the significance, there are
three relevant quantities for each object in the sum.
The first of these is  $\vethati{i}$, the true transverse momentum of the object. The significance
        provides a measure of whether an event is consistent with the null hypothesis of zero
        genuine total transverse momentum.  Under this hypothesis, $\sum_{i\,\in\, X } \vethati{i} = 0$.
The second is  $\veceti{i}$, the measured transverse momentum of the object, which is distributed
         according to $P_{i}(\veceti{i}|\vethati{i})$, the probability density function (pdf) for
         observing the measured transverse momentum given the true transverse momentum of the object.
The third is $\vet_{i}=\veceti{i}-\vethati{i}$.
For convenience, we define an equivalent pdf in terms of
        this difference:
        $p_{i}(\vet_{i}|\vethati{i}) \equiv P_{i}(\vet_{i}+\vethati{i}|\vethati{i})$.  Given the
        null hypothesis, $\sum\veceti{i}=\sum\vet_{i}$, so that the $i^{\rm th}$
        reconstructed object contributes $\vet_{i}$ to the measured total transverse momentum.

We first introduce the likelihood that we would observe a
total transverse momentum \vet under our null hypothesis. For the two object case, the likelihood function
is given by

\begin{eqnarray}
\like(\vet)
 & = &
\int P_{1}(\veceti{1}|\vethati{1})) P_2(\veceti{2}|\vethati{2}))
\delta(\vet-(\veceti{1}+\veceti{2}))
\,d\veceti{2}\,d\veceti{2} \nonumber \\
 & = & \int p_{1}(\vet_1|\vethati{1}) p_2(\vet_2|\vethati{2})
\delta(\vet-(\vet_1+\vethati{1}+\vet_2+\vethati{2}))
\,d\vet_1\,d\vet_2 \nonumber \\
 & = & \int p_{1}(\vet_1|\vethati{1}) p_2(\vet_2|\vethati{2})
\delta(\vet-(\vet_1+\vet_2))
\,d\vet_1\,d\vet_2,
\label{eq:metsig:basic}
\end{eqnarray}

since $0 = \sum_{i} \vethati{i} = \vethati{1}+\vethati{2}$.
For an arbitrary number of input objects, the full likelihood function can be generated
by a recursive application of Eq.~(\ref{eq:metsig:basic}).
The significance is defined as the log-likelihood ratio

\begin{equation}
\signif \equiv 2\ln\left(\frac
{\like(\vet=\sum\vet_{i})}
{\like(\vet=0)}
\right),
\label{eq:metsig:defn}
\end{equation}

which compares the likelihood of  measuring the total observed
$\vecet^{\rm \!\!\!total}=\sum\veceti{i}=\sum\vet_{i}$ to the
likelihood of the null hypothesis, $\vecet^{\rm \!\!\!total}=0$.

This formulation is completely general
and accommodates any probability distribution function.
In practice, however, we often employ Gaussian uncertainties for measured
quantities, for which the integrals of Eq.~(\ref{eq:metsig:basic})
can be done analytically.  The Gaussian probability density function is given
by

$$
p_i(\vet_i|\vethati{i})\sim \exp\left(-\frac{1}{2}(\vet_{i})^{\,T} {\mathbf V}_i^{-1} (\vet_{i})\right), \\
$$

where ${\mathbf V}_i$ is the $2\times 2$ covariance matrix associated with the $i^{\rm th}$ measurement.
The integration of Eq.~(\ref{eq:metsig:basic}) yields

$$
\like(\vet)\sim \exp\left(-\frac{1}{2}(\vet)^{\,T} \,{\mathbf V}^{-1}\, (\vet)\right)
$$

with ${\mathbf V}={\mathbf V}_1+{\mathbf V}_2$.
When many measurements contribute,
the expression generalizes to

\begin{equation}
\like(\vet)\sim \exp\left(-\frac{1}{2}(\vet)^{\,T} \left(\sum_i {\mathbf
            V}_i\right)^{-1} (\vet)\right).
\label{eq:metsig:like}
\end{equation}

The covariance matrix ${\mathbf U}_i$ for each reconstructed object in the $\vecet$ sum
is initially specified in a natural
coordinate system having one axis aligned with the measured
$\veceti{i}$ vector, $\veceti{i}\equiv (E_{T_{i}}\,\cos\phi_{i},\,E_{T_{i}}\,\sin\phi_{i})$:

\begin{equation}
{\mathbf U}_i =
\left(\begin{array}{cc}
        \sigma^2_{E_{T_i}}& 0     \\
        0&  {E^2_{T_i}}\,\sigma^2_{\phi_i}
\end{array}\right).
\label{eq:metsig:res}
\end{equation}

(We adopt the simplifying assumption that \et\ and $\phi$ measurements are
uncorrelated.) This matrix is rotated into the
standard CMS $x-y$ reference frame  to give the error matrix

\begin{equation}
{\mathbf V}_i = R(\phi_i) {\mathbf U}_i R^{-1}(\phi_i),
\label{eq:metsig:rot}
\end{equation}
where $R(\phi_i)$ is the rotation matrix.
The \signif-matrix summation is then performed in this common reference frame.
Combining Eqs.~(\ref{eq:metsig:defn}), (\ref{eq:metsig:like}), and
(\ref{eq:metsig:rot}) yields

\begin{equation}
\signif =
\left(\sum_{i\,\in\, X } \veceti{i} \right)^T
\left(
\sum_{i\,\in\, X } R(\phi_i) {\mathbf U}_i R^{-1}(\phi_i)
\right)^{-1}
\left(\sum_{i\,\in\, X } \veceti{i} \right).
\label{eq:metsig:impl}
\end{equation}

Equation~(\ref{eq:metsig:impl}) makes explicit
the dependence of \signif\ and $\vecmet$ on the set of objects $X$
over which the vectors and matrices are summed.
In general \signif\ is small when the
\MET  can be attributed to measurement resolution, and large otherwise.

In the Gaussian case,
\signif\ is simply a $\chi^2$ with two degrees of freedom. If we rotate into a
coordinate system with the $x$ axis parallel to the $\vecmet$ axis,
instead of the CMS horizontal axis, then Eq.~(\ref{eq:metsig:impl}) is
simplified to $\signif = E^2_T/(\sigma^2_{E_T}(1-\rho^{2}))$, where $\sigma^2_{E_T}$
is the variance of the magnitude of \vecmet, and $\rho$ is
the correlation coefficient between the variances parallel to and perpendicular to
the measured \vecmet.
This form emphasizes the
essential meaning of \signif, but obscures the important
feature that, through its denominator, \signif\ embodies the full
topological information in the event.  Essential features such as the
angles between the measured \vecmet and the reconstructed objects
in the event are embedded in the definition of
the denominator. This form also makes apparent the relationship between
the true significance (in the Gaussian limit) and the more naive measure
$\Sigma=\met/\sqrt{\sumet}$.

The specialization to a Gaussian probability density function is less restrictive
than it may appear, as any probability density function expressible as a linear combination
of Gaussians is accommodated by the formalism presented here.

To apply Eq.~(\ref{eq:metsig:impl}) to \pfmet significance, we note
that the Gaussian pdf only accommodates measurement resolution.  Using only reconstructed
PF particles to determine the covariance matrix
would neglect fluctuations
in the measured PF particle content itself. These fluctuations arise from finite detection and reconstruction
efficiencies, and provide a non-negligible contribution to the \pfmet resolution.
These fluctuations, however, also affect the PF jet resolutions.  We can therefore
substitute the  PF jet resolutions for the combined measurement resolutions of
the PF particles that have been clustered into jets.
Hence the sum of covariance matrices in Eq.~(\ref{eq:metsig:impl}) includes contributions from
PF jets,
PF particles that were not considered during jet finding (e.g. isolated leptons),
and
PF particles that are not clustered into any jet.
This approach inherently takes into account the contributions both from measurement resolution and from
fluctuations in the reconstructed particle content.

The covariance matrices ${\mathbf U}_{i}$ of Eq.~(\ref{eq:metsig:res}) are obtained from
our knowledge of the response of each type of PF particle or jet as a function
of \pt and $\eta$.  The charged hadron and muon resolutions are obtained on a
particle-by-particle basis from the error matrix from the final track fit, and the
resolutions for electrons are those obtained from studies of data samples of
known resonances such as neutral pions, the Z, etc.
The jet and photon resolutions are from
simulation.
No input resolutions were tuned based on the behaviour of
the significance distribution itself.

\subsection{Performance of $\signifpf$ in dijet events}

Because \signifpf is $\chi^{2}$ distributed, it should exhibit a flat
probability of $\chi^{2}$, $\probchi$, for two degrees of freedom
in an event sample
that nominally has no genuine \MET.
(That is, $1-\probchi$
is the standard cumulative distribution function of the $\chi^{2}$ statistic for
two degrees of freedom.)
Dijet samples from pp
collisions are dominated by such events.

We select dijet events by requiring
at least two jets satisfying  $|\eta|<2.3$ and $p_{\rm T}>p_{\rm T}^{\rm min}$,
with thresholds $p_{\rm T}^{\rm min}$ of 30 or 60 GeV.  One of the
jets above threshold must have been responsible for the event
passing an HLT single-jet trigger. We use
data collected with a 15 GeV trigger threshold for our 30 GeV dijet sample,
and a 30 GeV trigger threshold for the 60 GeV dijet sample
(Because of different prescale factors applied to the two trigger
streams, the 60 GeV dijet sample is not a direct subset of the 30 GeV sample.)

\begin{figure}[htbp]
\begin{center}
\subfigure{ \includegraphics[width=0.48\textwidth]{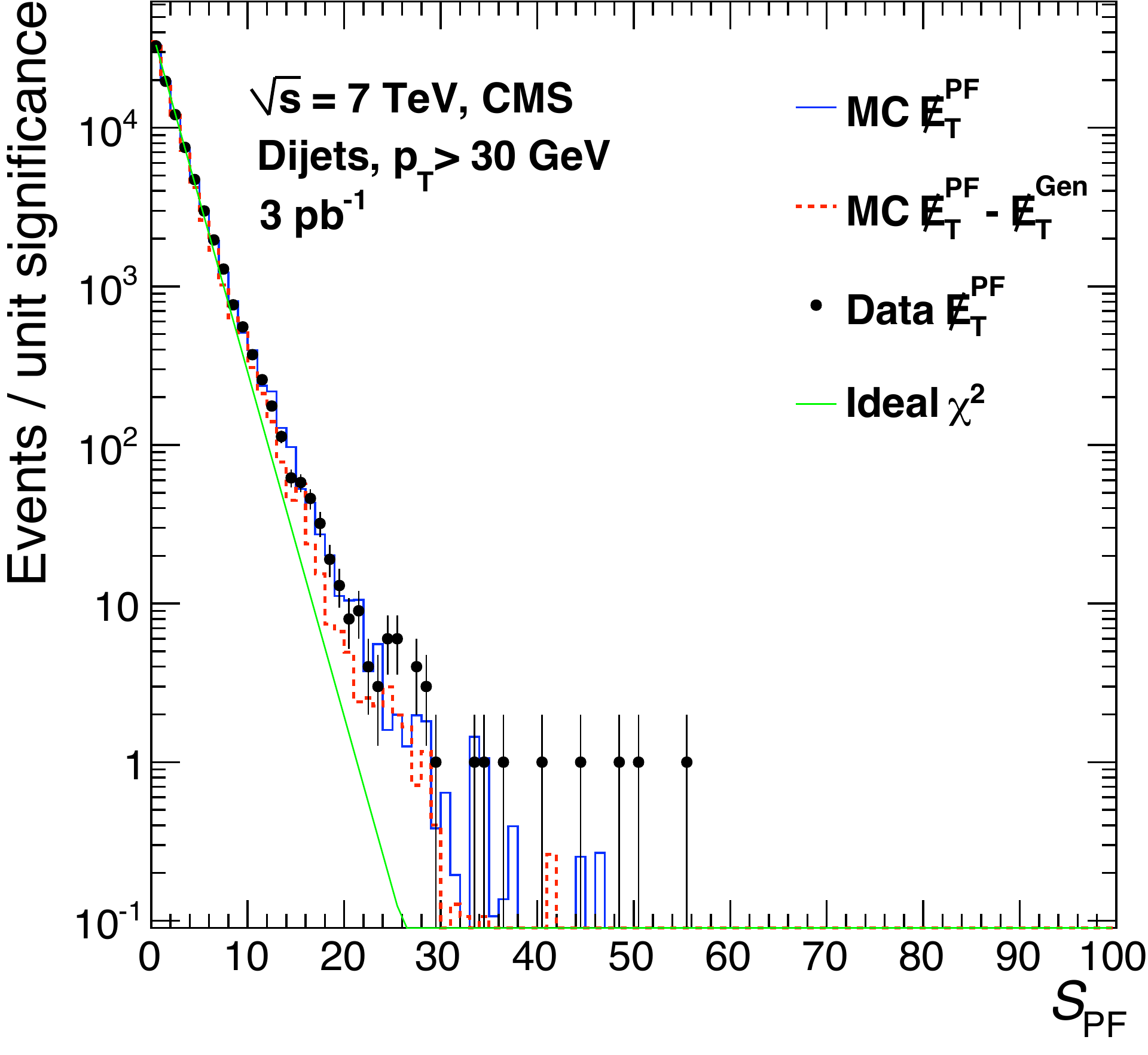} }
\subfigure{ \includegraphics[width=0.48\textwidth]{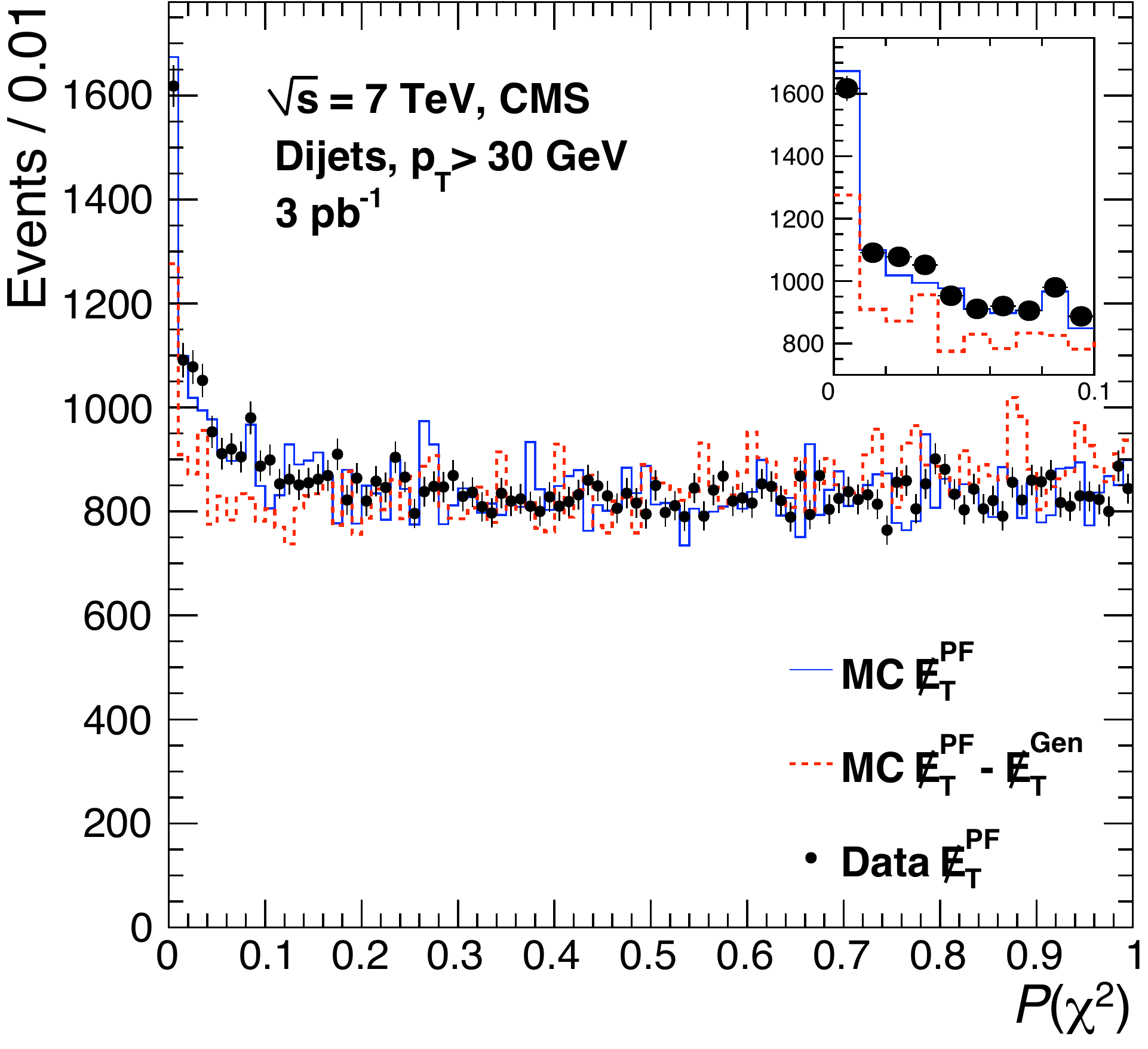} }\\
\subfigure{ \includegraphics[width=0.48\textwidth]{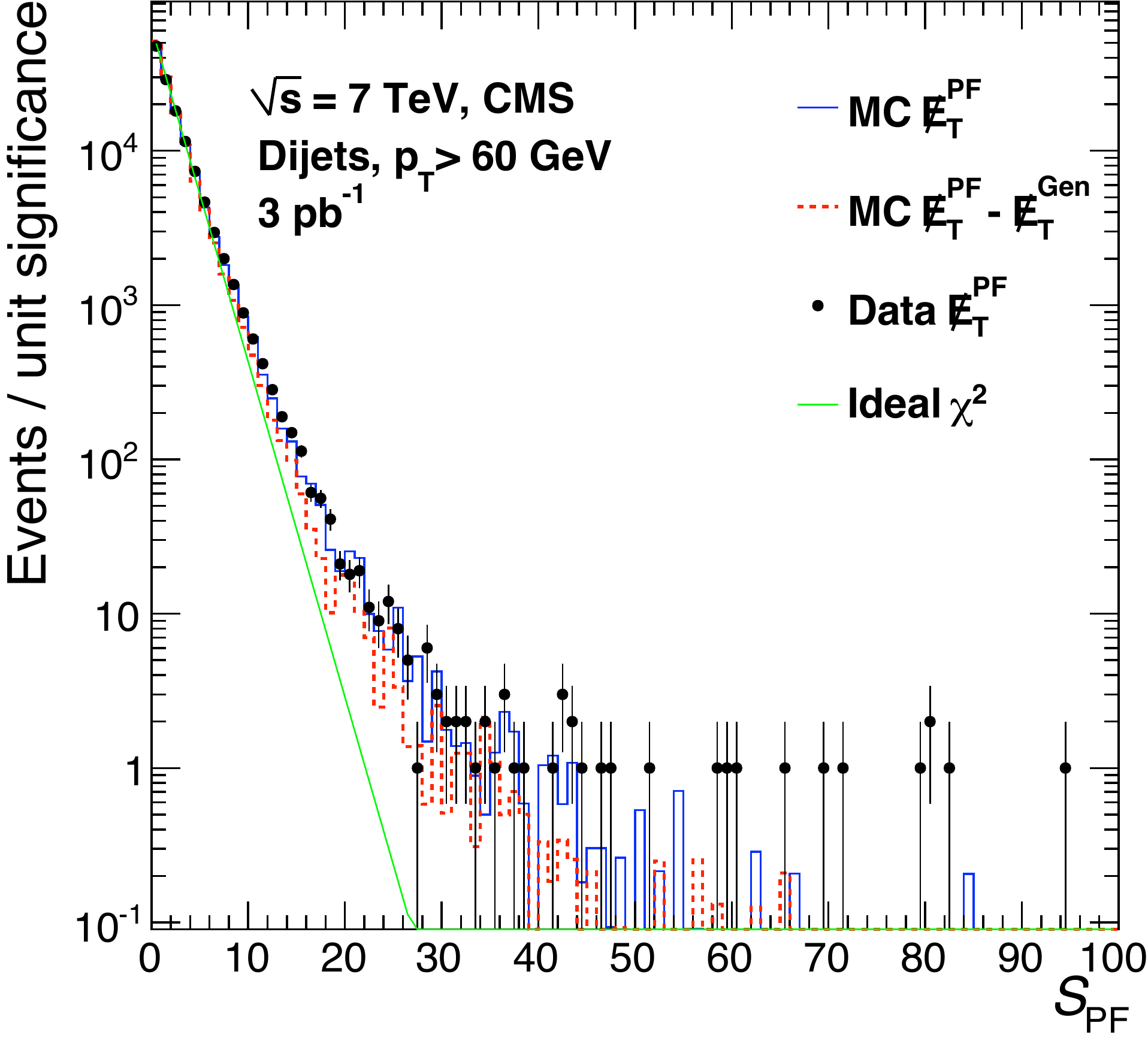} }
\subfigure{ \includegraphics[width=0.48\textwidth]{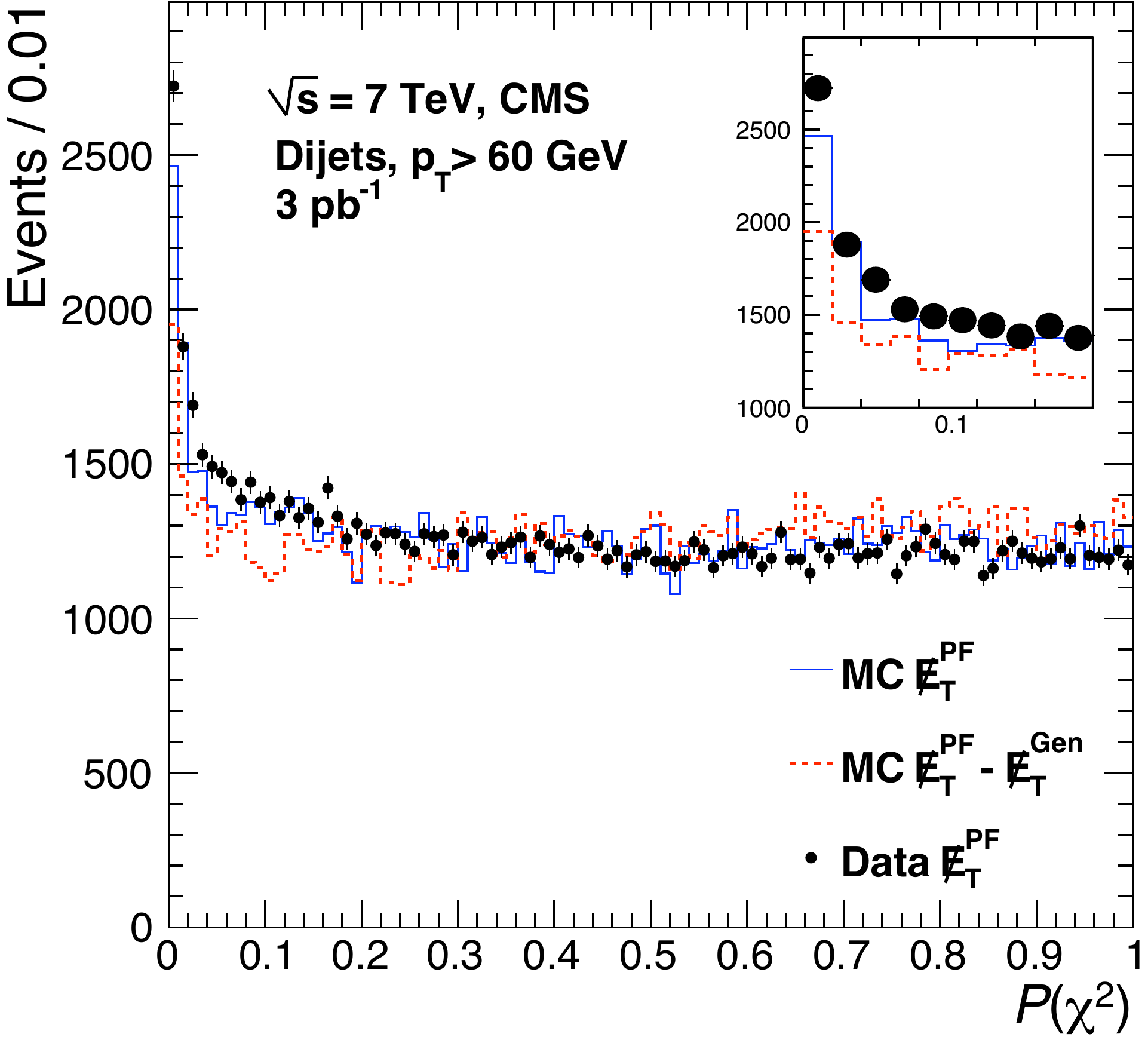} }
\end{center}
\caption{\label{sigfig:dijetDataMC} The \MET significance \signifpf distributions (left)
and the corresponding probability of $\chi^{2}$, \probchi, distributions (right) for
dijet event samples in data (points) and simulation (solid histograms) with 30~\gevc (top) and 60~\gevc (bottom) jet $p_{T}$ thresholds.
The dashed histograms show the
simulation distributions with true \MET contributions, from physics and finite acceptance
effects, subtracted event-by-event.  The dotted line overlaid on the \signifpf distributions shows a
reference pure exponential function.  Each inset expands the small \probchi region.}
\end{figure}

We compare the \signifpf distributions, as well as their
corresponding \probchi distributions, in
data and simulation in Fig.~\ref{sigfig:dijetDataMC} for
both values of the $p_{\rm T}^{\rm min}$ threshold.
The significance distribution very closely follows a
pure exponential, and the \probchi distribution is populated
quite uniformly between zero and unity in both data and simulation.
There is a small peak at zero in \probchi; simulation
(Fig.~\ref{sigfig:dijetDataMC}) indicates
that about half of this peak results from genuine \MET in the
event sample.  This \MET arises from a combination of sources such as
the semileptonic decays of heavy quarks and the finite $\eta$ acceptance of
the detector.  The data and simulation distributions match
well in the 30 GeV threshold sample. MC studies show that the remainder of
the excess of low probability events after accounting for genuine \MET
typically have at least one high-$p_{T}$
jet whose response is in the non-Gaussian tail of the response function.

To probe the stability of the \signifpf behaviour, we have studied dijet samples
with different $p_{T}^{\rm min}$ thresholds, which changes the relative contributions
of different detector regions in the covariance matrix calculations.  We find that,
overall, the \signifpf distributions for the bulk of the data
continues to  exhibit near-ideal behaviour independent of threshold.  As
the 60 GeV sample shown here demonstrates, though, the higher threshold data does begins to develop a
larger tail in the significance, and a correspondingly larger peak at zero in \probchi, than
we find in the simulation. The discrepancy between data and MC is below the $0.2$\% level.
Visual examination of the events with low probability reveal that the discrepancy
arises from a combination of events with a residual anomalous energy contamination and
other events with a high-$p_{T}$ jet with activity straddling the endcap (HE) and forward
(HF) calorimetry, for which the non-Gaussian tails are not yet perfectly modeled.

For the $\signifpf$ distributions shown here, the transition point for use of
resolutions based on PF jets rather than resolutions from unclustered PF particles
in the \signifpf calculation (Eq.~(\ref{eq:metsig:impl})) occurs at a jet $p_{T}$ of 3 \gevc.
The \signifpf distributions are insensitive to the variation of this
transition point between jets and individual particles over the
range of 1 to approximately 6 \gevc.  By 10 \gevc, a slope in
the \probchi distribution has clearly appeared, indicating that
we no longer account sufficiently for contributions to the
\MET resolution from fluctuations in the reconstructed particle
content.

\begin{figure}[htbp]
\begin{center}
\subfigure{ \includegraphics[width=0.48\textwidth]{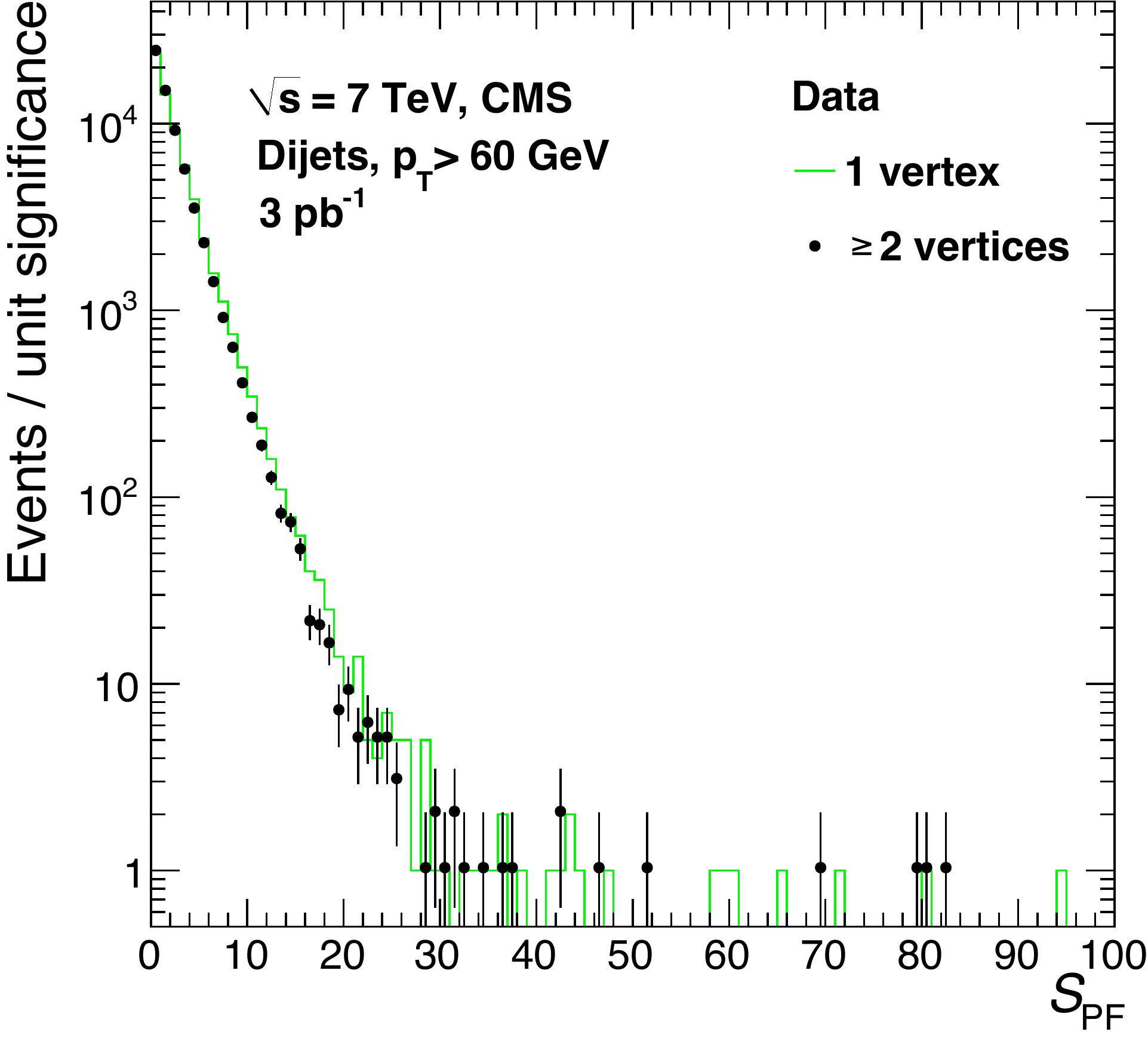} }
\subfigure{ \includegraphics[width=0.48\textwidth]{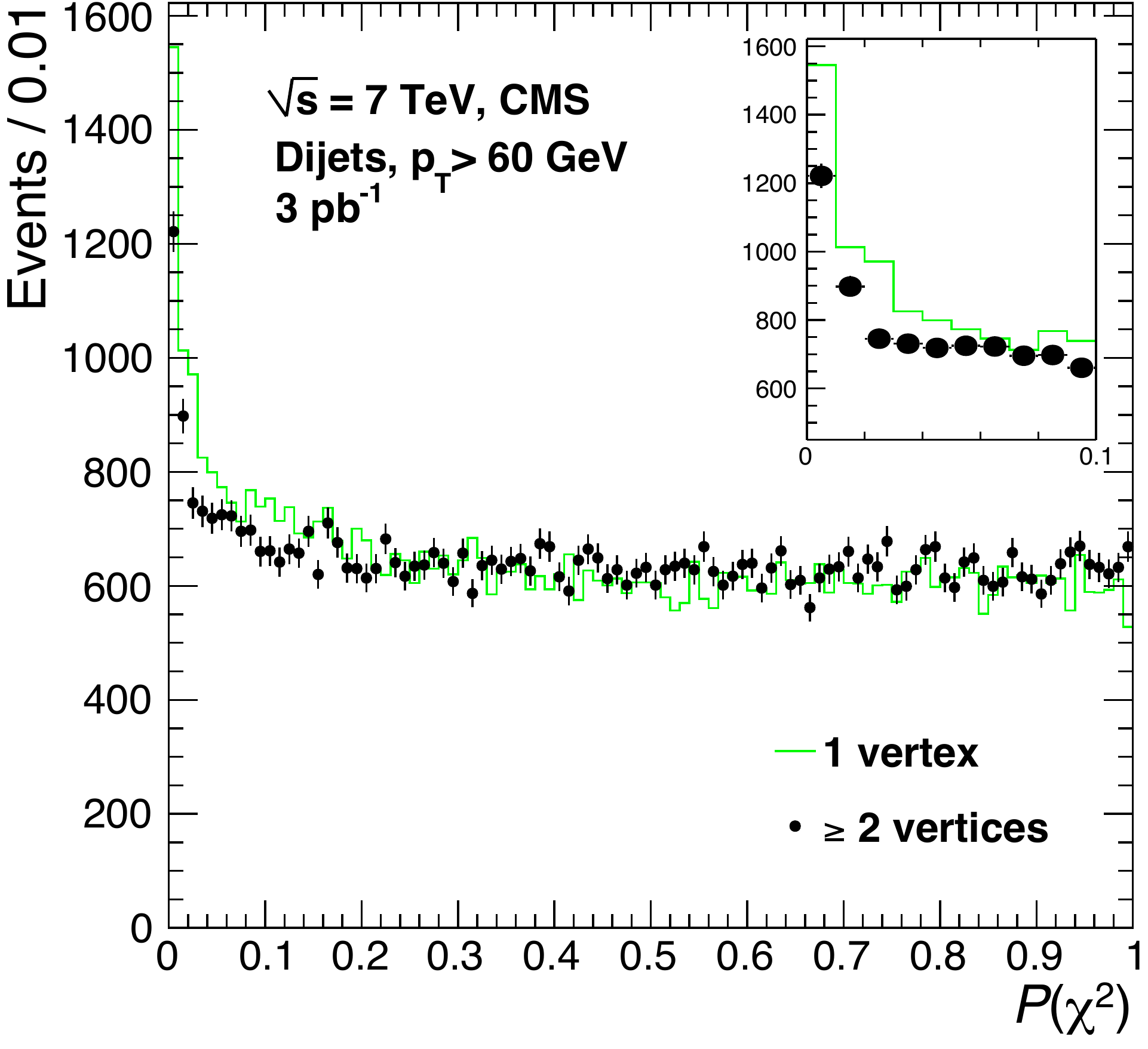} }
\end{center}
\caption{\label{sigfig:dijetPileup} The \MET significance (left) and \probchi (right)
distributions for events with a single interaction vertex (histogram) and multiple interaction vertices (points)
in the 60 GeV threshold dijet sample. The inset expands the small \probchi region.}
\end{figure}

A powerful feature of the \MET significance is that its distribution
is insensitive to pile-up (for events with no
genuine \MET).  As long as the correct resolutions
are input, the significance should still have a pure exponential
behaviour with a uniformly distributed \probchi.  In Fig.~\ref{sigfig:dijetDataMC},
no restrictions were made on the number of interaction vertices
in the data, while the simulation has no pile-up.
In Fig.~\ref{sigfig:dijetPileup}, we compare the shapes of the single-vertex
and multiple-vertex significance and \probchi distributions
in data.  The shapes are  very similar,
as expected.  The main difference arises in the low probability region, where the
multiple interaction data exhibits behaviour closer to the ideal -- an example of
the central limit theorem.  With the additional contributions to the \MET resolution,
the roles of the non-Gaussian response tails and genuine \MET are diminished.  The
overall insensitivity can be useful, for example, when extrapolating backgrounds
dominated by samples with nominally zero genuine \MET.

\subsection{Application to \texorpdfstring{\Wen}{W to e nu} events}
As a case study, we examine the potential gain of introducing the significance
variable into the selection criteria for \Wen analyses.
The set of criteria employed is that of the recent
measurement by the CMS collaboration of the W cross section~\cite{Collaboration:2010xn},
for which backgrounds were controlled using a stringent, 80\% efficient, electron isolation
criterion.  Signal and background yields in that analysis were determined by a fit
to the reconstructed \MET distribution, though because of the large backgrounds at
small values of \MET, the signal level is largely determined from the $\MET>20$~GeV region.

\begin{figure}[hbpt]
    \begin{center}
        \includegraphics[width=0.48\textwidth]{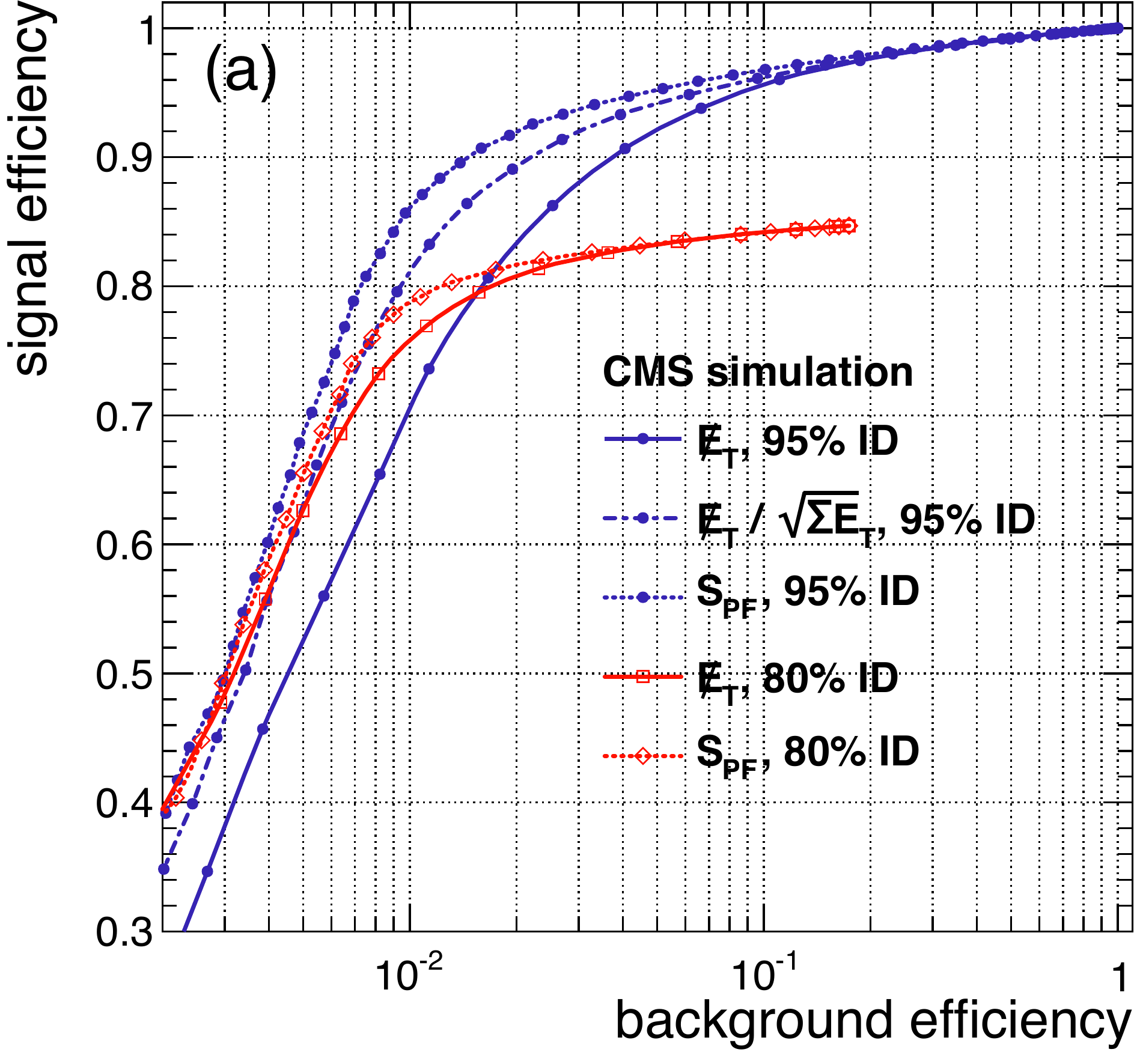}
        \includegraphics[width=0.48\textwidth]{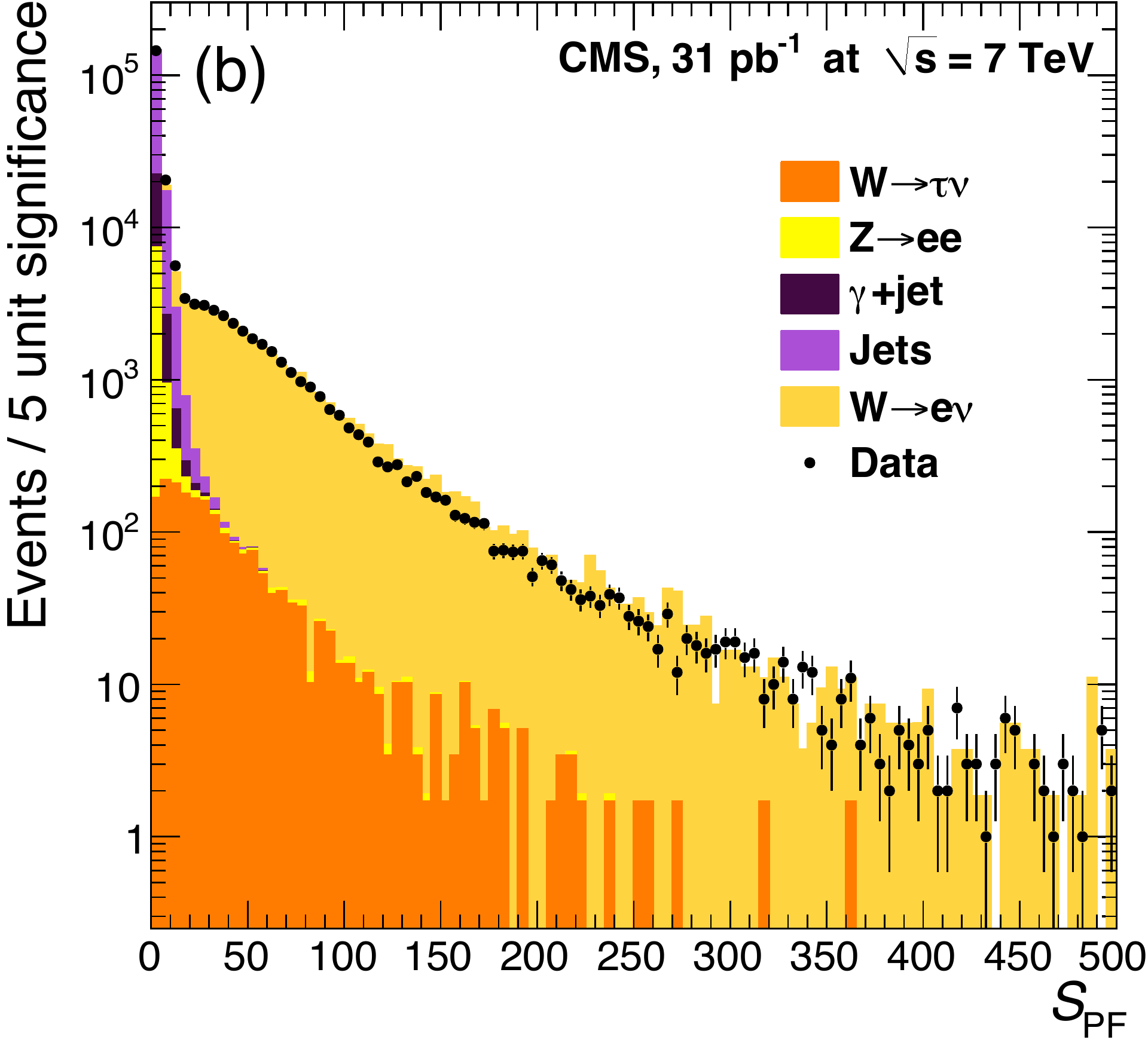}
       \caption{(left) Efficiency curves for \Wen signal versus backgrounds varying the minimum value
       of $\etmiss$ (solid lines), of \signifpf (dotted lines), and of
$\etmiss/\sqrt{\sum E_{T_{i}}}$
       (dot-dash line), with the 95\% efficient (blue) or 80\% efficient (red) electron isolation criterion applied.
       (right) Distributions for \signifpf in candidate \Wen  events from data (points) and simulation (stacked histograms).
       The simulation components, from top to bottom, are signal (mustard) and backgrounds from
       jets (purple), $\gamma +$ jets (black), \Zee (yellow), and
       $\PW^\pm\to \Pgt^\pm\Pgngt$ (orange).
       The simulation is scaled by a fit to the data with floating normalizations for the signal and the
       total background.
       }
        \label{fig:wenuSig}
    \end{center}
\end{figure}

One analysis option would be to relax the electron isolation from an 80\% to a 95\% efficient
criterion and introduce \MET significance to help control backgrounds.
Figure~\ref{fig:wenuSig} compares the efficiency for signal
versus background in simulation for increasing minimum thresholds on:
$\etmiss$, with both the 80\% and 95\% electron isolation criteria applied;
\signifpf, again with both isolation criteria; or
$\etmiss/\sqrt{\sum E_{T_{i}}}$ with the 95\% isolation criteria.
All efficiencies are measured relative to the signal or background yield obtained
with the looser 95\% electron isolation criterion applied.  (As a result, the tighter 80\% criterion
has an asymptotic value of
approximately 84\%.) Application of the tighter criterion changes the relative
signal and background distributions for \MET and \MET significance compared to the
looser criterion.  When a minimum \MET threshold is applied, the tighter isolation criterion
provides a better signal to background ratio at low background levels than the looser
criterion.  Application of a minimum \signifpf threshold with the looser criterion,
however, outperforms all the other combinations for background rejection
at a given signal efficiency.

We note that in the calculation of the significance, the isolated signal electron candidate enters
as an electron, and in particular with the resolution associated with an electron.  This
approach was found to outperform the option where each event was treated as electron-free
(as is the case for the dominant background).

\lkgHideText{
One would most likely employ both $\MET$ and its significance in
an analysis.  Consider, for example, a $\MET > 20$~GeV threshold. Taking the looser
electron isolation, one could require a \signifpf threshold that would result in the
same signal efficiency as the tighter electron isolation.  The background from events with
nominally zero \MET would, however, be reduced by a factor of 45 relative to that with
the tighter isolation.  A significantly reduced background would provide not only a
statistical advantage, but potential systematic gains as well, since the same
fractional uncertainty on a smaller background means a smaller absolute uncertainty.
Relaxation of the electron isolation criteria is just one potential re-optimization route with
the introduction of the \MET significance.  Other analyses might, for example, gain sensitivity
by relaxing the minimum electron energy threshold. This example does, however, illustrate that the
\MET significance adds a powerful criterion for optimizing the use of our
data.
}  

Figure~\ref{fig:wenuSig} also shows that the \signifpf
distributions for \Wen in data and simulation agree well. As expected, the backgrounds without genuine $\met$
are compressed towards low values of \signifpf while signal events having
real $\met$ extend to high values of \signifpf.

\begin{figure}[htbp]
    \begin{center}
        \includegraphics[width=0.48\textwidth]{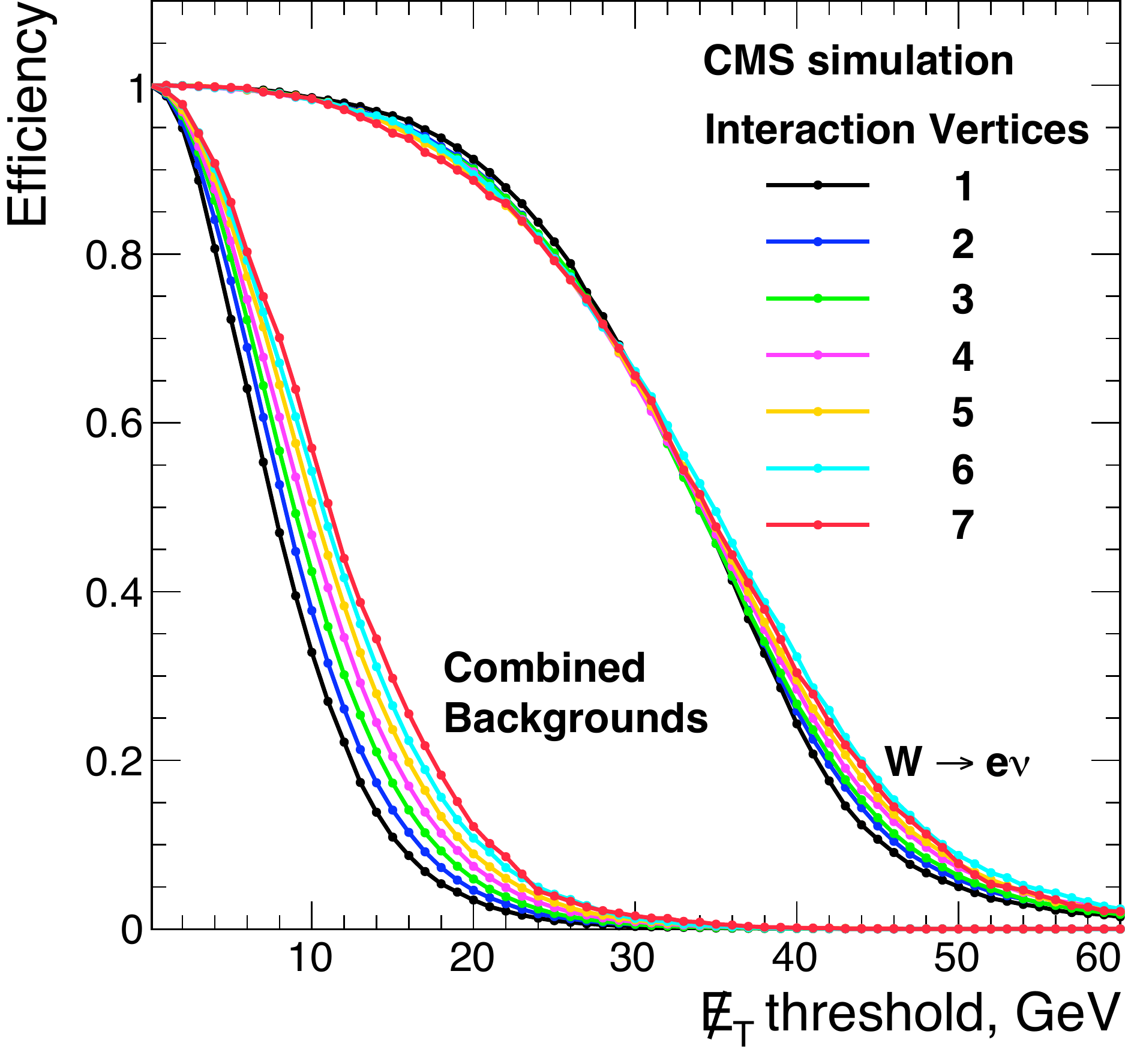}
        \includegraphics[width=0.48\textwidth]{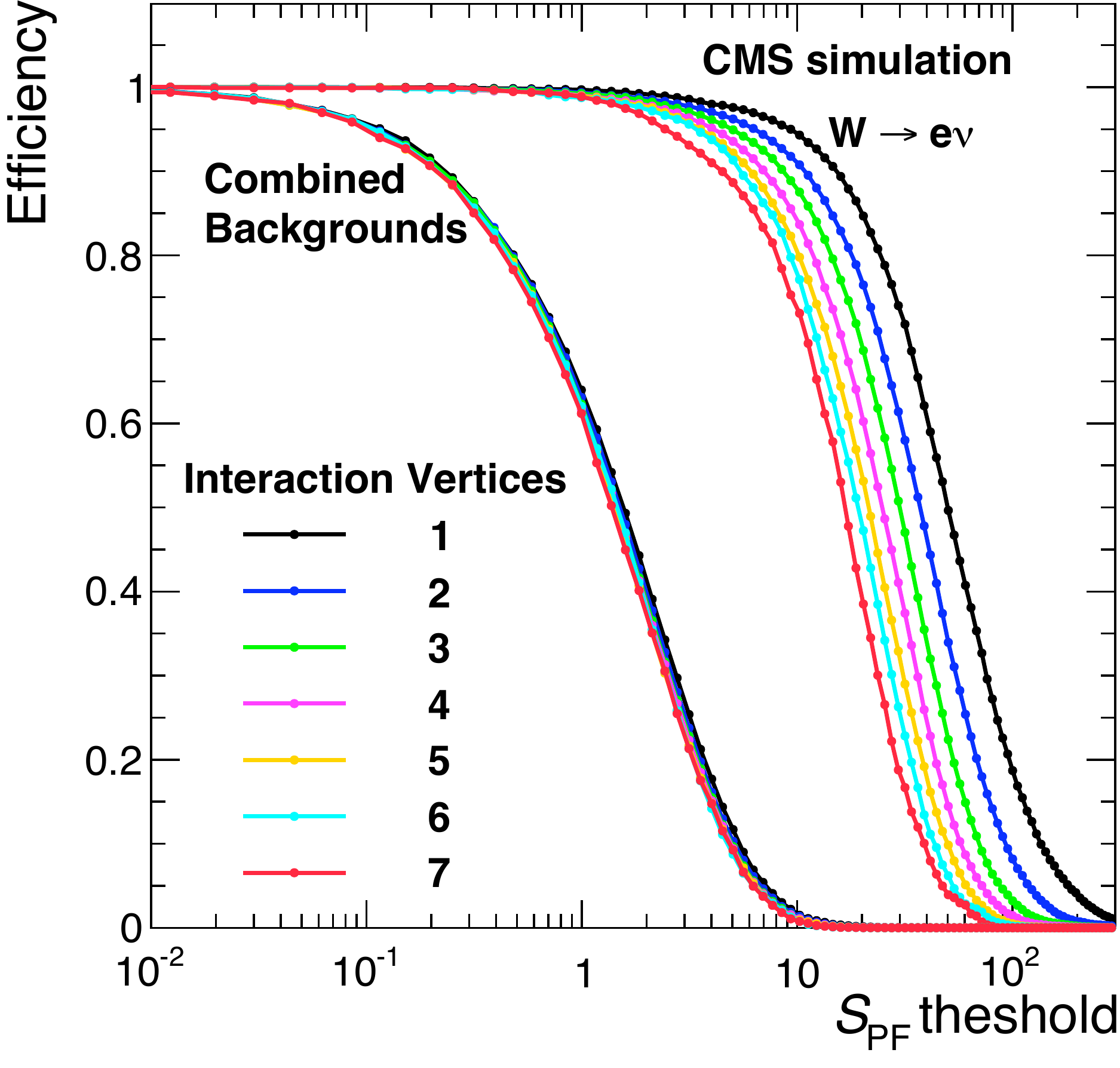}
       \caption{Efficiency versus minimum threshold curves for \Wen signal and for total background
       for different numbers of interaction vertices with a minimum applied \MET threshold (left) and a
       minimum applied \signifpf threshold (right).
}
        \label{fig:wenuSigPU}
    \end{center}
\end{figure}

\begin{figure}[htbp]
    \begin{center}
        \includegraphics[width=0.48\textwidth]{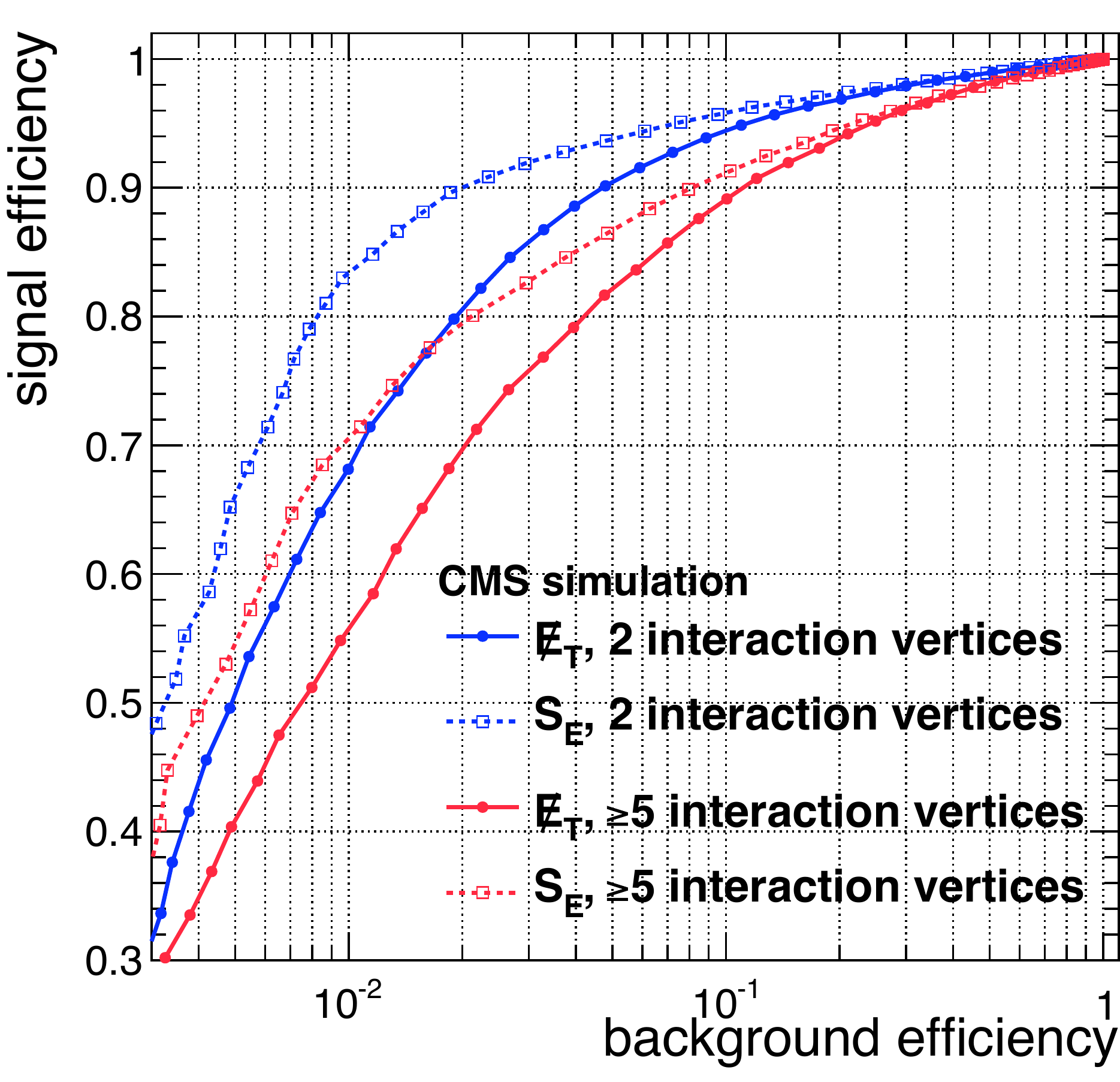}
       \caption{
Efficiency curves for \Wen signal versus backgrounds varying the minimum value
       of $\etmiss$ (solid lines) and of \signifpf (dotted lines) for events with two interaction vertices (blue) or at least five
       interaction vertices (red).}
        \label{fig:wenuSigPU2}
    \end{center}
\end{figure}

Figures~\ref{fig:wenuSigPU} and ~\ref{fig:wenuSigPU2} contrast the behaviour of signal and total background efficiencies for
minimum \MET or \signifpf thresholds for different numbers of interaction vertices (pile-up) in
simulation.  The jets and  $\gamma +$ jets backgrounds, which have no genuine \met,
dominate.
The background contribution at higher \MET grows
as pile-up increases, while the \signifpf levels remain quite stable.  As a result, a background
subtraction based on extrapolation of \MET will be sensitive to the modeling of pile-up,
while one based on extrapolation of \signifpf would not.
As one can see from the signal versus background efficiency curves shown in Fig.
~\ref{fig:wenuSigPU2},
differentiation of signal from background degrades for both \MET and \signifpf as pile-up increases.  Regardless of the
amount of pile-up, however,
\signifpf always provides a superior signal to background ratio compared to \MET.

\clearpage

\section{Conclusions}
\label{sec:conclusions}
In conclusion, we studied
\vecmet as measured by the CMS detector with pp collisions at $\sqrt{s}=7$ TeV
during the 2010 run of the LHC.  
We have determined the \vecmet scale, resolution, tails, and the degradation of the \vecmet performance
due to pile-up.
We find that we are able to remove
most sources of anomalous energies that produce artificial, large \met.
The measured \met\ scale agrees with the 
expectations of the detector simulation, but the resolution is degraded by 10\%  in data.
CMS has three different algorithms for calculating \met.
Algorithms using tracker information have an improved resolution, and
the use of a global particle-flow event reconstruction gives the best resolution.
We find that pile-up interactions
contribute to the degradation of the \met\ resolution but have
little effect on the scale of a component of \met.  We also find that we can model its effects
with a simple parameterization.

One of the most important uses of \met is to distinguish between
genuine \met produced by weakly interacting particles and artificial \met
from detector resolutions.
An algorithm, called \met significance, for 
separating genuine \met from artificial \met,
is shown to perform better than traditional variables
such as \met  alone or \met divided by the square root of the \sumet .

Because of the demonstrated good measurement of \vecmet, 
the CMS detector is ready to be used for a variety
of precision physics measurements, such as studies of the W boson, the top quark, and
searches for new neutral, weakly interacting particles.

\section*{Acknowledgements}
\hyphenation{Bundes-ministerium Forschungs-gemeinschaft Forschungs-zentren} We wish to congratulate our colleagues in the CERN accelerator departments for the excellent performance of the LHC machine. We thank the technical and administrative staff at CERN and other CMS institutes. This work was supported by the Austrian Federal Ministry of Science and Research; the Belgium Fonds de la Recherche Scientifique, and Fonds voor Wetenschappelijk Onderzoek; the Brazilian Funding Agencies (CNPq, CAPES, FAPERJ, and FAPESP); the Bulgarian Ministry of Education and Science; CERN; the Chinese Academy of Sciences, Ministry of Science and Technology, and National Natural Science Foundation of China; the Colombian Funding Agency (COLCIENCIAS); the Croatian Ministry of Science, Education and Sport; the Research Promotion Foundation, Cyprus; the Estonian Academy of Sciences and NICPB; the Academy of Finland, Finnish Ministry of Education, and Helsinki Institute of Physics; the Institut National de Physique Nucl\'eaire et de Physique des Particules~/~CNRS, and Commissariat \`a l'\'Energie Atomique et aux \'Energies Alternatives~/~CEA, France; the Bundesministerium f\"ur Bildung und Forschung, Deutsche Forschungsgemeinschaft, and Helmholtz-Gemeinschaft Deutscher Forschungszentren, Germany; the General Secretariat for Research and Technology, Greece; the National Scientific Research Foundation, and National Office for Research and Technology, Hungary; the Department of Atomic Energy, and Department of Science and Technology, India; the Institute for Studies in Theoretical Physics and Mathematics, Iran; the Science Foundation, Ireland; the Istituto Nazionale di Fisica Nucleare, Italy; the Korean Ministry of Education, Science and Technology and the World Class University program of NRF, Korea; the Lithuanian Academy of Sciences; the Mexican Funding Agencies (CINVESTAV, CONACYT, SEP, and UASLP-FAI); the Pakistan Atomic Energy Commission; the State Commission for Scientific Research, Poland; the Funda\c{c}\~ao para a Ci\^encia e a Tecnologia, Portugal; JINR (Armenia, Belarus, Georgia, Ukraine, Uzbekistan); the Ministry of Science and Technologies of the Russian Federation, and Russian Ministry of Atomic Energy; the Ministry of Science and Technological Development of Serbia; the Ministerio de Ciencia e Innovaci\'on, and Programa Consolider-Ingenio 2010, Spain; the Swiss Funding Agencies (ETH Board, ETH Zurich, PSI, SNF, UniZH, Canton Zurich, and SER); the National Science Council, Taipei; the Scientific and Technical Research Council of Turkey, and Turkish Atomic Energy Authority; the Science and Technology Facilities Council, UK; the US Department of Energy, and the US National Science Foundation.

Individuals have received support from the Marie-Curie programme and the European Research Council (European Union); the Leventis Foundation; the A. P. Sloan Foundation; the Alexander von Humboldt Foundation; the Associazione per lo Sviluppo Scientifico e Tecnologico del Piemonte (Italy); the Belgian Federal Science Policy Office; the Fonds pour la Formation \`a la Recherche dans l'Industrie et dans l'Agriculture (FRIA-Belgium); and the Agentschap voor Innovatie door Wetenschap en Technologie (IWT-Belgium).

\clearpage
\bibliography{auto_generated}   
\clearpage
\appendix
\section{Appendix: Optimization of \texorpdfstring{\bigmet}{MET} Corrections}
\label{sec:Corrections}
In this section, we describe the optimization of the parameters used
in the type-I and type-II corrections to \vecmet.

\label{sec:typeICorrections}

For
type-I \vecmet, the \pt\ threshold used
to select the jets that receive the correction
was optimized by examining its effect on the \vecmet resolution and scale,
and led to our choice of
10 GeV 
for \pfmet.
We have optimized the \pt\ threshold to obtain the best \met\ scale and the best \met\ resolution at the same time, under the constraint that
very low-\pt\ jets should not be included in the calculation because their energy corrections have large uncertainties and can degrade the \met\ performances. The use of low \pt\ jets also makes the measurement more sensitive to the pile-up.

\begin{figure}[!h]
    \begin{center}
      { \includegraphics[height=0.4\textwidth]{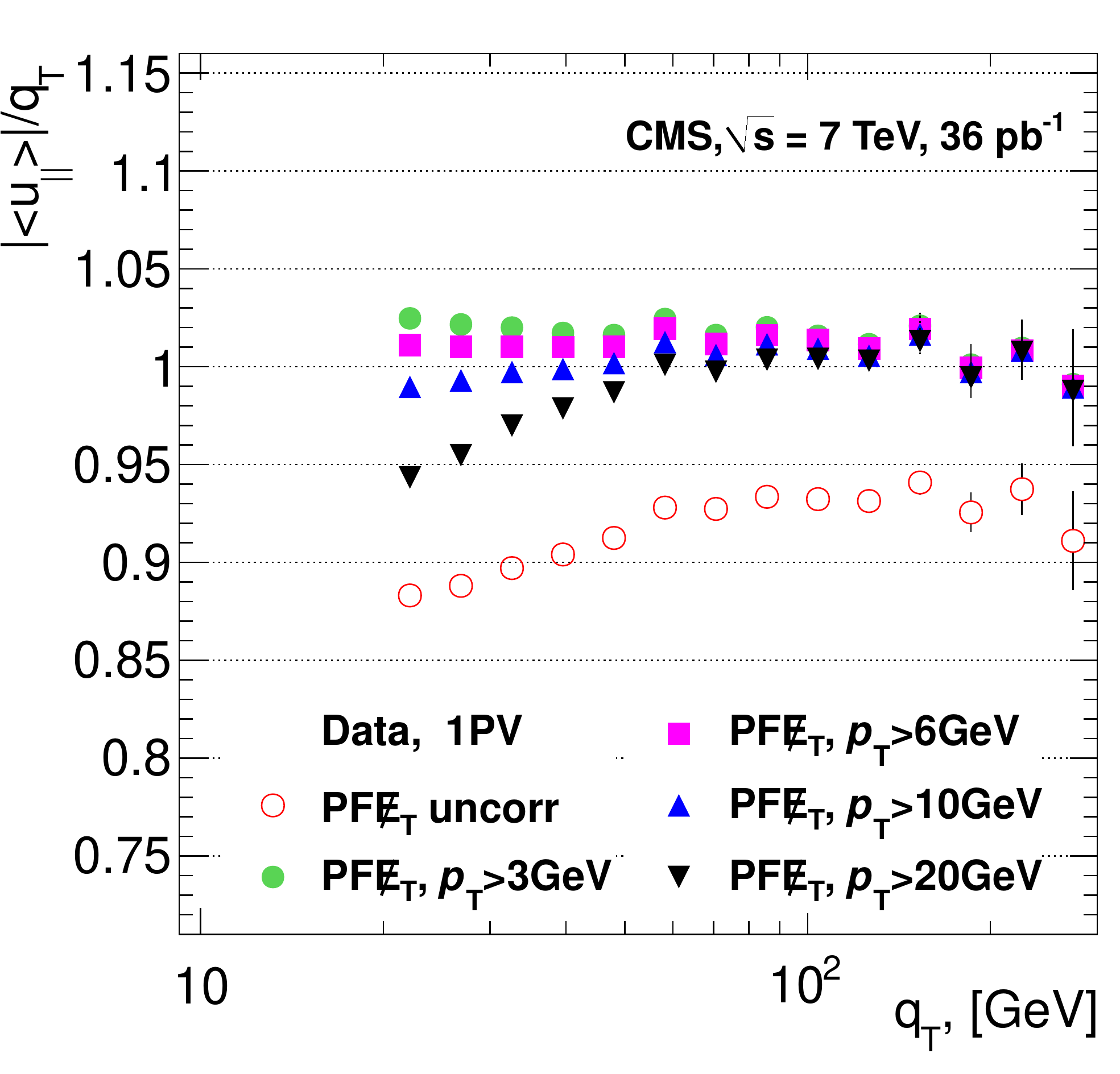} }
      { \includegraphics[height=0.4\textwidth]{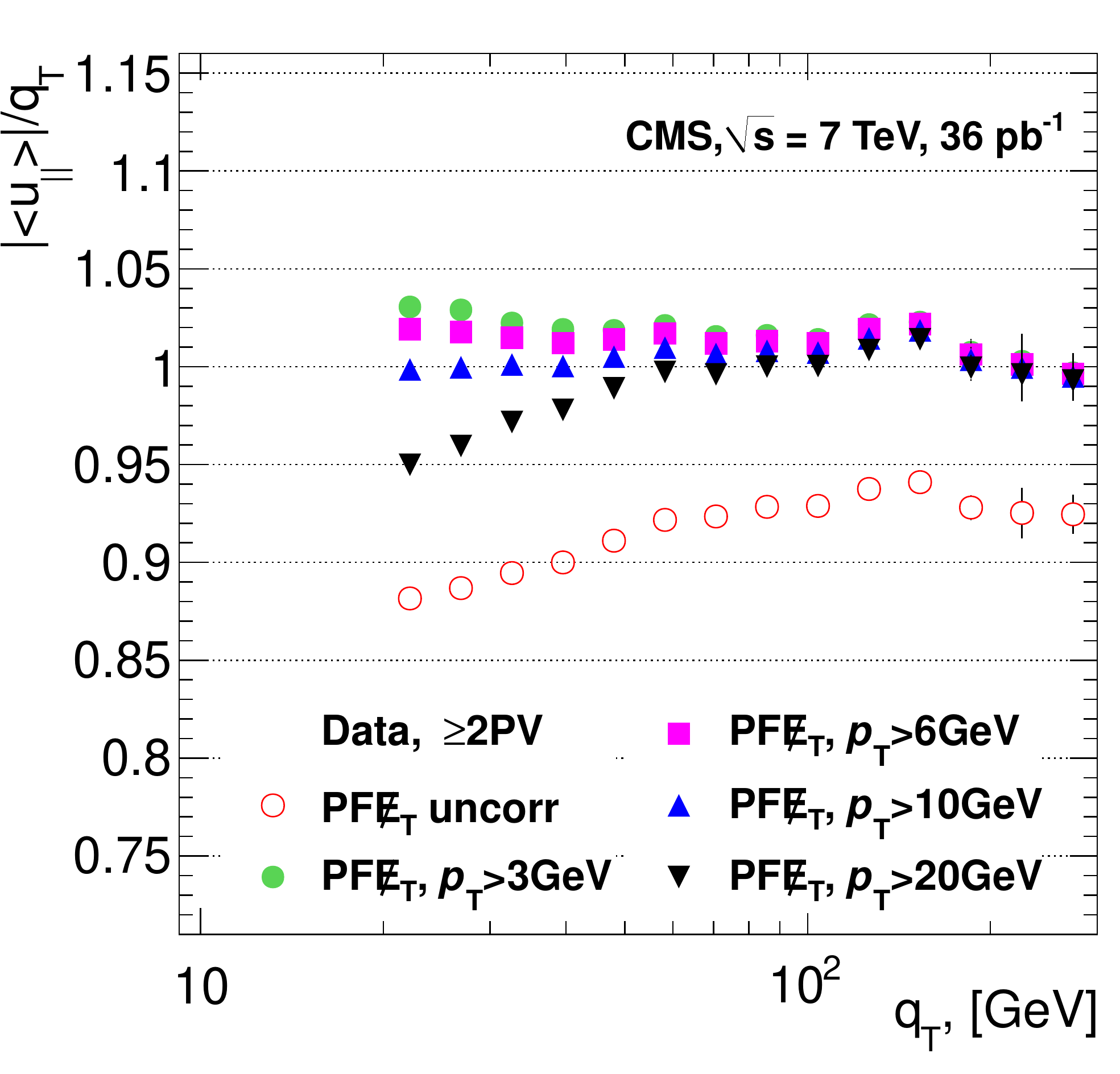} }
      \end{center}
	\caption{Response for \pfmet\ versus $\gamma$ \qt for
various jet corrections thresholds for $\gamma$ candidate 
events with 1 primary vertex
(left) and more than 1 vertex (right).
}
       \label{f:TIgam1}
    \end{figure}

Figure ~\ref{f:TIgam1}
shows the \pfmet\ response versus \qt for $\gamma$ candidate events
for various jet thresholds for events with and without pile-up.
As can be seen from these distributions, the type-I corrections substantially
improve the \met scale, the dependence on the threshold is small
for thresholds $\le$ 10 GeV, and is independent of the number of additional
pile-up interactions.

\begin{figure}[!h]
    \begin{center}
      { \includegraphics[angle=0,height=0.4\textwidth]{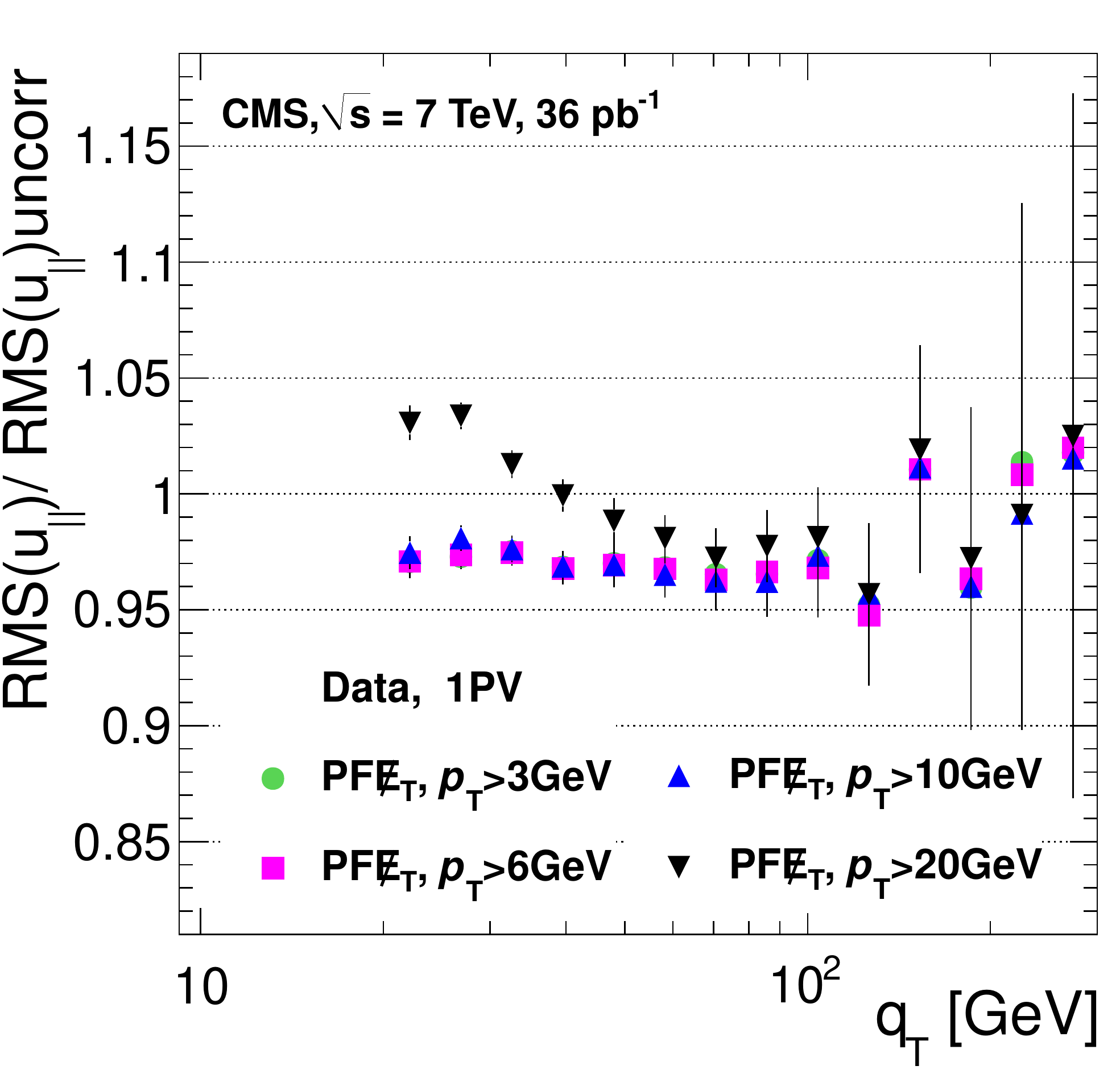} }
      { \includegraphics[angle=0,height=0.4\textwidth]{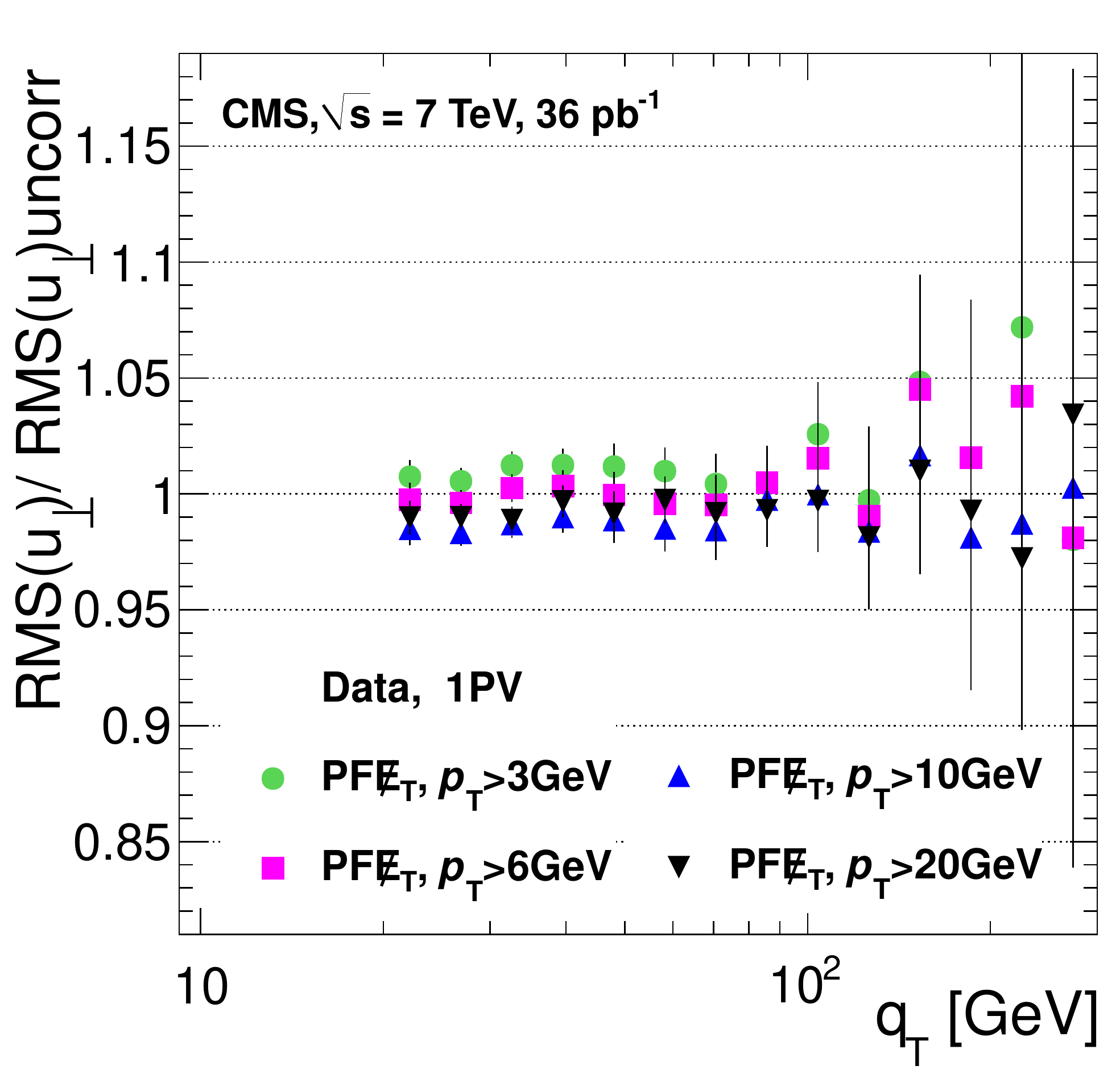} }\\
      { \includegraphics[angle=0,height=0.4\textwidth]{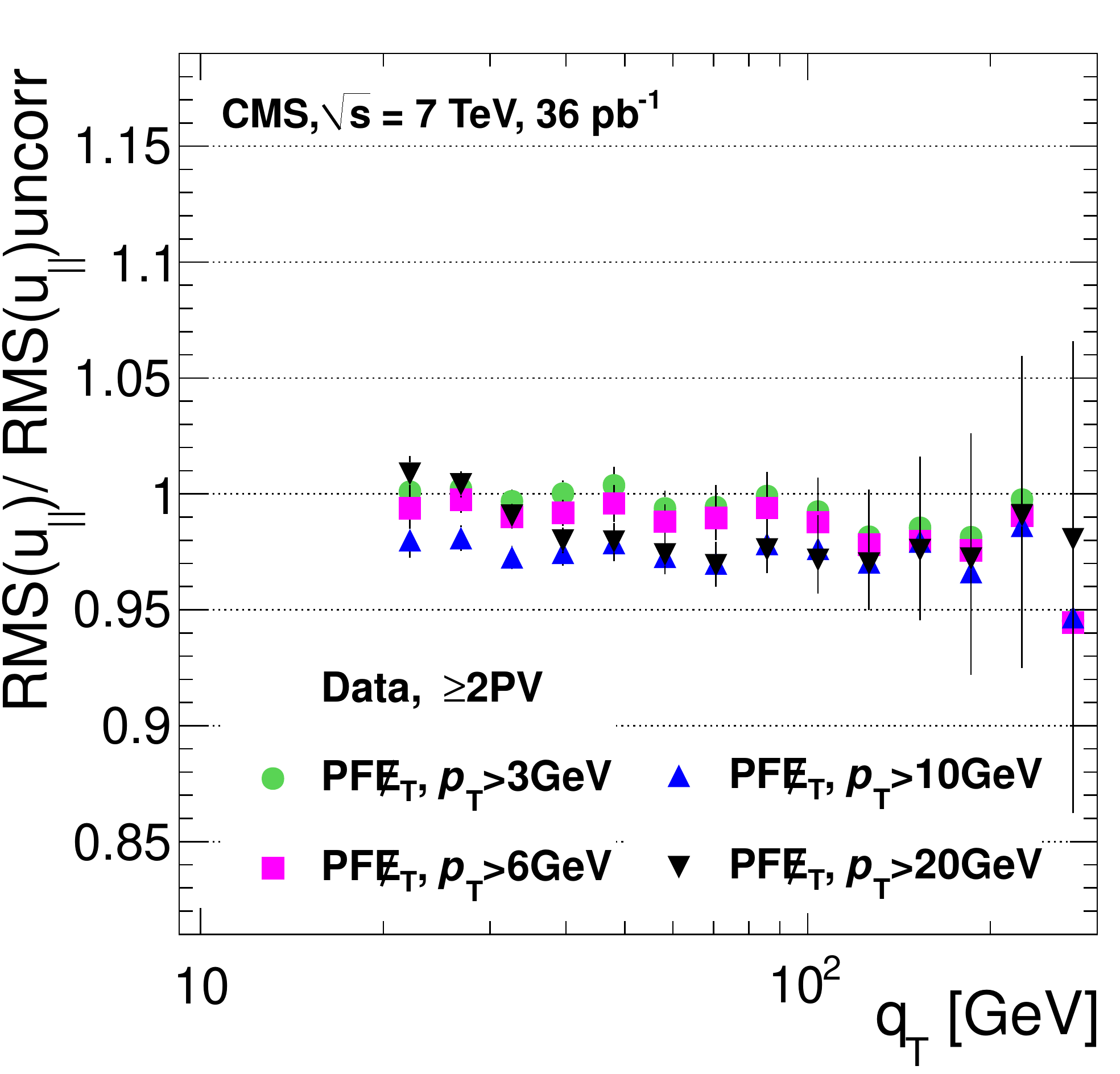} }
      { \includegraphics[angle=0,height=0.4\textwidth]{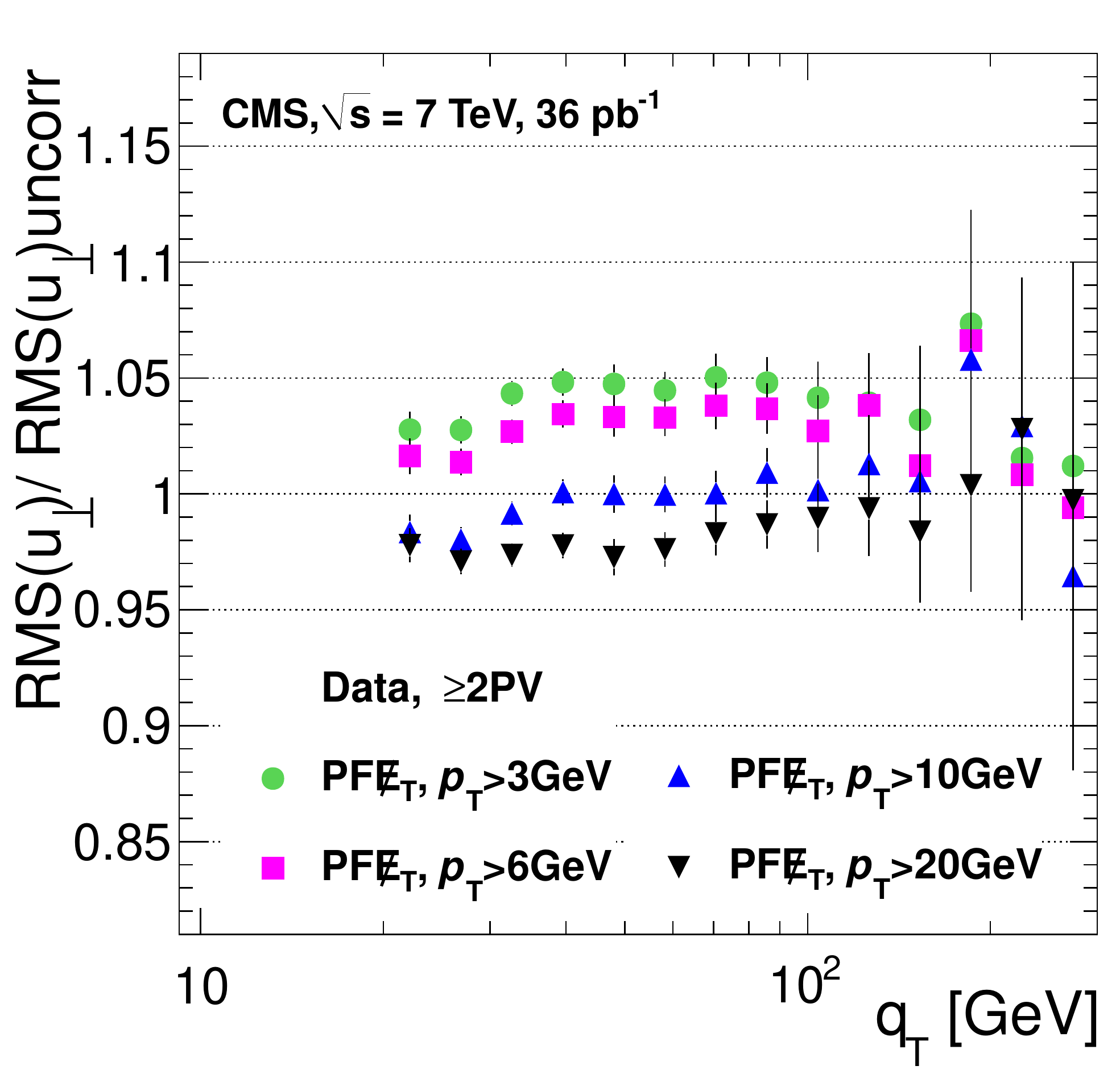} }
      \end{center}
	\caption{Ratio of \pfmet resolution with type-I corrections to the
the \pfmet resolution without any type-I corrections
for parallel (left) and perpendicular (right)
components of \pfmet\ versus $\gamma$ \qt for
various jet corrections thresholds for $\gamma$ candidate 
events with 1 primary vertex
(top) and more than 1 vertex (bottom) after correction for scale, using
scale corrections similar to those shown in Fig. \ref{f:PJresponse}.
}
       \label{f:TIgam2}
    \end{figure}

Figure \ref{f:TIgam2} shows the ratio 
of the \pfmet resolution with type-I corrections
to the \pfmet resolution without any type-I corrections
for the parallel and perpendicular
components of \pfmet\ versus \qt for
$\gamma$ candidate events for various jet thresholds
and for events with and without pile-up.
The optimal resolution for 1 vertex
events is obtained for the chosen 10 GeV threshold, and for this threshold,
the degradation with pile-up is not large.

The same study has been performed in \Zmm candidate events.
The 10 GeV threshold is also 
the best threshold to combine 
the improvement of the \met\ scale and \met\ resolution
in these events.

\label{sec:typeIICorrections}

Type-II corrections were not used for analyses done using the 2010 data.
However, studies were done to optimize the parameters for future use.
The type-II correction is obtained using a $Z\to ee$ data sample.
$Z$ events constitute an ideal sample for determining
the type-II correction, as $Z$'s are generally produced with low \qt
and the recoil is often dominated by unclustered energy.
We take the vector sum of momenta defined for all calorimeter towers
or PF particles
not corrected by the type-I correction as a single object denoted
$\vec{U}$.  The measured value of $\vec{U}$ is defined as:

\begin{equation}
\vec{U}_{\rm T, meas}=-\metvec_{\rm uncorr}
-\sum{\vec{p}_{\rm T, jet,uncorr}}-\sum{\vec{p}_{\rm T, ele,meas}},
\end{equation}

where
$\metvec_{\rm uncorr}$ is the uncorrected \met,
$\vec{p}_{\rm T, jets,uncorr}$ is the momentum of uncorrected jets, and
$\vec{p}_{\rm T, ele,meas}$ is the momentum of the measured electrons. The
sum is over all jets with the corrected $p_{\rm T}>20$ GeV for \calomet and $p_{\rm T}>10$
GeV for \pfmet.

The type-II correction for the unclustered energy $\vec{U}_{\rm meas}$ is
obtained by selecting events without any reconstructed jets
using the
correlation between $\vec{U}_{\rm meas}$ and the \qt of the Z
measured from the electrons.

As the direction of the $\vec{U}_{\rm meas}$ may differ from the
direction of \qt due to noise, underlying event, etc.,
the parallel component of $\vec{U}_{\rm meas}$ projected on the
direction of \qt, $\vec{U}_{\rm meas,||}$, is used for
the derivation of the correction. The response of the unclustered
energy is then defined as $R(U_{\rm T,  meas})=U_{\rm T,  meas,||}  / \qt$.
The obtained correction factor, $U_{\rm scale}$, is
parametrized as $2.3 + 1.9 \, \exp(-0.2 \cdot U_{\rm T,  meas})$ for
\calomet. For \pfmet\ the obtained correction leads to a $U_{\rm
scale}$ of $1.4$. These values obtained from data correspond well to the MC
expectation.

\begin{figure}[!htbp]
    \begin{center}
      \includegraphics[height=0.4\textwidth]{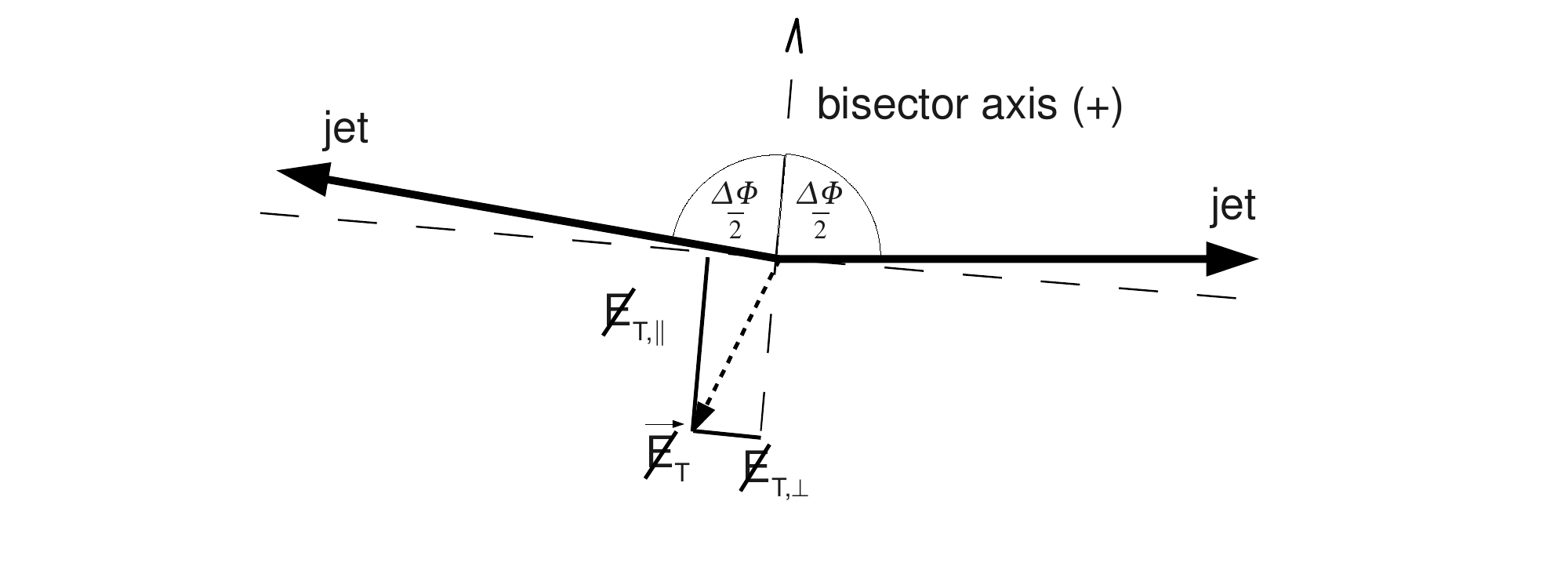}
      \end{center}
        \caption{Kinematic definitions for dijet events. }
       \label{fig:typeII_bisectordefinition}
    \end{figure}

\begin{figure}[htbp]
    \begin{center}
      {\includegraphics[angle=90,height=0.4\textwidth]{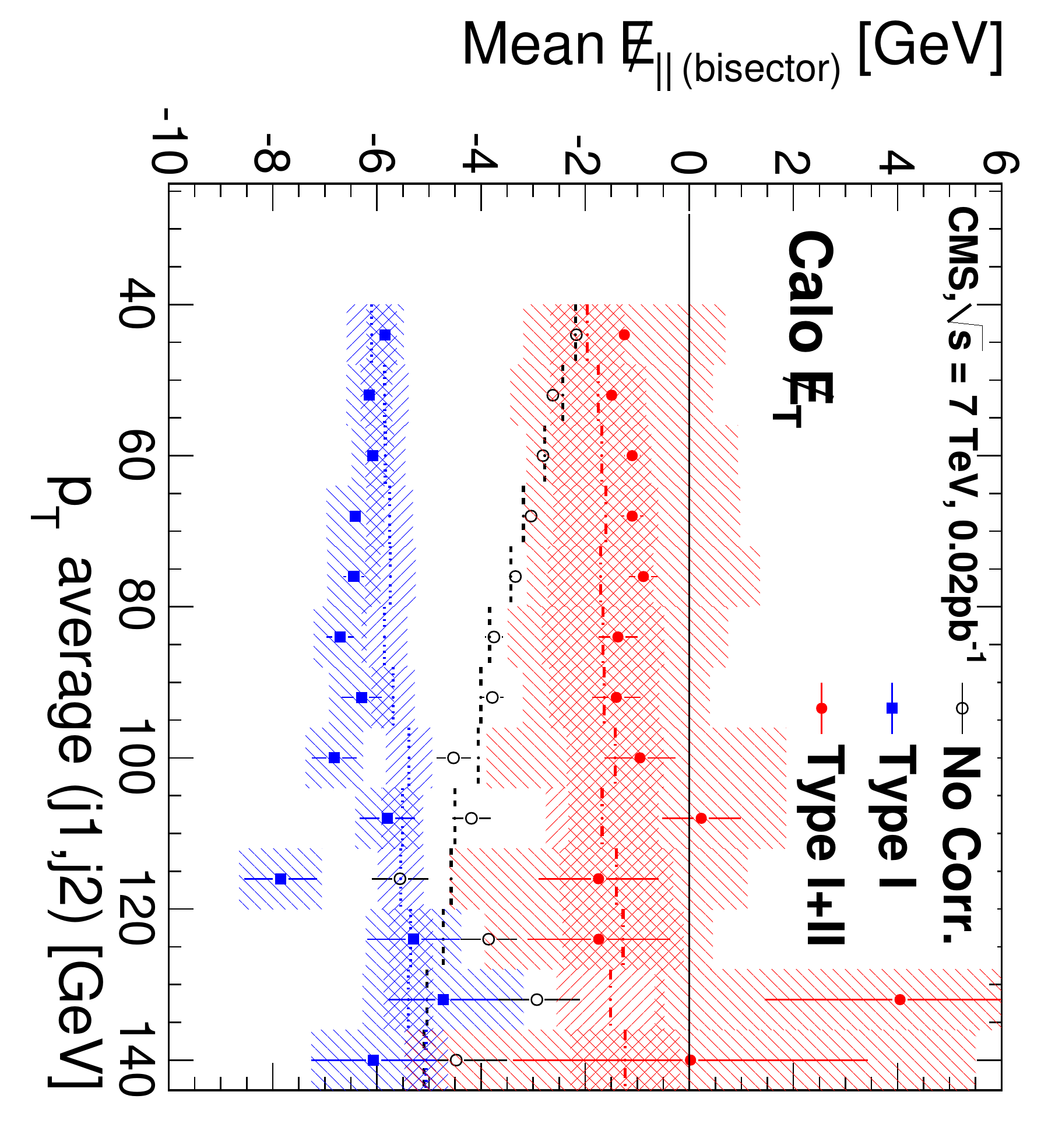}
      }
       {
       \includegraphics[angle=90,height=0.4\textwidth]{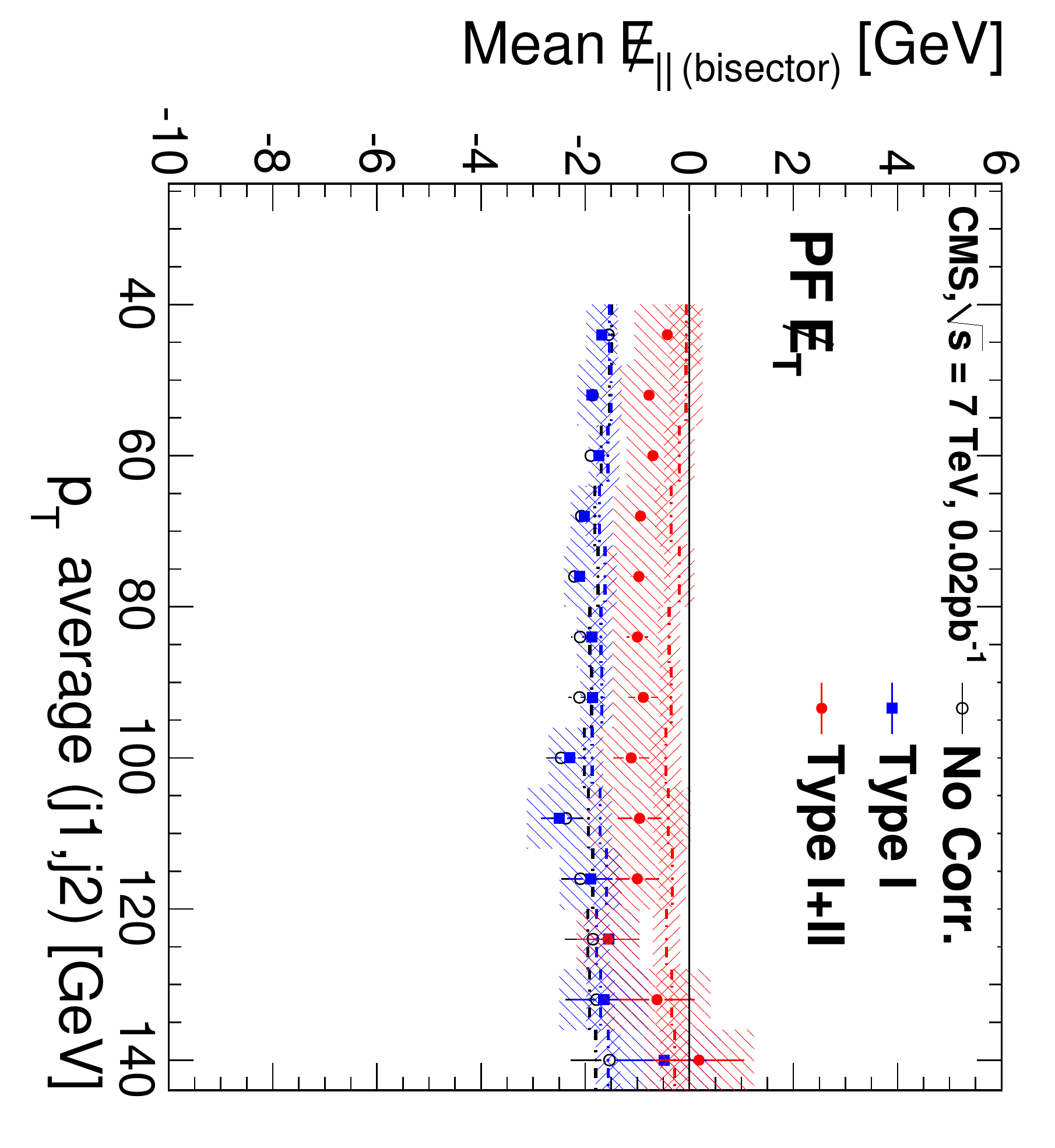}
       }
      \end{center}
        \caption{Mean
values of $\met_{||}$ as a function of average $p_{\rm T}$ of the leading two
jets ($p_{\rm T}^{\rm avg}$) in
dijet events compared with Monte Carlo simulation (dashed lines) for
uncorrected, type-I corrected, and type-II corrected (left) \calomet\ and (right) \pfmet\ .
The blue band on the distribution with type-I corrections represents
the uncertainty due to uncertainties on the jet energy scale.
The red band on the distribution with type-II corrections represents
uncertainties due to uncertainties on the jet energy scale and
statistical uncertainties due to the size of the \Zee sample from which the
correction for unclustered energies
was derived.
}
       \label{fig:typeIandII_bisector}
    \end{figure}

One of the ways to validate the type-II corrected \met\ is to look at \metvec\
decomposed into two components in dijet
events. The decomposition is based on the dijet bisector axis, which divides the
azimuthal opening angle between the leading two jets in two as illustrated in
Fig.~\ref{fig:typeII_bisectordefinition}. Positive $\met_{||}$ points towards
the smaller opening angle of the leading two jets, and positive $\met_{\perp}$ points towards the more central jet
(smaller $|\eta|$).
The distribution of $\met_{\perp}$ is symmetric for all correction levels,
while by definition there is a slight asymmetry in $\met_{||}$. This is due to
the fact that the bisector axis always points towards the opening angle of the leading
two jets, while the \met\ tends to point in the opposite direction.  The type-I
correction introduces a more significant asymmetry in the $\met_{||}$ distribution, because
it produces artificial \met\ in the direction opposite to the dijet
opening angle.  The type-II correction, however, calibrates the rest
of the calorimeter energies, and makes the $\met_{||}$ distribution
nearly symmetric again.
This trend may
be seen in Fig.~\ref{fig:typeIandII_bisector} which shows the mean
values of $\met_{||}$ as a function of $p_{\rm T}^{\rm avg}
=(p_{\rm T}^{\rm jet1}+p_{\rm T}^{\rm jet2})/2$ for uncorrected, type-I
corrected, and type-II corrected Calo- and \pfmet\ in events containing at
least two jets with $p_{\rm T}>40$ GeV and $|\eta|<3$.

\cleardoublepage\section{The CMS Collaboration \label{app:collab}}\begin{sloppypar}\hyphenpenalty=5000\widowpenalty=500\clubpenalty=5000\textbf{Yerevan Physics Institute,  Yerevan,  Armenia}\\*[0pt]
S.~Chatrchyan, V.~Khachatryan, A.M.~Sirunyan, A.~Tumasyan
\vskip\cmsinstskip
\textbf{Institut f\"{u}r Hochenergiephysik der OeAW,  Wien,  Austria}\\*[0pt]
W.~Adam, T.~Bergauer, M.~Dragicevic, J.~Er\"{o}, C.~Fabjan, M.~Friedl, R.~Fr\"{u}hwirth, V.M.~Ghete, J.~Hammer\cmsAuthorMark{1}, S.~H\"{a}nsel, M.~Hoch, N.~H\"{o}rmann, J.~Hrubec, M.~Jeitler, W.~Kiesenhofer, M.~Krammer, D.~Liko, I.~Mikulec, M.~Pernicka, H.~Rohringer, R.~Sch\"{o}fbeck, J.~Strauss, A.~Taurok, F.~Teischinger, P.~Wagner, W.~Waltenberger, G.~Walzel, E.~Widl, C.-E.~Wulz
% ...
\vskip\cmsinstskip
% ...
1:~~Also at CERN, European Organization for Nuclear Research, Geneva, Switzerland\\
% ...

\end{sloppypar}
\end{document}